\documentclass{aa}  
\usepackage[english]{babel}
\usepackage{amssymb}
\usepackage{amsmath}
\usepackage{bm}
\usepackage{subfigure}
\usepackage{natbib}
\usepackage{graphicx}
\usepackage{times}
\usepackage{hyperref}
\usepackage{color}
\usepackage{pifont}

\newcommand{\EQ}{\begin{equation}}
\newcommand{\EE}{\end{equation}}
\newcommand{\EQA}{\begin{eqnarray}}
\newcommand{\EEA}{\end{eqnarray}}

\newcommand{\pd}{\partial}
\newcommand{\DIV}{\bm{\nabla} \cdot }

\newcommand{\urms}{u_{\rm rms}}
\newcommand{\Urms}{U_{\rm rms}}
\newcommand{\brms}{B_{\rm rms}}

\newcommand{\kef}{k_{\rm f}}
\newcommand{\Beq}{B_{\rm eq}}

\newcommand{\Pm}{{\rm Pm}}
\newcommand{\Pra}{{\rm Pr}}
\newcommand{\Ray}{{\rm Ra}}
\newcommand{\Tay}{{\rm Ta}}

\newcommand{\Roc}{{\rm Ro_{\rm c}}}
\newcommand{\Co}{{\rm Co}}

\newcommand{\Rey}{{\rm Re}}

\newcommand{\Ek}{{\rm Ek}}

\def\onehalf{{\textstyle{1\over2}}}
\def\onethird{{\textstyle{1\over3}}}

\begin{document} 

\title{Large-scale dynamos in rapidly rotating plane layer convection}
\titlerunning{Large-scale dynamos in rapidly rotating convection}

\author{P. J. Bushby\inst{1}
\and P. J. K\"apyl\"a\inst{2,3,4,5}
\and Y. Masada\inst{6}
\and A. Brandenburg \inst{5,7,8,9}
\and B. Favier\inst{10}
\and C. Guervilly\inst{1}
\and M. J. K\"apyl\"a\inst{4,3}}

\institute{School of Mathematics and Statistics, Newcastle University, Newcastle upon Tyne, NE1 7RU, UK \\ \email{paul.bushby@ncl.ac.uk}
\and Leibniz-Institut f\"ur Astrophysik Potsdam, An der Sternwarte 16, D-11482 Potsdam, Germany
\and ReSoLVE Centre of Excellence, Department of Computer Science, Aalto University, PO Box 15400, FI-00076 Aalto, Finland
\and Max-Planck-Institut f\"ur Sonnensystemforschung, Justus-von-Liebig-Weg 3, D-37077 G\"ottingen, Germany
\and Nordita, KTH Royal Institute of Technology and Stockholm University, Roslagstullsbacken 23, SE-10691 Stockholm, Sweden
\and Department of Physics and Astronomy, Aichi University of Education; Kariya, Aichi 446-8501, Japan
\and JILA and Department of Astrophysical and Planetary Sciences, Box 440, University of Colorado, Boulder, CO 80303, USA
\and Department of Astronomy, AlbaNova University Center, Stockholm University, SE-10691 Stockholm, Sweden
\and Laboratory for Atmospheric and Space Physics, 3665 Discovery Drive, Boulder, CO 80303, USA
\and Aix Marseille Univ, CNRS, Centrale Marseille, IRPHE UMR 7342, Marseille, France
}

\date{\today}

\abstract
{Convectively-driven flows play a crucial role in the dynamo processes
that are responsible for producing magnetic activity in stars and planets.
It is still not fully understood why many astrophysical magnetic
fields have a significant large-scale component.}
{Our aim is to investigate the dynamo properties of compressible convection 
in a rapidly rotating Cartesian domain, focusing upon a parameter regime in 
which the underlying hydrodynamic flow is known to be unstable to a
large-scale vortex instability.}
{The governing equations of three-dimensional nonlinear magnetohydrodynamics 
(MHD) are solved numerically. Different numerical schemes are compared 
and we propose a possible benchmark case for other similar codes.}
{In keeping with previous related studies, we find that convection in this
parameter regime can drive a large-scale dynamo. 
The components of the mean horizontal magnetic field oscillate, leading 
to a continuous overall rotation of the mean field.
Whilst the large-scale vortex instability dominates the early 
evolution of the system, it is suppressed by the magnetic field
and makes a negligible contribution to the mean electromotive force that is
responsible for driving the large-scale dynamo. 
The cycle period of the dynamo is comparable to the ohmic decay time, 
with longer cycles for dynamos in convective systems that are 
closer to onset. 
In these particular simulations, large-scale dynamo action is found only when
vertical magnetic field
boundary conditions are adopted at the upper and lower boundaries.
Strongly modulated large-scale dynamos 
are found at higher Rayleigh numbers, with periods of reduced activity 
(``grand minima''-like events) occurring during transient phases in which the 
large-scale vortex temporarily re-establishes itself, before being 
suppressed again by the magnetic field.}
{}


\maketitle

\section{Introduction}
\label{sect:intro}
In a hydromagnetic dynamo the motions in an electrically conducting
fluid continuously sustain a magnetic field against the action of ohmic
dissipation.
Most astrophysical magnetic fields, including those in stars and planets,
are dynamo-generated.
In numerical simulations, turbulent motions almost invariably produce an
intermittent magnetic field distribution that is correlated on the scale
of the flow.
However, most astrophysical objects exhibit magnetism that is organised
on much larger scales.
These large-scale fields might be steady or they may exhibit some 
time dependence.
In the case of the Sun, for example, observations of surface magnetism
\citep[e.g.][]{Stix2002} indicate the presence of a large-scale
oscillatory magnetic field in the solar interior that changes sign
approximately every 11 years.
Depending upon their age and spectral type, similar
magnetic activity cycles can also be observed in other stars
\citep[e.g.][]{Brandenburgetal1998}. 
Whilst appearing to be comparatively steady on these time-scales, the 
Earth's predominantly dipolar field does exhibit long-term variations, 
occasionally even reversing its magnetic polarity \citep[although 
rather irregular, a typical time span between reversals is of the order 
of $3\times 10^5$ years, see, e.g.,][]{Jones2011}.
Our understanding of the physical processes that are responsible for the
production of large-scale magnetic fields in astrophysical objects relies
heavily on mean-field dynamo theory \citep[e.g.][]{Moffatt1978}.
This approach, in which the small-scale physics is parameterised in
a plausible way, has had considerable success.
However, despite much recent progress in this area, there are
apparently contradictory findings, indicating that we still do not
completely understand why large-scale magnetic fields appear to be so
ubiquitous in astrophysics.

Convectively-driven flows are a feature of stellar and planetary
interiors, where the effects of rotation can often play an
important dynamical role via the Coriolis force. In the rapidly rotating
limit, convective motions tend to be helical, leading to the expectation 
of a strong $\alpha$-effect (an important regenerative term for large-scale 
magnetic fields in mean-field dynamo theory, usually a parameterised 
effect in simplified mean-field models).
Many theoretical studies have therefore been motivated 
by the question of whether
rotationally influenced convective flows can drive a large-scale dynamo
in a fully self-consistent (i.e.\ non-parameterised) manner.

Using the Boussinesq approximation, in which the electrically conducting
fluid is assumed to be incompressible, \citet{ChildressSoward1972}
were the first to demonstrate that a rapidly rotating plane layer
of weakly convecting fluid was capable of sustaining a large-scale
dynamo \citep[see also][]{Soward1974}.
To be clear about terminology (here, and in what follows), we 
describe a plane layer dynamo as ``large-scale'' if it produces a magnetic
field with a significant horizontally averaged component
\citep[such a magnetic field can also be described as ``system-scale'',
see, e.g.,][]{Tobiasetal2011}.
Building on the work of \citet{ChildressSoward1972} and \citet{Soward1974},
many subsequent studies have explored
the dynamo properties of related Cartesian Boussinesq models
\citep[][]{FautrelleChildress1982,MeneguzziPouquet1989,StPierre1993,
JonesRoberts2000,RotvigJones2002,StellmachHansen2004,
CattaneoHughes2006,FavierProctor2013,Calkinsetal2015}
as well as the corresponding weakly-stratified system
\citep[][]{MizerskiTobias2013}.
However, whilst near-onset rapidly rotating convection does produce
a large-scale dynamo \citep[see, e.g.,][]{StellmachHansen2004}, only
small-scale dynamo action is observed in the rapidly rotating turbulent
regime \citep[][]{CattaneoHughes2006}, contrary to the predictions
of mean-field dynamo theory.
Indeed, \citet{Tilgner2014} was able to identify an approximate parametric
threshold (based on the Ekman number and the magnetic Reynolds number)
above which small-scale dynamo action is preferred: low levels of
turbulence and a rapid rotation rate are found to be essential for a 
large-scale convectively-driven dynamo.

There have been many fewer studies of the corresponding fully compressible 
system, so parameter space has not yet been explored to the same 
extent in this case. 
\citet{Kapylaetal2009} were the first to demonstrate that it is possible to
excite a large-scale dynamo in rapidly rotating compressible convection
at modestly supercritical values of the Rayleigh number (the key parameter
controlling the vigour of the convective motions).
However, it again appears to be much more difficult to drive a large-scale 
dynamo further from convective onset, with small-scale dynamos
typically being reported in this comparatively turbulent regime 
\citep[][]{FavierBushby2013}.
This is in agreement with
the Boussinesq studies, such as \citet{CattaneoHughes2006}.
Moreover, as noted by \citet{Guervillyetal2015}, the
transition from large-scale to small-scale dynamos in these compressible
systems appears to occur in a similar region of parameter space to the
Boussinesq transition that was identified by \citet{Tilgner2014}.

In rapidly rotating Cartesian domains, hydrodynamic convective flows
just above onset are characterised by a small horizontal spatial scale
\citep[][]{Chandrasekhar1961}.
However, at slightly higher levels of convective driving, these
small-scale motions can become unstable, leading to a large-scale vortical
flow \citep[][]{Chan2007}.
The width of these large-scale vortices is limited by the size of the
computational domain; in such simulations, the corresponding flow field
has a negligible horizontal average (unlike the large-scale magnetic
fields described above).
With increasing rotation rate, \citet{Chan2007} observed a
transition from cooler cyclonic vortices to warmer anticyclones,
and subsequent fully compressible studies have found similar behaviour
\citep[][]{Mantereetal2011,Kapylaetal2011,ChanMayr2013}.
In corresponding Boussinesq calculations, 
\citet{Favieretal2014} and \citet{Guervillyetal2014} found a
clear preference for cyclonic vortices, 
and dominant anticyclones are never observed, although
\citet{Stellmachetal2014} did find states consisting
of cyclones and anticyclones (of comparable magnitude) 
at higher rotation rates.
As noted by \citet{Stellmachetal2014} and \citet{Kunnenetal2016}, 
this large-scale vortex instability is inhibited when
no-slip boundary conditions are adopted at the upper and lower boundaries,
so the formation of large-scale vortices  
depends to some extent upon the use of stress-free boundary conditions.

From the point of view of the convective dynamo problem, this 
large-scale vortex
instability can play a very important role.
In the rapidly rotating Boussinesq dynamo of \citet{Guervillyetal2015},
the large-scale vortex leads directly to the production of
magnetic fields at a horizontal wavenumber comparable to that of the
large-scale vortex.
Although the total magnetic energy in this case is less than $1\%$
of the total kinetic energy, the resultant magnetic field is locally
strong enough to inhibit the large-scale flow.
This temporarily suppresses the dynamo until the magnetic field becomes
weak enough for the vortical instability to grow again.
In some sense, the dynamo in this case switches on and
off as the energy in the large-scale flow fluctuates.

In the fully compressible regime, the large-scale vortex
instability has been shown to produce a different type of dynamo
\citep[][]{Kapylaetal2013,MasadaSano2014a,MasadaSano2014b}.
As in the Boussinesq case
that was considered by \citet{Guervillyetal2015},
the large-scale flow produces a large-scale
magnetic field which exhibits some time-dependence.
However, whilst the large-scale vortex is again suppressed once
the magnetic field becomes dynamically significant, these dynamos
are able to persist without the subsequent regeneration of these
vortices \citep[suggesting that these dynamos may be a compressible
analogue of the dynamo considered by][albeit operating in the strong field
limit]{ChildressSoward1972}.
Once established, the magnetic energy in these dynamos is comparable
to the kinetic energy of the system, with the horizontal components of
the large-scale magnetic field oscillating in a regular manner, with a
phase shift of approximately $\pi/2$ between the two components, 
leading to a net rotation of the mean horizontal magnetic field.
Although each component of the mean magnetic field certainly
oscillates, because the temporal variation in the mean field
essentially takes the form of a global rotation of the field
orientation, it should be kept in mind that in a suitably corotating
frame the mean field would appear statistically stationary.

The large-scale dynamo that was found by
\citet{Kapylaetal2013} and \citet{MasadaSano2014a,MasadaSano2014b} is
arguably the simplest known example of a 
moderately supercritical
convectively-driven dynamo in a rapidly-rotating Cartesian domain.
However, to achieve this nonlinear magnetohydrodynamical state,
any numerical code must successfully reproduce the large-scale
vortex instability of rapidly rotating hydrodynamic convection in order
to amplify a weak seed magnetic field.
In the nonlinear regime of the dynamo, the resultant
large-scale magnetic field must then be sustained at a level that
is (approximately) in equipartition with the local convective
motions.
As a result, this dynamo is an excellent candidate
for a benchmarking exercise.
Corresponding benchmarks exist for convectively-driven dynamos in
spherical geometry, both for Boussinesq \citep[][]{Christensenetal2001,
Martietal2014} and for anelastic fluids \citep[][]{Jonesetal2011}.
To the best of our knowledge, there is no similar benchmark
for a fully compressible, turbulent, large-scale dynamo.

The main aim of this paper is to further investigate the properties of this
large-scale dynamo, focusing particularly upon the effects of 
varying the rotation rate and the convective driving, as well as the size of
the computational domain. 
We will establish the regions of parameter space in which this dynamo can 
be sustained, looking at the ways in which the dynamo
amplitude and cycle period depend upon the key parameters of the system. 
Most significantly, we will show that it is possible to induce 
strong temporal
modulation in large-scale dynamos of this type by increasing the level of
convective driving at fixed rotation rate. 
Finally, we carry out a preliminary code comparison 
(confirming the accuracy and validity of one particular solution via three 
independent codes) to assess 
whether or not this system could form the basis of a nonlinear
Cartesian dynamo benchmark, possibly involving
broader participation from the dynamo community.
In the next section, we set out our model and describe the numerical codes.
Our numerical results, are discussed in Sect.~\ref{Results}. 
In the final section, we present our conclusions.
The strengths and weaknesses of the proposed benchmark solution are
described in Appendix~\ref{BenchmarkComparison}.

\section{Governing equations and numerical methods}

\subsection{The model setup}
\label{sect:model}

We consider a plane layer of electrically conducting, compressible fluid,
which is assumed to occupy a Cartesian domain of dimensions $0 \le x \le
\lambda d$, $0 \le y \le \lambda d$ and $0\le z \le d$, where $\lambda$ is
the aspect ratio.
This layer of fluid is heated from below, and the whole 
domain rotates rigidly about the vertical axis with constant angular velocity
$\bm{\Omega}=\Omega_0\mathbf{e}_z$.
We define $\mu$ to be the dynamical viscosity of the fluid,
whilst $K$ is the
radiative heat conductivity, $\eta$ is the magnetic diffusivity, $\mu_0$
is the vacuum permeability, whilst $c_{\rm P}$ and $c_{\rm V}$ are the
specific heat capacities at constant pressure and volume respectively (as
usual, we define $\gamma=c_{\rm P}/c_{\rm V})$.
All of these parameters are assumed to be constant, as is the
gravitational acceleration $\bm{g} = -g\bm{e}_{z}$ ($z$ increases upwards).
The evolution of this system is then determined by the equations of
compressible magnetohydrodynamics, which can be expressed in the 
following form,
\begin{equation}
\frac{\pd \bm A}{\pd t} = {\bm U}\times{\bm B} - \eta \mu_0 {\bm J},
\label{equ:AA}
\end{equation}
\begin{equation}
\frac{D \ln \rho}{Dt} = -\bm\nabla\cdot\bm{U},
\end{equation}
\begin{equation}
\frac{D\bm{U}}{Dt} = \bm{g} -2\bm\Omega\times\bm{U}+\frac{1}{\rho}
\left(2\mu \bm\nabla \cdot \bm{\mathsf{S}}-\bm\nabla p + 
{\bm J}\times{\bm B}\right),
\end{equation}
\begin{equation}
T\frac{D s}{Dt} = \frac{1}{\rho}\left(K\bm\nabla^2 T 
+ 2\mu \bm{\mathsf{S}}^2 + \mu_0 \eta{\bm J}^2 \right),
\label{equ:ss}
\end{equation}
where ${\bm A}$ is the magnetic vector potential, $\bm{U}$ is the
velocity, ${\bm B} =\bm\nabla\times{\bm A}$ is the magnetic field,
${\bm J} =\bm\nabla\times{\bm B}/\mu_0$ is the current density,
$\rho$ is the density, $s$ is the specific entropy,
$T$ is the temperature, $p$ is the pressure and
$D/Dt = \pd/\pd t + \bm{U} \cdot \bm\nabla$ denotes the
advective time derivative.
The fluid obeys the ideal gas law with $p=(\gamma-1)\rho e$, where
$e=c_{\rm V} T$ is the internal energy.
The traceless rate of strain tensor $\mbox{\boldmath ${\sf S}$}$ is given by
\begin{equation}
{\sf S}_{ij} = \onehalf (U_{i,j}+U_{j,i}) - \onethird \delta_{ij} \DIV \bm{U},
\end{equation}
whilst the magnetic field satisfies
\begin{equation}
\nabla\cdot\bm{B}=0.
\end{equation}
Stress-free impenetrable boundary conditions are used for the velocity,
\begin{equation}
U_{x,z} = U_{y,z} = U_z = 0 \quad \text{on}\quad z=0,d,
\end{equation}
and vertical field conditions for the magnetic field, i.e.\
\begin{eqnarray}
B_x = B_y = 0 \quad \text{on}\quad z=0,d,
\end{eqnarray}
respectively ($\nabla\cdot{\bm B}=0$ then
implies $B_{z,z}=0$ at $z=0,d$).
The temperature is fixed at the upper and lower boundaries.
We adopt periodic boundary conditions for all variables 
in each of the two horizontal directions.

\subsection{Nondimensional quantities and parameters}
\label{Nondimensional}

Dimensionless quantities are obtained by setting
\begin{eqnarray}
d = g = \rho_{\rm m} = c_{\rm P} = \mu_0 = 1\;,
\end{eqnarray}
where $\rho_{\rm m}$ is the initial density at $z=z_{\rm m}=0.5d$.
The units of length, time, velocity, density, entropy, and magnetic field are
\begin{eqnarray}
&& [x] = d\;,\;\; [t] = \sqrt{d/g}\;,\;\; [U]=\sqrt{dg}\;,\;\;
[\rho]=\rho_{\rm m}\;,\\ \nonumber &&[s]=c_{\rm P}\;,\;\; [B]=\sqrt{dg\rho
_{\rm m}\mu_0}\;. 
\end{eqnarray}
Having non-dimensionalised these equations, the behaviour of the system
is determined by various dimensionless parameters. 
Quantifying the two key diffusivity ratios, the fluid and 
magnetic Prandtl numbers are given by
\begin{eqnarray}
\Pra=\frac{\nu_{\rm m}}{\chi_{\rm m}}\;,\;\; \Pm=\frac{\nu_{\rm m}}{\eta},
\end{eqnarray}
where $\nu_{\rm m} = \mu/\rho_{\rm m}$ is the mean kinematic viscosity
and $\chi_{\rm m} = K/(\rho_{\rm m} c_{\rm P})$ is the mean thermal 
diffusivity.
Defining $H_{\rm P}$ to be the pressure scale height at $z_m$, the midlayer 
entropy gradient in the absence of motion is  
\begin{eqnarray}
\bigg(-\frac{1}{c_{\rm P}}\frac{{\rm d}s}{{\rm d}z}\bigg)_{\rm m} = \frac{\nabla-\nabla_{\rm ad}}{H_{\rm P}}\;,
\end{eqnarray}
where $\nabla-\nabla_{\rm ad}$ is the superadiabatic temperature
gradient with $\nabla_{\rm ad} = 1-1/\gamma$ and $\nabla = (\pd \ln T/\pd
\ln p)_{z_{\rm m}}$. 
The strength of the convective driving can then be characterised by the 
Rayleigh number,
\begin{eqnarray}
\Ray=\frac{gd^4}{\nu_{\rm m} \chi_{\rm m}} \bigg(-\frac{1}{c_{\rm P}}\frac{{\rm d}s}{{\rm d}z
} \bigg)_{\rm m} = \frac{gd^4}{\nu_{\rm m} \chi_{\rm m}}\bigg(\frac{\nabla-\nabla_{\rm ad}}{H_{\rm P}} \bigg)\;.
\end{eqnarray}
The amount of rotation is quantified by the Taylor number,
\begin{eqnarray}
\Tay = \frac{4\Omega_0^2 d^4}{\nu_{\rm m}^2}\;
\end{eqnarray}
(which is related to the Ekman number, $\Ek$, by 
$\Ek = \Tay^{-1/2}$).
Since the critical Rayleigh number for the onset of 
hydrodynamic convection
is proportional to $\Tay^{2/3}$ in the rapidly rotating 
regime \citep{Chandrasekhar1961},
it is also useful to consider the quantity
\begin{eqnarray}
\widetilde{\Ray} = \frac{\Ray}{\Tay^{2/3}} \left(=\Ray\,\Ek^{4/3}\right)
\end{eqnarray}
\citep[see, e.g.,][]{Julienetal2012}. 
This rescaled Rayleigh number is a measure of the 
supercriticality of the convection that takes 
account of the stabilising influence of rotation.
We also quote the convective Rossby number
\begin{eqnarray}
\Roc = \left(\frac{\Ray}{\Pra\Tay}\right)^{1/2},
\end{eqnarray}
which is indicative of the strength of the thermal forcing compared to 
the effects of rotation.

Whilst $\Ray$, $\Tay$ and $\Pra$ are input parameters that must be 
specified at the start of each simulation, it is possible 
to measure a number of useful quantities based on system outputs.
These are expressed here in dimensional form for ease of reference.
We define the fluid and magnetic Reynolds numbers via
\begin{eqnarray}
\Rey = \frac{\urms}{\nu_{\rm m} \kef}\;,\quad
{\rm Rm} = \frac{\urms}{\eta \kef}\;,
\end{eqnarray}
where $\kef = 2\pi/d$
is indicative of the vertical scale of variation of the convective
motions, and $\urms$ is the 
time-average of the rms velocity during the saturated phase of the dynamo.
The time-evolution of the rms velocity, $\Urms(t)$, is also considered, but
only its constant time-averaged value will be used to define other diagnostic 
quantities.
The quantity $\urms \kef$ is therefore an estimate of the
inverse convective turnover time in the nonlinear phase of the dynamo.
The Coriolis number, an alternative measure of the importance of rotation
(compared to inertial effects) is given by
\begin{eqnarray}
{\rm Co} = \frac{2\,\Omega_0}{\urms \kef} \equiv\frac{\Tay^{1/2}}{4\pi^2\Rey}\;. \label{equ:Co}
\end{eqnarray}
All of the simulations described in this paper have ${\rm Co}\gtrsim4$
(and $\Roc<0.4$) so are in a rotationally dominated
regime\footnote{Note that $(2\pi\Co)^{-1}=\urms/2\Omega_0 d$ is 
equivalent to the standard Rossby number, based on the layer depth 
and the rms velocity.
We use $\Co$ here so as not to confuse this 
quantity with the convective Rossby number, $\Roc$. 
}.
Finally, the equipartition magnetic field strength is defined by 
\begin{equation}
\Beq \equiv \langle\mu_0\rho\bm{U}^2\rangle^{1/2},\label{equ:Beq}
\end{equation}
where angle brackets denote volume averaging.

\subsection{Initial conditions}

All of the simulations in this paper are initialised from a hydrostatic 
state corresponding to a polytropic layer, for which $\displaystyle{p
\propto \rho^{1+1/m}}$, where $m$ is the polytropic index. 
Assuming a monatomic gas with $\gamma=5/3$ we adopt a polytropic index
of $m=1$ throughout. 
This gives a superadiabatic temperature gradient of 
$\nabla-\nabla_{\rm ad}=1/10$, so the layer is convectively unstable, as
required.
The degree of stratification is determined by specifying a density 
contrast of $4$ across the layer. 
To be as clear as possible about our proposed benchmark case
(see Appendix~\ref{BenchmarkComparison} for details), 
it is useful to 
provide explicit functional forms for the initial density, pressure and 
temperature profiles in our dimensionless units.
Recalling that the layer has a unit depth and that $\rho_{\rm m}=1$, the
initial density profile is given by
\begin{displaymath}
\rho(z) = \frac{2}{5}(4-3z),
\end{displaymath}
\noindent whilst the initial pressure and temperature profiles are given
by
\begin{displaymath}
p(z) = \frac{1}{15}(4-3z)^2
\end{displaymath}
and
\begin{displaymath}
T(z) = \frac{5}{12}(4-3z),
\end{displaymath}
respectively. 
The fixed temperature boundary conditions imply that $T(0)=5/3$ and 
$T(1)=5/12$ (independent of $x$ and $y$, for all time). 
These profiles are consistent with a dimensionless pressure 
scale height of $H_{\rm P}=1/6$ at the top of the domain, which is an
alternative way of specifying the level of stratification. 
Since the governing equations are formulated in terms of the specific 
entropy, it is worth noting that these initial conditions are consistent
with an initial entropy distribution of the form
\begin{displaymath}
s = \ln \left(\frac{T(z)}{T_{\rm m}}\right) - \frac{2}{5}\ln\left(\frac{p(z)}{p_{\rm m}}\right),
\end{displaymath}
where $T_{\rm m}=25/24$ and $p_{\rm m}=5/12$ are the midlayer values of
the temperature and pressure respectively (note that $s_{\rm m}=0$ with 
this normalisation). 

In all simulations, convection is initialised by weakly perturbing this 
polytropic state in the presence of a low amplitude seed magnetic field 
(which varies over short length scales, with zero net flux across the 
domain). 
The precise details of these initial perturbations do not strongly influence
the nature of the
final nonlinear dynamos, although it goes without saying that the early 
evolution of this system does depend upon the initial conditions 
that are employed (as illustrated in 
Appendix~\ref{BenchmarkComparison}).

\subsection{Numerical methods}
\label{subsec:methods}

The {\sc Pencil Code}\footnote{\url{https://github.com/pencil-code}}
(Code~1) is a tool for solving partial differential equations on
massively parallel architectures.
We use it in its default configuration in which the MHD equations
are solved as stated in Eqs.~(\ref{equ:AA})--(\ref{equ:ss}).
First and second spatial derivatives are computed using
explicit centred sixth-order finite differences.
Advective derivatives of the form $\bm{U} \cdot \bm\nabla$
are computed using a fifth-order upwinding scheme,
which corresponds to adding a sixth-order hyperdiffusivity
with the diffusion coefficient $|\bm{U}|\delta x^5/60$
\citep{Dobler_etal06}.
For the time stepping we use the low-storage Runge-Kutta scheme
of \cite{Williamson}.
Boundary conditions are applied by setting ghost zones outside
the physical boundaries and computing all derivatives on and
near the boundary in the same fashion.
The non-dimensionalising scalings that have been described above correspond
directly to those employed by Code~1. 
Whilst the other codes that have been used (see below) employ different 
scalings, we have rescaled results from these to ensure direct comparability
with the results from the {\sc Pencil Code}.

The second code that shall be used in this paper (Code 2) is an updated 
version of the code that was originally described by \citet{Matthewsetal1995}.
The system of equations that is solved is entirely equivalent to those
presented in Sect.~\ref{sect:model}, but instead of time-stepping 
evolution equations for the magnetic vector potential, the logarithm 
of the density, the velocity field and the specific entropy, this code 
solves directly for the density, velocity field and temperature 
\citep[see, e.g.,][]{Matthewsetal1995}, whilst a poloidal-toroidal 
decomposition is used for the magnetic field. 
Due to the periodicity in both horizontal directions, horizontal
derivatives are computed in Fourier space using standard fast Fourier
transform libraries.
In the vertical direction, a fourth-order finite difference
scheme is used, adopting an upwind stencil for the advective terms.
The time-stepping is performed by an explicit third-order Adams-Bashforth
technique, with a variable time-step. 
As noted above, this code adopts a different set of
non-dimensionalising scalings to those described in 
Sect.~\ref{Nondimensional} \citep[see][for more details]{
FavierBushby2012}.

Code 3 is based upon a second-order Godunov-type finite-difference 
scheme that employs an approximate MHD Riemann solver with operator splitting
\citep[][]{Sanoetal1999,MasadaSano2014a,MasadaSano2014b}. 
The hydrodynamical part of the equations is solved by a Godunov method, 
using the exact solution of the simplified MHD Riemann problem. 
The Riemann problem is simplified by including only the tangential component 
of the magnetic field. 
The characteristic velocity is then that of the magneto-sonic wave alone, 
and the MHD Riemann problem can be solved in a way similar to the 
hydrodynamical one \citep[][]{ColellaWoodward1984}. 
The piecewise linear distributions of flow quantities are calculated with 
a monotonicity constraint following the method of \citet{vanLeer1979}. 
The remaining terms, the magnetic tension component of the equation of 
motion and the induction equation, are solved by the Consistent MoC-CT 
method \citep{Clarke1996}, guaranteeing $\nabla\cdot{\bm B} = 0$ to within 
round-off error throughout the calculation 
\citep{EvansHawley1988,StoneNorman1992}.

\section{Results}
\label{Results}

\begin{table*}
\caption{\label{table:1}Summary of the simulations in this paper.} 
\centering
\begin{tabular}{cccccccccccccc}
\hline
\hline
\rule{0pt}{2.7ex} 
Case & Code & Grid & $\lambda$ & $\Pra$ & $\Pm$ & $\Tay$ & $\Ray$ & $\widetilde{\Ray}$ &
$\Roc$ & $\Rey$ & ${\rm Rm}$ & $\Co$ & Large-scale\\
\hline
A1 & 1 & $256^3$ & 2 & 1 & 1 & $5\cdot10^8$ & $2.4\cdot10^7$ & 38.1 & 0.219 & 57 & 57 & 10.0 & Yes \\
A2 & 2 & $192^3$ & 2 & 1 & 1 & $5\cdot10^8$ & $2.4\cdot10^7$ & 38.1 & 0.219 & 57 & 57 & 9.9 & Yes \\
A3 & 3 & $256^3$ & 2 & 1 & 1 & $5\cdot10^8$ & $2.4\cdot10^7$ & 38.1 & 0.219 & 58 & 58 & 9.8 & Yes\\
\hline
B1 & 2& $192^3$ & 2 &1 & 1.33 & $5\cdot10^8$ & $2.4\cdot10^7$ & 38.1 & 0.219 & 56 & 74 & 10.2 & Yes \\ 
B2 & 2& $192^3$ & 2 & 1 & 1 & $5\cdot10^8$ & $3\cdot10^7$ & 47.6 & 0.245  & 73 & 73 & 7.7 & Yes\\
B3a & 2& $256^2\times 224$ & 2 & 1 & 1 & $5\cdot10^8$ & $4\cdot10^7$ & 63.5 & 0.283  & 111 & 111 & 5.1 & Intermittent \\
B3b &1& $288^3$ & 2 & 1 & 1 & $5\cdot10^8$ & $4\cdot10^7$ & 63.5 & 0.283 & 106 & 106 & 5.3 & Intermittent\\ 
B4 & 1 & $288^3$ & 2 & 1 & 1 & $5\cdot10^8$ & $5\cdot10^7$ & 79.4 & 0.316 & 132 & 132 & 4.3 & Intermittent \\ 
B5 & 1 & $288^3$ & 2 & 1 & 1 & $5\cdot10^8$ & $6\cdot10^7$ & 95.2 & 0.346 & 146 & 146 & 3.9 & Intermittent \\  
B6 & 2 & $128^2\times 192$ & 1 & 1 & 1 & $5\cdot10^9$ & $8\cdot10^7$ & 27.4 & 0.126 & 92 & 92 & 19.5 & No\\ 
B7 & 2& $128^2\times 192$ & 1 & 1 & 1.33 & $5\cdot10^9$ & $8\cdot10^7$ & 27.4 & 0.126  & 88 & 117 & 20.3 & No\\
B8 & 2 & $128^2\times 192$ & 1 & 1 & 1 & $5\cdot10^9$ & $1\cdot10^8$ & 34.2 & 0.141 & 109 & 109 & 16.4 & No\\
\hline 
C1 & 2 & $128^2\times 192$ & 2 & 1 & 1 & $5\cdot10^8$ & $2.1\cdot10^7$ & 33.3 &0.205 & 49 & 49 & 11.7 & Yes\\
C2 & 2& $192^3$ & 2 & 1 & 1 & $4\cdot10^8$ & $2.4\cdot10^7$ & 44.2 & 0.245 & 65 & 65 & 7.8 & Yes\\
C3 & 2&$128^2\times 192$ & 2 & 1 & 1 & $6\cdot10^8$ & $2.4\cdot10^7$ & 33.7 & 0.2  & 51 &51 & 12.2 & Yes\\
C4 & 2&$128^2\times 192$ & 2 & 1 & 1 & $7\cdot10^8$ & $2.4\cdot10^7$ & 30.4 & 0.185  & 47 & 47 & 14.3 & Yes\\
C5 & 2&$128^2\times 192$ & 2 & 1 & 1 & $7.5\cdot10^8$ & $2.4\cdot10^7$ & 29.1 & 0.179  & 46 & 46 & 15.0 & ?\\
\hline
D1 & 1& $288^3$ & 2 & 1 & 1 & $5\cdot10^8$ &  $2.4\cdot10^7$ & 38.1 & 0.219 &  57 &  57 & 10.0 & Yes \\
D2 & 1& $288^3$ & 2 & 1 & 1 & $2\cdot10^9$ &  $9.6\cdot10^7$ & 60.5 & 0.219 & 120 & 120 & 9.5 & Yes \\
D3 & 1& $576^3$ & 2 & 1 & 1 & $8\cdot10^9$ & $3.84\cdot10^8$ & 96.0 & 0.219  & 237 & 237 & 9.6 & Yes \\
\hline
E1 & 2& $192^3$ & 2 & 1 & 1 & $5\cdot10^8$ &  $2.4\cdot10^7$ & 38.1 & 0.219 & 60 & 60  & 9.5 & Yes \\
\hline
\end{tabular}
\tablefoot{A1, A2 and A3 correspond to the reference solution (a detailed 
comparison of these cases is presented in Appendix~\ref{BenchmarkComparison}).
Simulations B1--8 were initialised from a polytropic state, whilst simulations
C1--5 were started from case A2, gradually varying parameters along this 
solution branch.
In Sets~D and E the convective Rossby number is fixed at
$\Roc=0.219$.
In E1 the density contrast is $1.2$ in comparison to
$4$ in the other sets.
All input parameters are defined in the text.
We recall that the Reynolds numbers, $\Rey$ and ${\rm Rm}$,
and the Coriolis number, $\Co$,
are measured during the saturated phase of the dynamo (and the 
quoted values are time-averaged over this phase). 
The final column indicates whether or not there is a
large-scale dynamo.}

\end{table*}

\subsection{Reference solution}

In this section, we present a detailed description of a typical dynamo run 
exhibiting large-scale dynamo action (carried out using Code 2).
This case, which will henceforth be referred to as the reference solution, 
forms the basis for the proposed benchmark calculation that is discussed 
in Appendix~\ref{BenchmarkComparison}
 (with the corresponding calculations denoted by cases A1, A2 
and A3 in the upper three rows of Table~\ref{table:1}, where 
our parameter choices are also summarised).   
For numerical convenience, we fix $\Pra=\Pm=1$ for the reference solution, 
whilst our choice of $\Tay=5\cdot10^8$ ensures that rotation plays a 
significant role in the ensuing dynamics.
Defining $\Ray_{\rm crit}$ to be the critical Rayleigh number at which the 
hydrostatic polytropic state becomes linearly unstable to convective 
perturbations, it can be shown that $\Ray_{\rm crit}=6.006\cdot10^6$ 
for this set of parameters.
This critical value (here quoted to three decimal places) was 
determined using an independent Newton-Raphson-Kantorovich 
boundary value solver for the corresponding linearised system. 
At onset, the critical horizontal wavenumber for the convective 
motions is very large (in these dimensionless units, the preferred
horizontal wavenumber at onset is given by $k \approx 36$), 
indicating a preference for narrow convective cells. 
Our choice of Rayleigh number, $\Ray = 2.4\cdot10^7\approx 4\,\Ray_{\rm crit}$, 
is moderately supercritical.
However, we would still expect small-scale convective motions during the
early phases of the simulation. 
Our choice of $\lambda=2$ for the aspect ratio of the domain is small
enough to enable us to properly resolve these motions, whilst also being
large enough to allow the expected large-scale vortex instability to grow.

\subsubsection{A large-scale dynamo}

\begin{figure}[t]
\resizebox{\hsize}{!}{\includegraphics{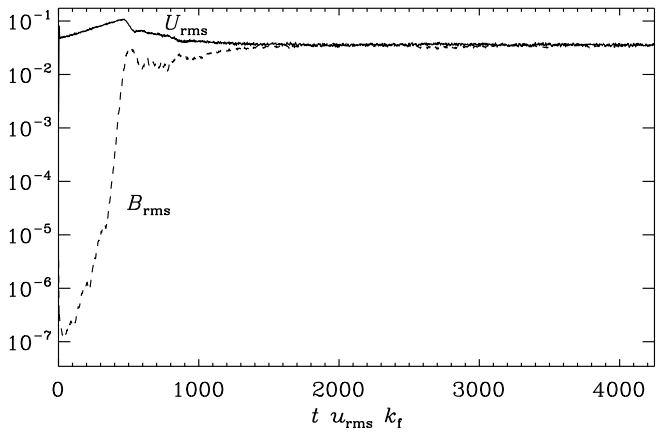}}
\caption{Time evolution of $\Urms(t)$ and $\brms(t)$ for the reference solution. 
Time has been normalised by the inverse convective turnover time in the 
final nonlinear state. }
\label{fig:fig1}
\end{figure}

\begin{figure}[t] 
\begin{center}
\resizebox{0.83\hsize}{!}{\includegraphics{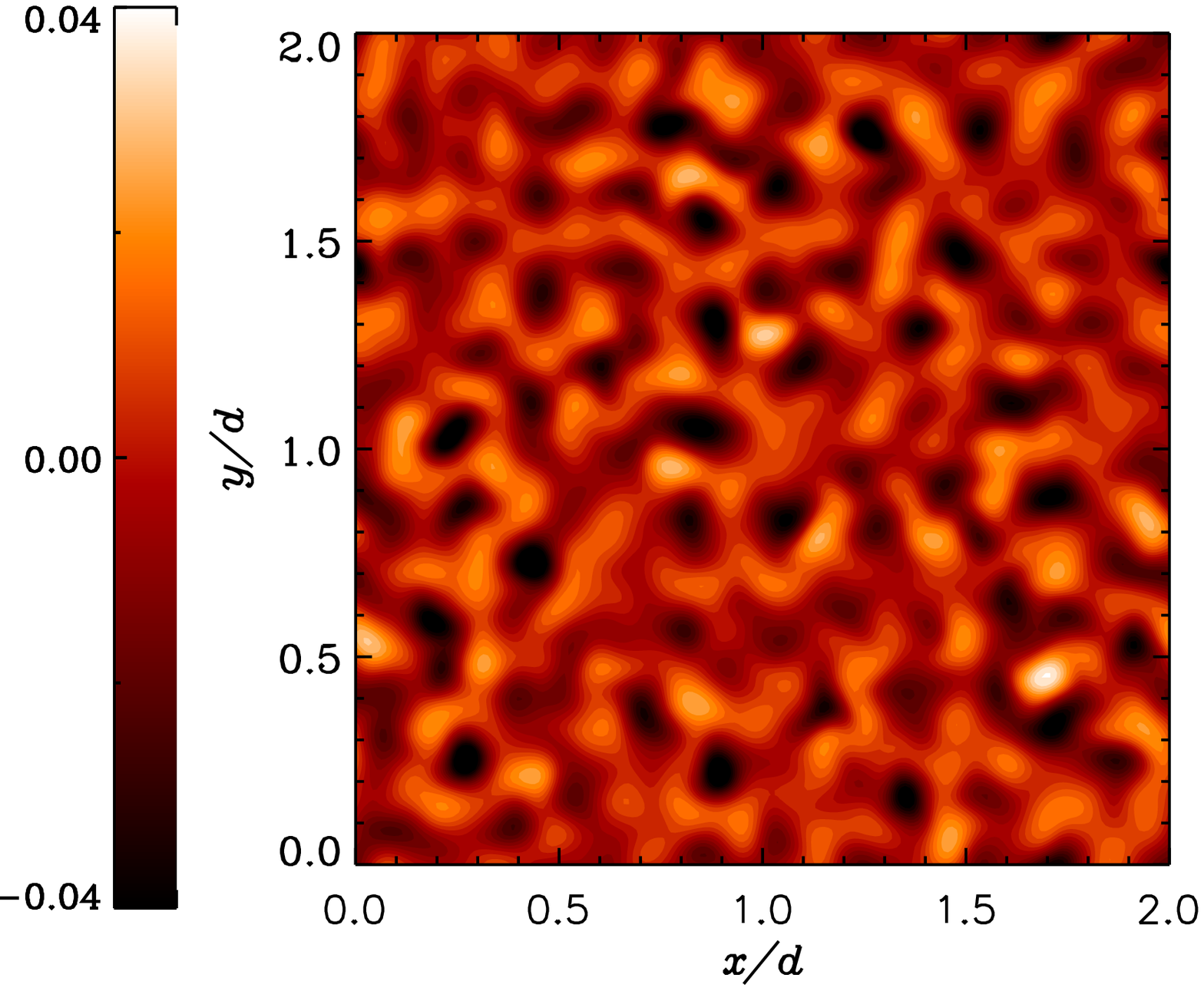}}
\resizebox{0.83\hsize}{!}{\includegraphics{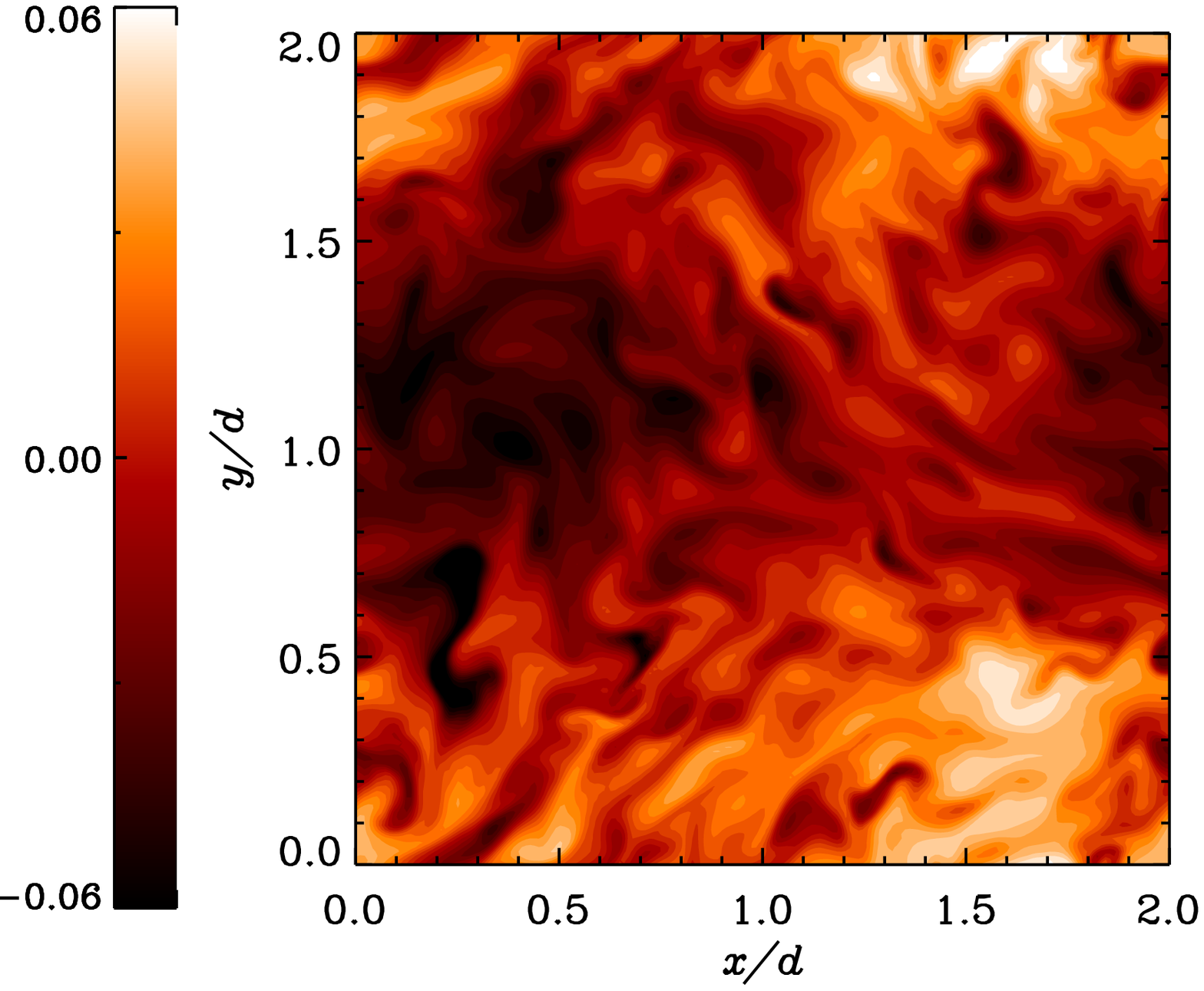}}
\resizebox{0.93\hsize}{!}{\includegraphics{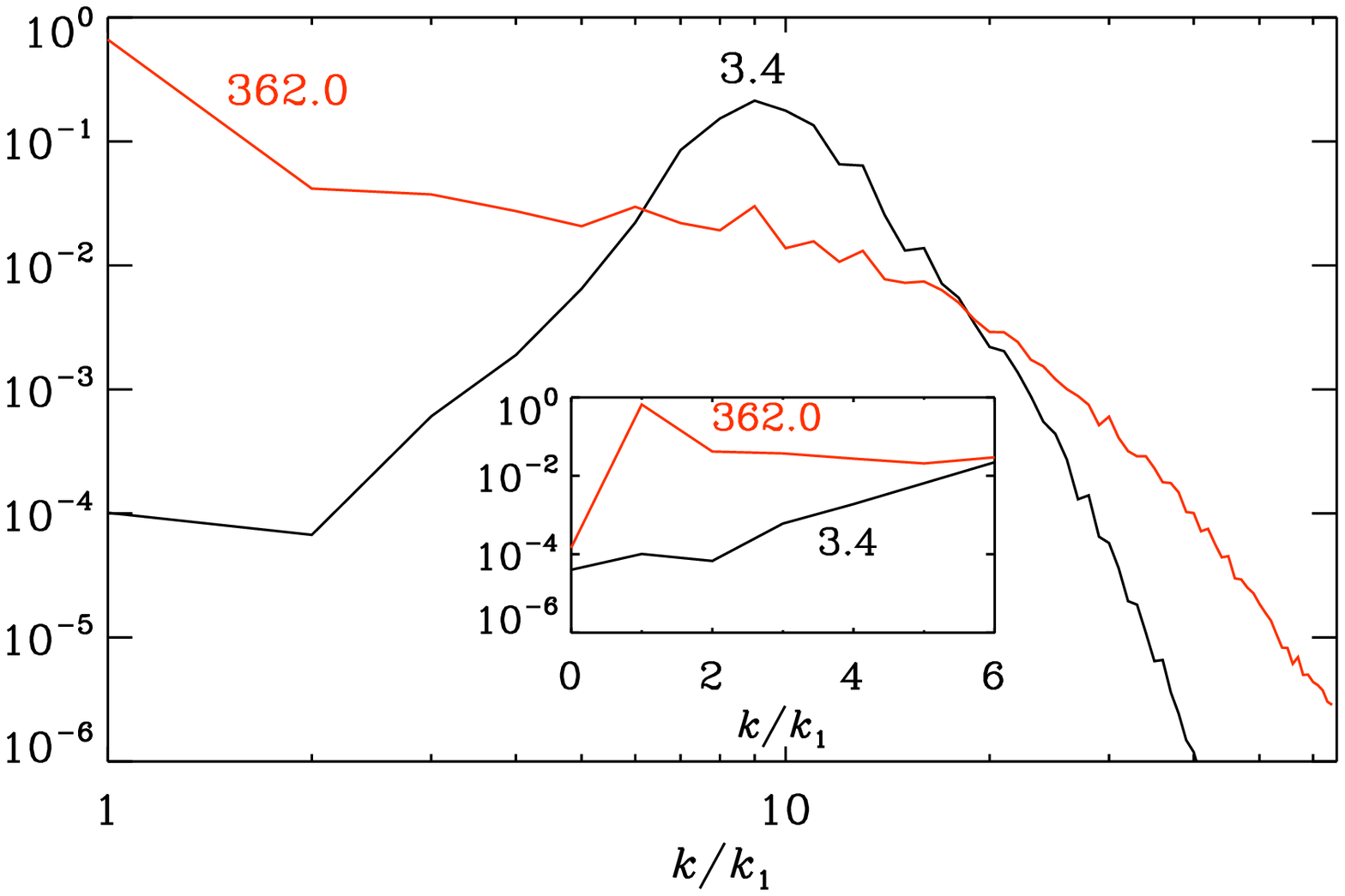}}
\end{center}
\caption{Early stages of the reference solution.
Mid-layer temperature distributions (at $z_{\rm m}$) 
at times $t\urms\kef=3.4$ ({\it top}) and $t\urms\kef=362.0$ 
({\it middle}), plotted as functions of $x$ and $y$. 
In each plot, the mid-layer temperature is normalised by its mean
(horizontally averaged) value,
with the contours showing the deviations from the mean.
{\it Bottom:} mid-layer kinetic energy spectra at the same times, 
normalised by the average kinetic energy at $z_{\rm m}$ at 
each time.
Here $k$ is the horizontal wavenumber (in the $x$ direction), 
whilst $k_{\rm1}=2\pi/\lambda$ is the 
fundamental mode. 
The inset highlights the low wavenumber behaviour, using a non-logarithmic
abscissa to include $k/k_1=0$.}
\label{fig:fig2}
\end{figure}

Figure~\ref{fig:fig1} shows the time evolution of 
$\Urms(t)$ and $\brms(t)$, 
the rms velocity and magnetic field, 
for the reference solution.
Whilst the rms quantities have been left in dimensionless form, time has 
been normalised by the inverse convective turnover time during the final 
stages of the simulation, by which time the system is in a 
statistically steady state. 
Initially, there is a very brief period (spanning no more than a few 
convective turnover times) of exponential growth, followed by a similarly 
rapid reduction in the magnitude of $\Urms(t)$.
At these very early times, the convective motions are (as expected
from linear theory) 
characterised by a small horizontal length scale. 
The solution then enters a prolonged phase of more 
gradual evolution, lasting until time $t\urms \kef \approx 470$, during 
which time $\Urms(t)$ tends to increase. 
This period of growth coincides with a gradual transfer of energy from 
small to large scales, leading eventually to a convective state that is 
dominated by larger scales of motion.
This is an example of the large-scale vortex instability that was 
discussed in Sect.~\ref{sect:intro}.

The transition from small to large-scale convection is clearly illustrated
by the time evolution of the corresponding temperature distribution.
The upper two plots in Fig.~\ref{fig:fig2} show the mid-layer 
temperature distribution 
at time $t\urms\kef=3.4$, at which point the motions have a small 
horizontal length scale, and time $t\urms\kef=362.0$, at which the 
temperature distribution varies over the largest available scale.
The lower plot shows the corresponding kinetic energy 
spectra as a function of the horizontal wavenumber, $k$, at 
times $t\urms\kef=3.4$ and $t\urms\kef=362.0$. 
As expected from linear theory, most of the power at 
time $t\urms\kef=3.4$ is concentrated at relatively high 
wavenumbers.
Defining $k_{\rm 1}=2\pi/\lambda$ (in this domain, 
$k_{\rm 1}=\pi$) to be the 
fundamental mode, corresponding to one full oscillation across 
the width of the domain, the corresponding kinetic energy 
spectrum peaks at around $k/k_{\rm 1}=9$ at this early time. 
This implies a favoured horizontal wavenumber of
$k\approx 28.3$ in these dimensionless units.
This is comparable to the critical wavenumber at convective 
onset. 
At time $t\urms\kef=362.0$, 
there is still significant energy in the higher 
wavenumber, small-scale components of the flow (in fact, 
the spectrum is now broader, extending to higher 
wavenumbers than before). However,
most of the power in the kinetic 
energy spectrum has been transferred to the $k/k_{\rm 1}=1$ 
component. 
This corresponds to the large-scale vortex.
Even at $t\urms\kef=3.4$ there are some indications of 
non-monotonicity in the power spectrum at low wavenumbers,
so this instability sets in very rapidly as the system evolves.

\begin{figure}[t]
\resizebox{\hsize}{!}{\includegraphics{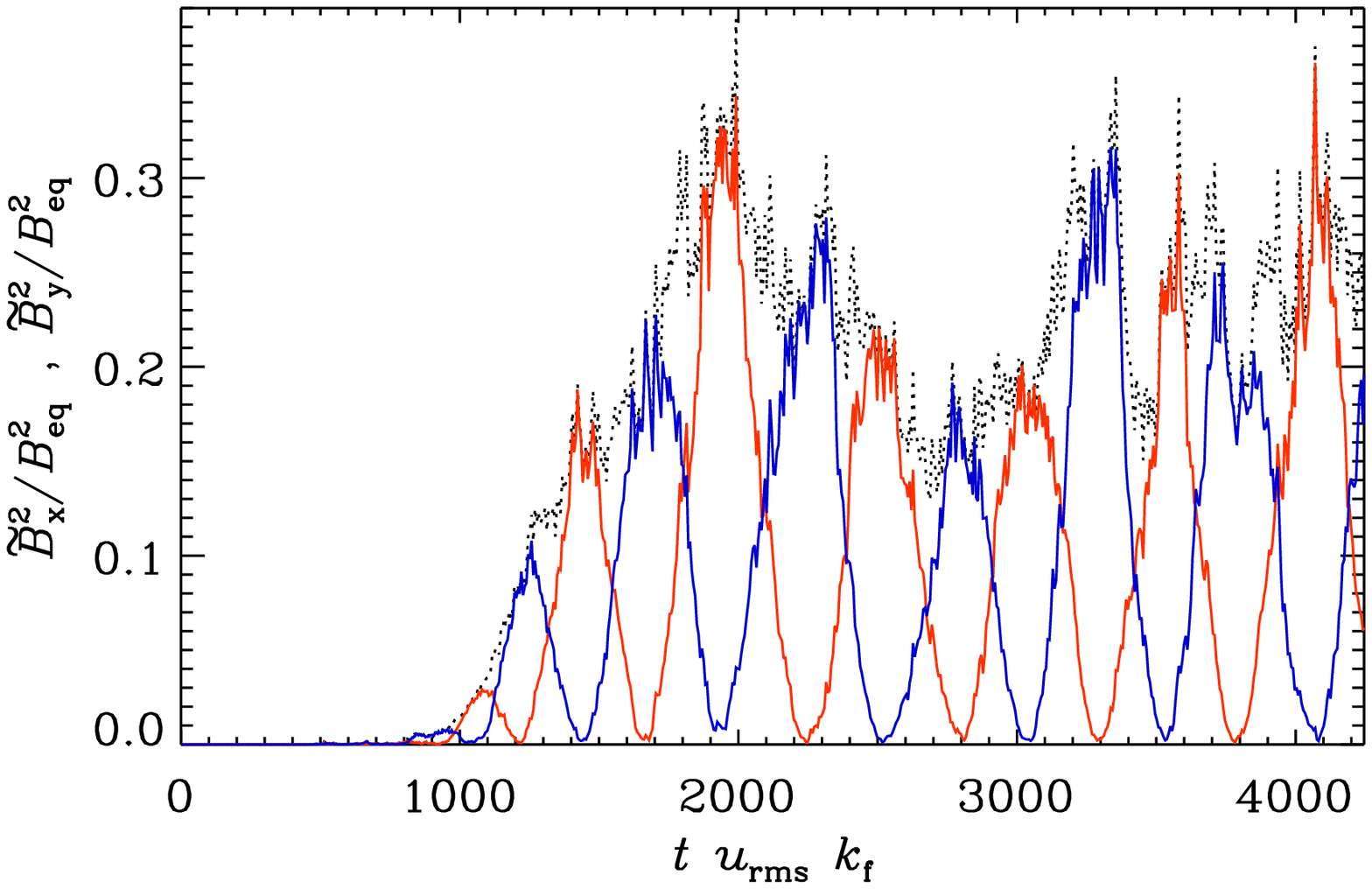}}
\resizebox{\hsize}{!}{\includegraphics{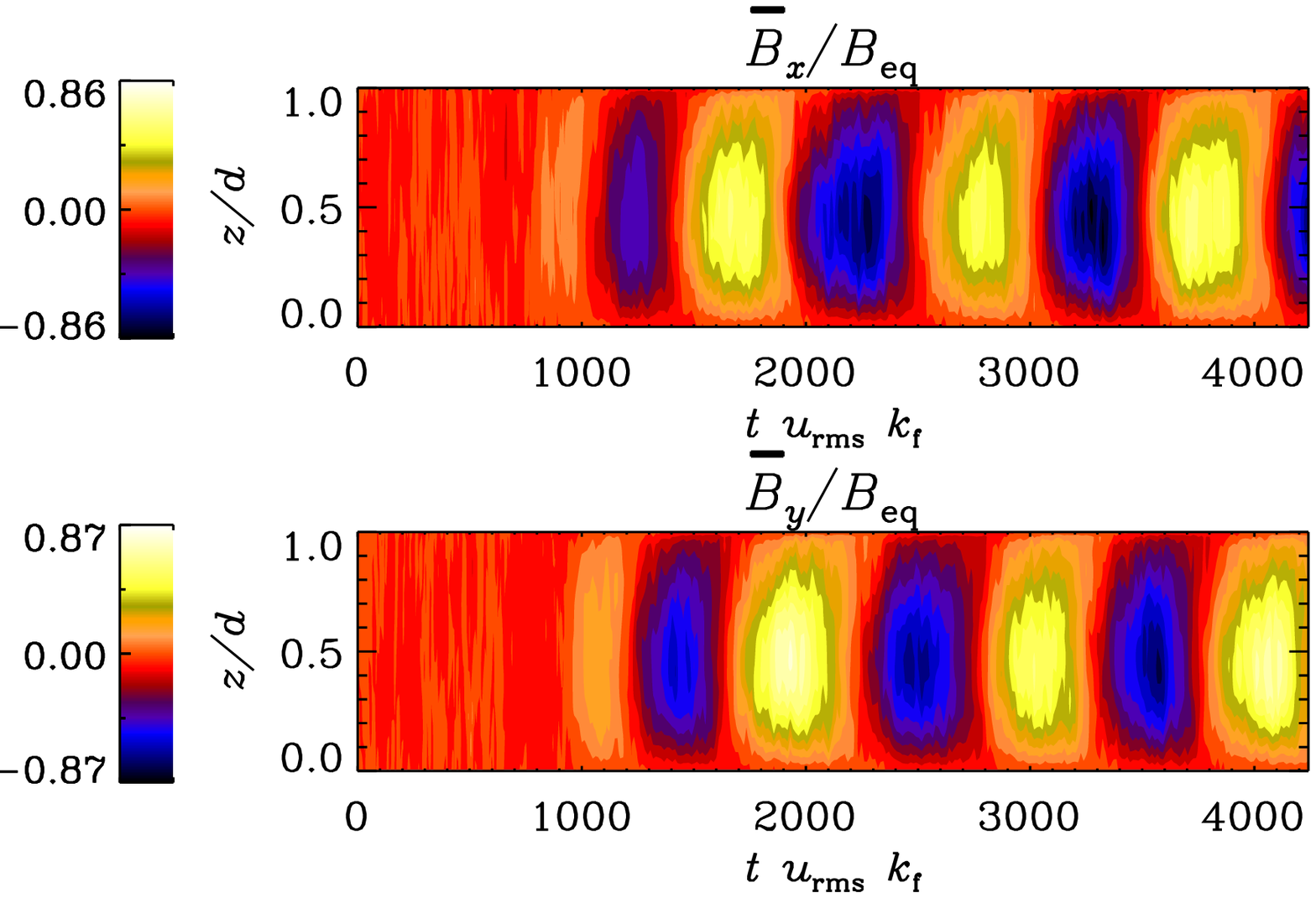}}
\resizebox{\hsize}{!}{\includegraphics{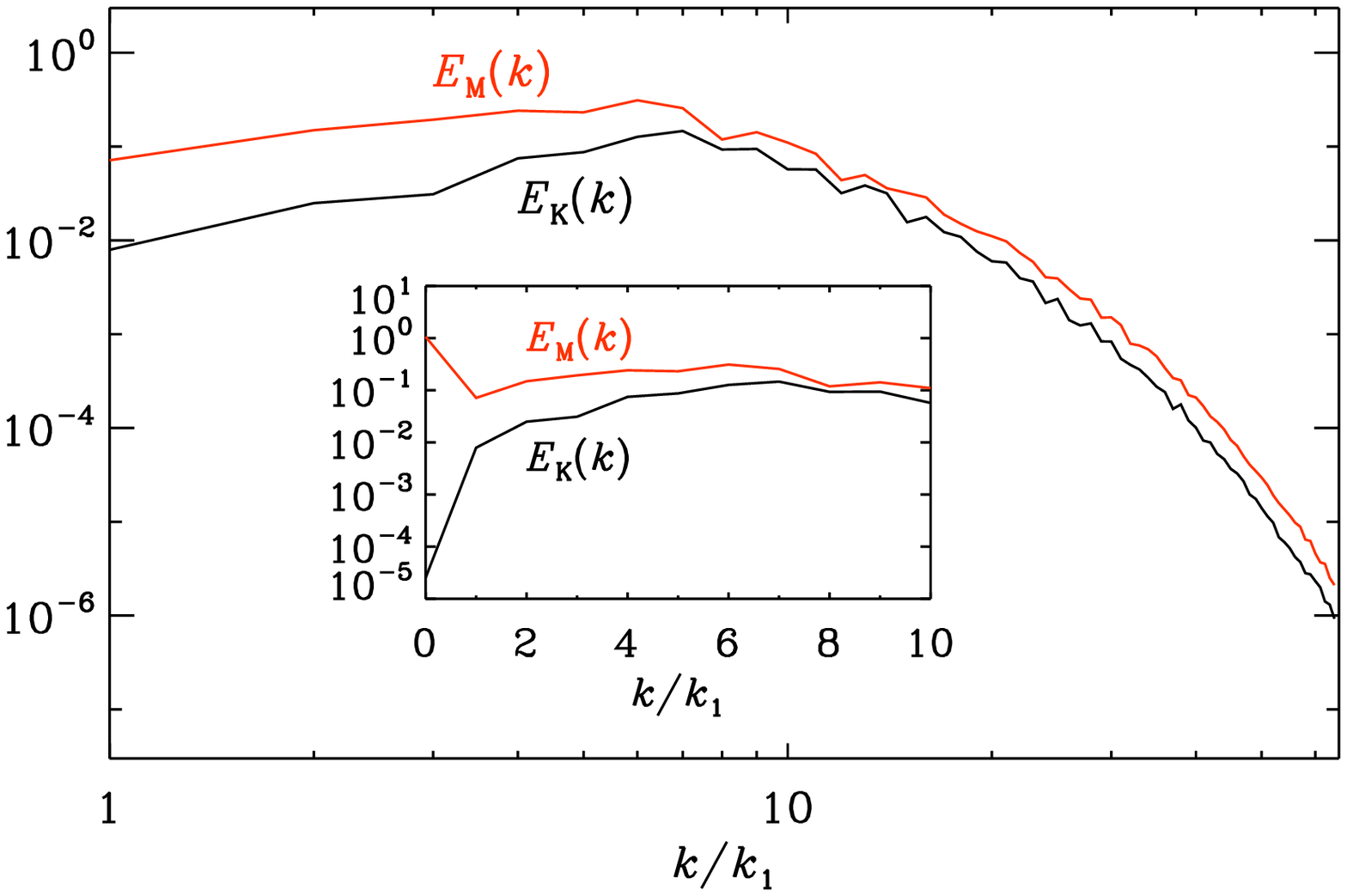}}
\caption{Reference solution dynamo. {\it Top}: normalised volume averages of
the squared horizontal magnetic field components, 
$\tilde{B}_x^2/\Beq^2$ (blue) and $\tilde{B}_y^2/\Beq^2$ (red) 
as functions of time; the black dotted line shows 
$(\tilde{B}_x^2+\tilde{B}_y^2)/\Beq^2$.
{\it Middle}: horizontally averaged horizontal magnetic 
fields (normalised by the equipartition field strength) as functions
of $z$ and time. {\it Bottom}: mid-layer kinetic (black) and magnetic (red) 
energy spectra, normalised by the average kinetic energy at 
$z_{\rm m}$ at the end of the dynamo run.
The inset highlights the low wavenumber behaviour of the spectra, 
using a non-logarithmic abscissa to include $k/k_1=0$.}
\label{fig:fig3}
\end{figure}

As $\Urms(t)$ (or, equivalently, the kinetic energy in the system) 
increases, it soon reaches a level above which the flow is sufficiently 
vigorous to excite a dynamo. 
The seed magnetic field is initially very weak, with $\brms(t)$ many 
orders of magnitude smaller than the typical values of $\Urms(t)$. 
However, as illustrated in Fig.~\ref{fig:fig1}, once the large-scale 
vortex instability starts to grow,  $\brms(t)$ also starts to increase.
With increasing $\Urms(t)$ (effectively, an ever-increasing magnetic 
Reynolds number during this growth phase), the growth of $\brms(t)$ 
accelerates until it reaches a point at which the magnetic field is 
strong enough to exert a dynamical influence upon the flow. 
The sharp decrease in $\Urms(t)$ at time $t\urms \kef \approx 470$ is an 
indicator of the effects of the Lorentz force upon the flow, with the 
large-scale mode being strongly suppressed by the magnetic field. 
After this point, 
which marks the start of the nonlinear phase of the dynamo,
a more gradual growth of $\brms(t)$ is accompanied 
by a gradual decrease in $\Urms(t)$.
By the time $t\urms \kef \approx 1600$, the dynamo appears to reach a 
statistically-steady state in which $\brms(t)$ is comparable to 
$\Urms(t)$, 
which itself is comparable to the rms velocity before the 
large-scale vortex started to develop. 
The ohmic decay time, $\tau_{\eta}$ (based on the depth of the layer), 
corresponds to $\approx2300$ convective turnover times. 
So having continued this calculation to $t\urms \kef \approx 4250$, 
observing no significant changes in $\brms$ or $\urms$ over the 
latter half of the simulation, we can be very confident that this is a 
persistent dynamo solution that will not decay over longer times. 

Having discussed the time evolution of the system, we now turn 
our attention to the form of the magnetic field. 
Figure~\ref{fig:fig3} shows the time- and $z$-dependence of the 
horizontally averaged profiles for $B_x$ and $B_y$, as well as 
the kinetic and magnetic energy spectra at the end of the dynamo
run. 
Here the $\overline{B}_i$ (for $i=x$ or $y$) represent 
horizontally averaged magnetic field components, whilst the
$\tilde{B}_i$ correspond to volume averages.
It is clear from these plots that the magnetic field distribution has 
no significant mean horizontal component
(compared to its equipartition value) when the dynamo
enters the nonlinear phase (at $t\urms \kef \approx 470$). 
However, as the dynamo evolves from $t\urms \kef \approx 470$, 
a cyclically-varying mean horizontal field gradually emerges, eventually
growing to a level at which the peak mean horizontal field almost 
reaches the equipartition level (although it should be noted that
the peak amplitude does vary somewhat, from one cycle to the next).
This mean magnetic field is approximately symmetric about the
mid-plane with no real indication of any propagation of activity
either towards or away from this mid-plane.
The cycle period, $\tau_{\rm cyc}$, is approximately $1000$ 
convective turnover times, which is of a similar order of magnitude 
to the ohmic decay time quoted above 
($\tau_{\rm cyc}\approx 0.5\tau_{\eta}$). 
The dominance of the large-scale magnetic field is confirmed by
the presence of a pronounced peak at $k/k_{\rm 1}=0$ in the 
magnetic energy spectrum that is shown in the lower part of 
Fig.~\ref{fig:fig3}.
It should be stressed again that the kinetic energy spectrum is
peaked at small scales at this stage, so the large-scale vortex
(which plays a critical role in initialising the dynamo) no longer
appears to have a significant role to play in this large-scale dynamo. 
This is emphasised by the lack of a significant $k/k_{\rm 1}=1$ 
peak in the kinetic energy spectrum.  

\subsubsection{Understanding the dynamo}

Defining $\boldsymbol{\mathcal{E}}= {\bm U}\times{\bm B}$, 
and recalling that an overbar denotes a horizontal average, 
it can be shown that
\begin{equation}
\frac{\partial\overline{B}_x}{\partial t} = - \frac{\partial\overline{\mathcal{E}}_y}{\partial z} 
+ \eta \frac{\partial^2\overline{B}_x}{\partial z^2},
\label{eq:bxmean}
\end{equation}
whilst
\begin{equation}
\frac{\partial\overline{B}_y}{\partial t} = \frac{\partial\overline{\mathcal{E}}_x}{\partial z} 
+ \eta \frac{\partial^2\overline{B}_y}{\partial z^2}.
\label{eq:bymean}
\end{equation}
With the given boundary and initial conditions, it is straightforward to 
show that $\overline{B}_z=0$ for all $z$ and $t$. 
In the absence of a significant mean flow, standard mean-field 
dynamo theory \citep[e.g.][]{Moffatt1978} suggests that there should 
be a linear relationship between the components of 
the mean electromotive force,
$\overline{\boldsymbol{\mathcal{E}}}$, and the components and $z$-derivatives 
of $\overline{\bm B}$. 
To be more specific, we might expect relations of the 
form 
\begin{eqnarray}
\overline{\mathcal{E}}_x(z) = \alpha(z)\overline{B}_x(z) 
+ \beta(z)\frac{\partial \overline{B}_y}{\partial z},\nonumber  \\ 
\overline{\mathcal{E}}_y(z) = \alpha(z)\overline{B}_y(z) 
- \beta(z)\frac{\partial \overline{B}_x}{\partial z},
\label{eq:mft}
\end{eqnarray}
where $\alpha(z)$ and $\beta(z)$ are scalar functions of 
height only.
Whilst the $\beta(z)$ term would dominate near the boundaries, 
where the mean magnetic field vanishes (and the field gradient
is large), we might expect the $\alpha(z)$ term to dominate 
in a non-negligible region in the vicinity of the mid-plane, 
where the mean field gradient is relatively small. 
Having said that, it should be stressed that the flow that is responsible for 
driving this large-scale dynamo is the product of a fully nonlinear 
magnetohydrodynamic state. 
Even before the large-scale magnetic field starts to grow, this flow
is strongly influenced by magnetic effects. 
As noted by \citet{Courvoisieretal2009}, this probably implies that
the relationship between $\overline{\boldsymbol{\mathcal{E}}}$ and 
$\overline{\bm B}$ is more complicated than that suggested by
Eq.~(\ref{eq:mft}).
We defer detailed considerations of this question to future work, 
focusing here upon the form of $\overline{\boldsymbol{\mathcal{E}}}$ 
and the 
components of the flow that are responsible for producing it. 

\begin{figure}[t]
\resizebox{\hsize}{!}{\includegraphics{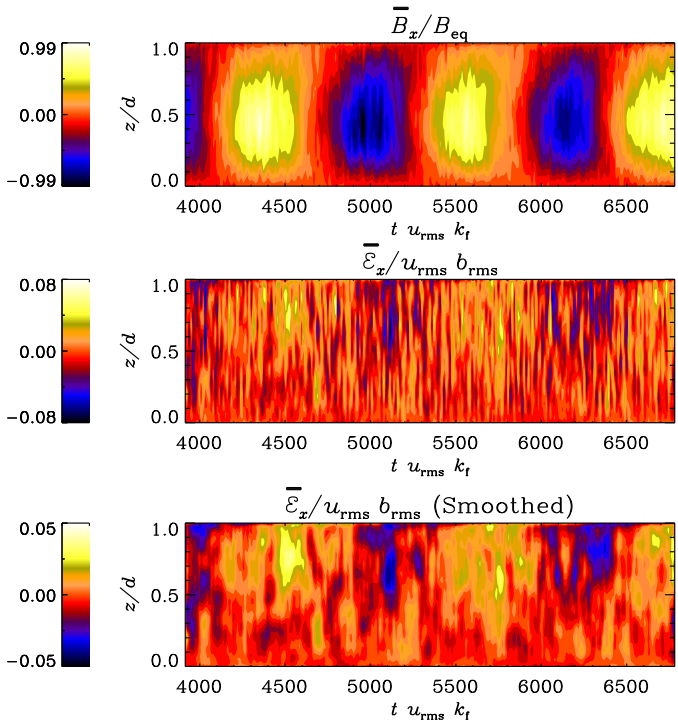}}
\caption{Time and $z$ dependence of $\overline{B}_x$ ({\it top}) and 
$\overline{\mathcal{E}}_x$ ({\it middle}) for a dynamo simulation with the 
reference solution parameters.
Note that  $\overline{\mathcal{E}}_x$ has been normalised by 
the product of $\urms$ ($0.035$) and 
the corresponding rms magnetic field, 
$b_{\rm rms}$ ($0.036$),
both of which have been time-averaged over the given time period.
{\it Bottom:} the results of temporally smoothing 
$\overline{\mathcal{E}}_x$, using a sliding (boxcar) average with a 
width of approximately $50$ convective turnover times.
}
\label{fig:emfx}
\end{figure}

Figure~\ref{fig:emfx} shows contour plots of $\overline{\mathcal{E}}_x$ 
and (for comparison) $\overline{B}_x$ as functions of $z$ and $t$, taken 
from a 
simulation that duplicates the reference solution parameters, but is evolved 
from a state with a weaker initial thermal perturbation. 
A quantitatively comparable solution is obtained, indicating 
that this large-scale dynamo is robust to such changes
to the initial conditions. 
Clearly $\overline{\mathcal{E}}_x$ exhibits much stronger temporal 
fluctuations than $\overline{B}_x$, whilst $\overline{\mathcal{E}}_x$
is more asymmetric about the mid-plane than $\overline{B}_x$.
Nevertheless, 
in purely qualitative terms, there are some 
indications of a simple temporal correlation 
between $\overline{\mathcal{E}}_x$ and $\overline{B}_x$, with a 
similar long time-scale variation apparent in both cases.
This is not inconsistent with the relation suggested by Eq.~(\ref{eq:mft}). 
Although not shown here, 
$\overline{B}_y$ and $\overline{\mathcal{E}}_y$ 
evolve in a similar way.

\begin{figure}[t]
\resizebox{\hsize}{!}{\includegraphics{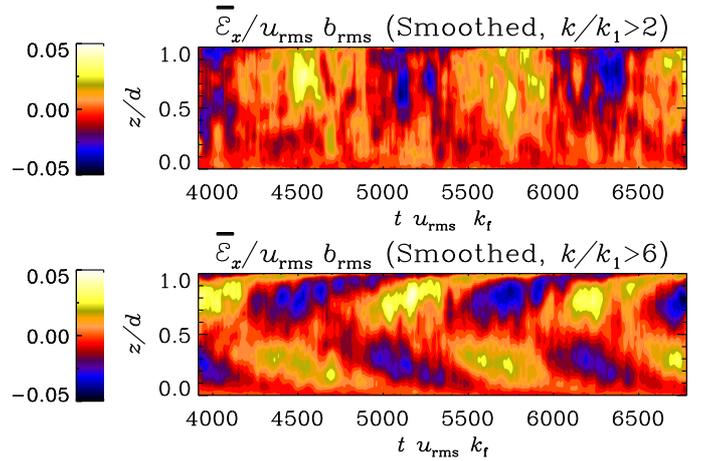}}
\caption{Contours of $\overline{\mathcal{E}}_x$ (again normalised
by $\urms b_{\rm rms}$) as a 
function of $z$ and time, derived 
(in post-processing)
from a filtered flow.
{\it Top:} horizontal wavenumbers with 
$k/k_{\rm 1}\le 2$ have been removed from the flow 
before calculating $\overline{\mathcal{E}}_x$.  
{\it Bottom:} the filtering threshold is set at 
$k/k_{\rm 1} \le 6$.
}
\label{fig:emffilt}
\end{figure}

\begin{figure}[t]
\resizebox{\hsize}{!}{\includegraphics{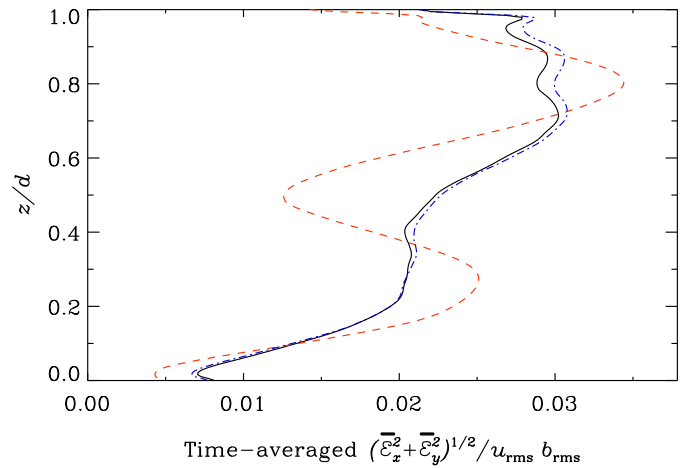}}
\caption{Time-averaged $(\overline{\mathcal{E}}^2_x+\
\overline{\mathcal{E}}^2_y)^{1/2}$, normalised by
$\urms b_{\rm rms}$, as a function of $z$ (solid black curve). 
The blue (dash-dotted) curve (very close to the black line)
shows the same quantity but 
having removed low horizontal wavenumber components 
($k/k_{\rm 1}\le 2$) from the flow before calculating 
$\overline{\mathcal{E}}_x$ and $\overline{\mathcal{E}}_y$. The 
red (dashed) curve shows the effects of removing more Fourier
modes ($k/k_{\rm 1} \le 6$).
}
\label{fig:emfbar}
\end{figure}

As we have already noted, the large-scale vortex is strongly
suppressed by the magnetic field. 
To confirm that the low wavenumber components of the flow make 
a negligible contribution to this large-scale dynamo, we have 
investigated the effects of removing the low wavenumber content 
from the flow before calculating $\overline{\boldsymbol{\mathcal{E}}}$.
It should be emphasised that we are not making any changes to the 
dynamo calculation itself -- all filtering is carried out in post-processing.
The upper part of Fig.~\ref{fig:emffilt} shows the effects of 
removing all Fourier modes with $k/k_{\rm 1} \le 2$ before calculating
$\overline{\mathcal{E}}_x$ (although not shown here, similar results 
are obtained for $\overline{\mathcal{E}}_y$).
At least in qualitative terms, it is clear that this is comparable to the 
corresponding unfiltered $\overline{\mathcal{E}}_x$ that is shown in
Fig.~\ref{fig:emfx}. 
A much more significant change is observed when modes with 
$k/k_{\rm 1} \le 6$ are removed: this filtered 
$\overline{\mathcal{E}}_x$ appears to be almost antisymmetric about
the mid-plane,
with some propagation of activity away
from the mid-plane that is not apparent in the reference solution. 
Intriguingly, there is a significant fraction of the domain over 
which this filtering process actually changes the sign of 
the resulting $\overline{\mathcal{E}}_x$ (compared to
the corresponding unfiltered case).
This sign change occurs gradually as the filtering threshold is 
increased from $k/k_{\rm 1}=2$ to $6$,
so there is no abrupt transition.
Further increments to this filtering threshold lead to no
further qualitative changes in the
distribution of $\overline{\mathcal{E}}_x$, although its 
amplitude obviously decreases as more components are 
removed from the flow. 

Figure~\ref{fig:emfbar} shows the time-averaged values of 
$(\overline{\mathcal{E}}^2_x+\overline{\mathcal{E}}^2_y)^{1/2}$ as a
function of $z$, for the filtered and unfiltered cases. 
Clearly the first filtered case, with only the $k/k_{\rm 1}\le 2$ modes 
removed, is quantitatively comparable to the unfiltered case, 
which reinforces the notion that the low wavenumber content of the 
flow plays no significant role in the dynamo.
On the other hand, the clear quantitative differences between the 
two filtered cases indicate that   
components of the flow with 
horizontal wavenumbers in the range $3 \le k/k_{\rm 1} \le 5$ play 
a crucial role in driving this large-scale dynamo. 
It is worth highlighting the fact that the small-scale motions alone 
seem to be capable of producing a coherent, 
systematically-varying $\boldsymbol{\mathcal{E}}$ 
with a peak value that is of comparable magnitude to the peak value 
of the unfiltered $\boldsymbol{\mathcal{E}}$. 
This suggests that it may be possible for the small-scale motions 
alone to sustain a dynamo.
Having said that, it should also be emphasised that these small-scale 
motions are not independent of the large-scale flows
and magnetic field in the system
(we stress again that all filtering is carried out in ``post-processing''), 
and so the small-scale motions are almost certainly strongly influenced by the fact that a 
large-scale dynamo is operating.
Whilst it is tempting to speculate that there may be some connection 
here with the near-onset dynamos of (e.g.) \citet{StellmachHansen2004}, 
which rely purely on small-scale motions, we have not yet been able to 
demonstrate this in a conclusive manner.

Having identified the components of the flow that are responsible for
driving the large-scale dynamo, we conclude this section
with a brief comment on the role of the magnetic boundary conditions.
Integrating Eqs.~(\ref{eq:bxmean}) and~(\ref{eq:bymean}) over $z$, it
is straightforward to verify that our magnetic boundary conditions 
(in which $B_x=B_y=0$ at $z=0$ and $z=1$) allow the net horizontal 
flux to vary in time
\citep[c.f. the near-onset study of][]{FavierProctor2013}.
In particular, these boundary conditions allow for the diffusive 
transport of magnetic flux out of the domain. 
Recalling that the dynamo oscillates on a timescale that is of the same 
order of magnitude as the ohmic decay time across the layer, it is
probable that the cycle period of the dynamo reflects the rate at which 
horizontal magnetic flux can be ejected from the domain. 
Assuming that the simple mean-field ansatz of Eq.~(\ref{eq:mft}) is 
a reasonable description of $\overline{\boldsymbol{\mathcal{E}}}$, this process 
would be accelerated by a positive $\beta(z)$ at the boundaries 
(which would enhance diffusion), so we should not be surprised to 
find examples of this dynamo with a considerably shorter cycle period.
Having said that, even a large $\beta(z)$ at the boundaries can only
help if the boundary conditions allow it to do so.
If we were to modify the magnetic boundary conditions to the 
Soward-Childress (perfect conductor) conditions of $B_z=\partial
B_y/\partial z = \partial B_x/\partial z=0$ at $z=0$ and $z=1$, that
would mean that
the total horizontal flux would be invariant (initially set to zero).
Under these circumstances, we have confirmed that a simulation with the reference dynamo 
parameters produces only a small-scale dynamo. 
Of course, we cannot exclude the possibility that there are 
regions of parameter space in which a large-scale dynamo can 
be excited via this large-scale vortex mechanism, with these
perfectly conducting magnetic boundary conditions.
However, we can say that these simulations suggest that such
a configuration would be less favourable for large-scale dynamo 
action than the vertical conditions that we have adopted here.

As a final remark, it is worth noting that one way of circumventing 
the dependence of this dynamo upon the magnetic boundary 
conditions is to include one or more convectively stable layers into 
the system \citep[][]{Kapylaetal2013}, which could reside 
either above or below the convective layer.
Even if perfectly conducting boundary conditions are applied, such a 
stable region can act as a ``flux repository''
for the system: if the convective 
layer can expel a sufficient quantity of magnetic flux into this 
region, the large-scale dynamo could continue to operate as 
described above \citep[as is the case for run 
D1e in][]{Kapylaetal2013}.
Whilst we emphasise that this is not a model for the solar
dynamo, the possible importance of an underlying stable
layer acting as a flux repository has also been recognised in
that context.
Certain solar dynamo models
\citep[e.g.][]{Parker1993,Tobias1996b, 
MacGregorCharbonneau1997} rely on the assumption
that the bulk of the large-scale toroidal magnetic flux in the solar 
interior resides in the stable layer just below the base of the 
convection zone.
This can therefore be regarded another example of a situation in 
which the addition of a stable layer can be beneficial for dynamo
action.

\subsection{Sensitivity to parameters}

\begin{figure}[t]
\resizebox{\hsize}{!}{\includegraphics{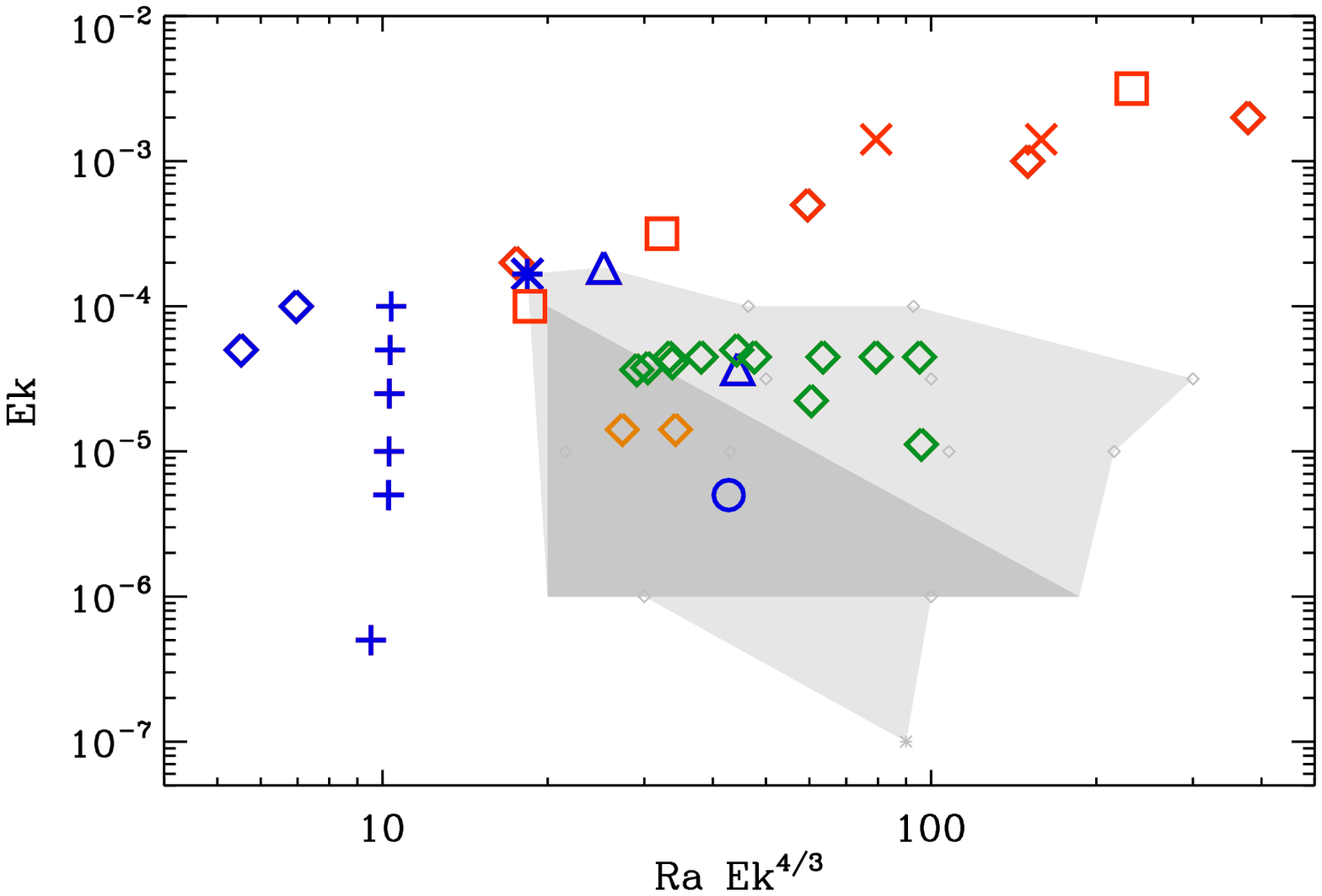}}
\caption{
$\Ek$ as a function of $\Ray\,\Ek^{4/3}(=\widetilde{\Ray})$ 
from
  various studies in the literature where large-scale dynamos were
  present (blue symbols) or absent (red). Data is shown from
  \cite{Kapylaetal2009} ($\diamond$), \cite{Kapylaetal2013}
   (\ding{83}),
  \cite{StellmachHansen2004} ($+$), \cite{FavierBushby2013}
  ($\square$), \cite{MasadaSano2014b,MasadaSano2016}
  ($\bigtriangleup$), \cite{CattaneoHughes2006} ($\times$), and
  \cite{Guervillyetal2015} ($\bigcirc$). The green and orange diamonds
  correspond to the present study with (green) or without (orange)
  large-scale dynamos. The shaded area shows the parameter
  region where large-scale vortices are present in the hydrodynamic
  regime. The darker area corresponds to the large-scale vortex
  region identified by
  \cite{Guervillyetal2015}, whereas the small grey diamonds and the star refer to
  the simulations of \cite{Favieretal2014} and
  \cite{Stellmachetal2014}, respectively.}
\label{fig:pekram}
\end{figure}

As indicated in Table~\ref{table:1}, we have carried out a range of 
simulations to assess the sensitivity of the reference solution 
dynamo to variations in the parameters. 
Cases B1--8 were all initialised from our standard polytropic state, 
whilst cases C1--5 were initialised from the reference solution (A2). 
Cases D1--3 investigated the effects of increasing $\Ray$ and 
$\Tay$ at fixed convective Rossby number, $\Roc$. 
Finally, case E1 is a pseudo-Boussinesq calculation with the initial
density varying linearly between $0.9$ and $1.1$; all other 
parameters (including the polytropic index) were identical to the 
reference solution.

When comparing dynamos at different rotation rates, it is 
convenient to consider the modified Rayleigh number,
$\widetilde{\Ray}=\Ray/\Tay^{2/3}$.
For ease of comparison with previous studies, it is worth
recalling that the Ekman number, $\Ek$, is related to the 
Taylor number by $\Ek = \Tay^{-1/2}$.
So $\widetilde{\Ray}$ is equivalent 
to $\Ray\,\Ek^{4/3}$, which tends to be the more usual form
of this parameter in the geodynamo literature. 
Figure~1 of \citet{Guervillyetal2015} suggests that 
$\widetilde{\Ray}$ must exceed a value of approximately $20$ 
for the large-scale  vortex instability to operate (obviously
rapid rotation is also required). 
For the reference solution, $\widetilde{\Ray}\approx 38$,
which is certainly consistent with that picture. 
The other point to note from Fig. 1 of \citet{Guervillyetal2015} 
is that 
(in the absence of a large-scale vortex instability)
large-scale dynamos tend to be restricted to a small 
region of parameter space in which the layer is rapidly-rotating
(i.e.\ $\Tay>10^8$; equivalently, $\Ek<10^{-4}$) 
and the convection is only weakly supercritical
(typically,  $\widetilde{\Ray}$ is $O(10)$). 

Following a similar approach to that of \citet{Guervillyetal2015}, 
Fig.~\ref{fig:pekram} shows how the simulations that are 
reported in this paper compare with others in the literature.
These dynamo simulations are classified
according to $\widetilde{\Ray}=\Ray\,\Ek^{4/3}$ and $\Ek$
\citep[the Ekman number has been used here, 
rather than the Taylor number, for ease of comparability 
with][]{Guervillyetal2015}.
The shaded area indicates the approximate region of parameter space 
where the large-scale vortex instability has been observed,
although the limits of this region do also depend to some extent
on the aspect ratio of the corresponding simulation domains.
The parameter regime in the lower right part of the plot is also
likely to support large-scale vortices but 
numerical simulations in
this regime are currently beyond the available computational
resources.
It should be stressed that there are many different types of
convective dynamos on this plot. Some simulations are
Boussinesq rather than compressible, others have 
perfectly-conducting magnetic boundary conditions, others
feature underlying and/or overlying stable layers.
These model differences become particularly important at
the edges of the large-scale vortex region.
Focusing on the upper left-hand corner of the shaded region,
the single-layer calculation of \citet{FavierBushby2013} was just
outside of the large-scale vortex region and so found a 
small-scale dynamo. 
The large-scale vortex instability was, however, present in the 
multiple-layer model of \citet{Kapylaetal2013}, which 
explains the observation of large-scale dynamo action
in that case.  
Having said that, regardless of the details of the model, 
it is clear that 
the large-scale vortex instability seems to provide a 
route by which large-scale dynamos can be found in 
moderately supercritical, rapidly rotating convection, 
outside of their normal operative region of parameter space.

It is worth emphasising at this stage that 
not all simulations in the large-scale vortex parameter region 
produce large-scale dynamos.
In particular, we only find small-scale dynamos at low aspect 
ratios (i.e.\ $\lambda<2$). For $\lambda=1$
(see cases B6, B7 and B8,
which correspond to the orange diamonds in Fig.~\ref{fig:pekram}),
this is true even at 
higher rotation rates, where there is a greater separation in scales
between the domain size and the horizontal scale of the near-onset 
convective motions. 
The failure of the large-scale dynamo in this case can probably 
be attributed 
to the fact that the large-scale vortex instability, which plays a crucial 
role in initialising the dynamo, is inhibited in smaller domains 
\citep[see also][]{Guervillyetal2015}.
Certainly, the large-scale vortex (measured, 
for example, by the rms velocity) stops growing long before the 
magnetic field becomes dynamically significant, which suggests
that geometrical effects are limiting its growth in these low
aspect ratio cases.
A strong suppression of the large-scale vortex will also
limit the production of motions at those 
intermediate scales that are responsible
for sustaining the large-scale dynamo.

\begin{figure}[t]
\resizebox{\hsize}{!}{\includegraphics{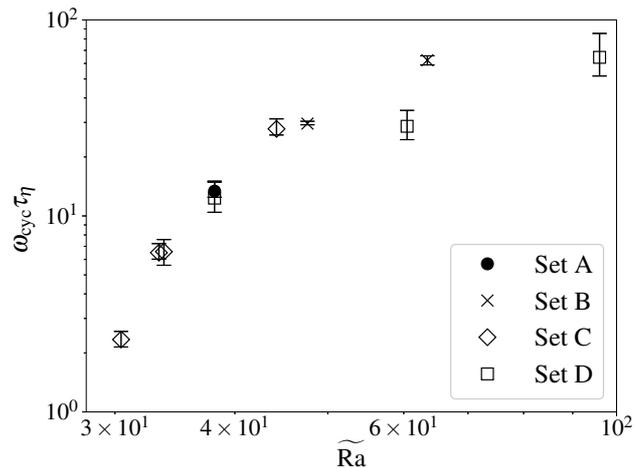}}
\caption{Cycle frequency, $\omega_{\rm cyc}$, of large-scale 
dynamo simulations, plotted against $\widetilde{\Ray}$.
In each case, the cycle 
frequency is normalised by the ohmic decay time, 
$\tau_{\eta}$.
}
\label{fig:period}
\end{figure}

Figure~\ref{fig:period} shows the cycle frequency, 
$\omega_{\rm cyc}=2\pi/\tau_{\rm cyc}$ (where 
$\tau_{\rm cyc}$ is the cycle period), for the 
successful large-scale dynamos, as a function of 
$\widetilde{\Ray}$.
To ensure some degree of comparability across the 
simulations, the cycle frequency has been normalised 
by the ohmic decay time, $\tau_{\eta}$, in each case. 
This choice of normalisation was motivated by the 
comparability of $\tau_{\eta}$ and $\tau_{\rm cyc}$ for
the reference solution (and reflects the important
role played by diffusive processes in the 
underlying dynamo mechanism). 
Case E1 has been excluded from this plot because
it has not been evolved for long enough to produce
an accurate determination of the cycle period, for
reasons that are discussed below.
Across the other simulations, a clear trend is observed
when moving closer to convective onset:
cycle frequencies tend to decrease
with decreasing $\widetilde{\Ray}$.
So, as we move closer to onset, the dynamo cycles become
longer \citep[c.f.][who found a similar result in 
a related quasi-geostrophic dynamo model]{Calkinsetal2016}.
In fact, this trend explains the uncertainty regarding 
whether or not C5 is a large-scale dynamo. If it were 
a large-scale dynamo, it would have an extremely long
cycle period, and we were unable to run this calculation
for long enough to establish whether or not this was 
the case.
At higher convective driving, modulational effects (as 
discussed in the next section) make it more difficult to 
determine cycle frequencies in some cases. 
However, it is clear that there is a greater degree of 
variability in the cycle frequencies at higher values of 
$\widetilde{\Ray}$.
For example, despite having rather similar values of 
$\widetilde{\Ray}$, the normalised cycle frequencies for 
cases D2 and B3a differ by more than a factor of two.
This suggests some dependence of the cycle frequency 
upon some of the other parameters in the system 
besides $\widetilde{\Ray}$ (a possible candidate
is the convective Rossby number, $\Roc$, which is 
fixed in cases D1--3 but is free to vary in cases B1--8).
These differences aside, although there is not a single 
best fit curve, there is still a clear tend of 
increasing cycle frequencies with increasing levels of
convective driving. 
The shortest period dynamos have a cycle
period that is about $10$\% of $\tau_{\eta}$.
Even in these cases, the cycle periods are still
long enough that diffusive effects are probably
playing a significant role.

\begin{figure}[t]
\resizebox{\hsize}{!}{\includegraphics{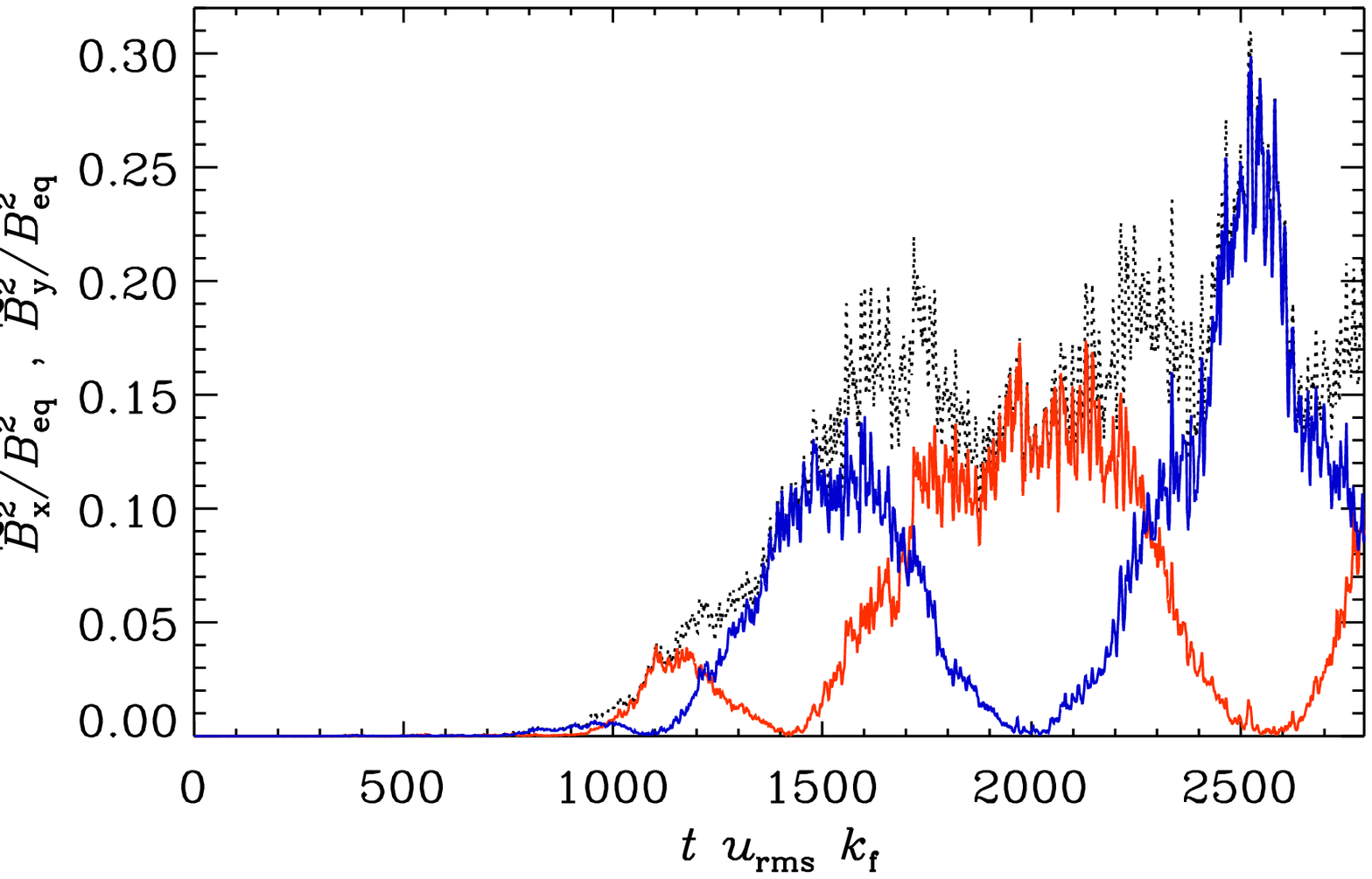}}
\resizebox{\hsize}{!}{\includegraphics{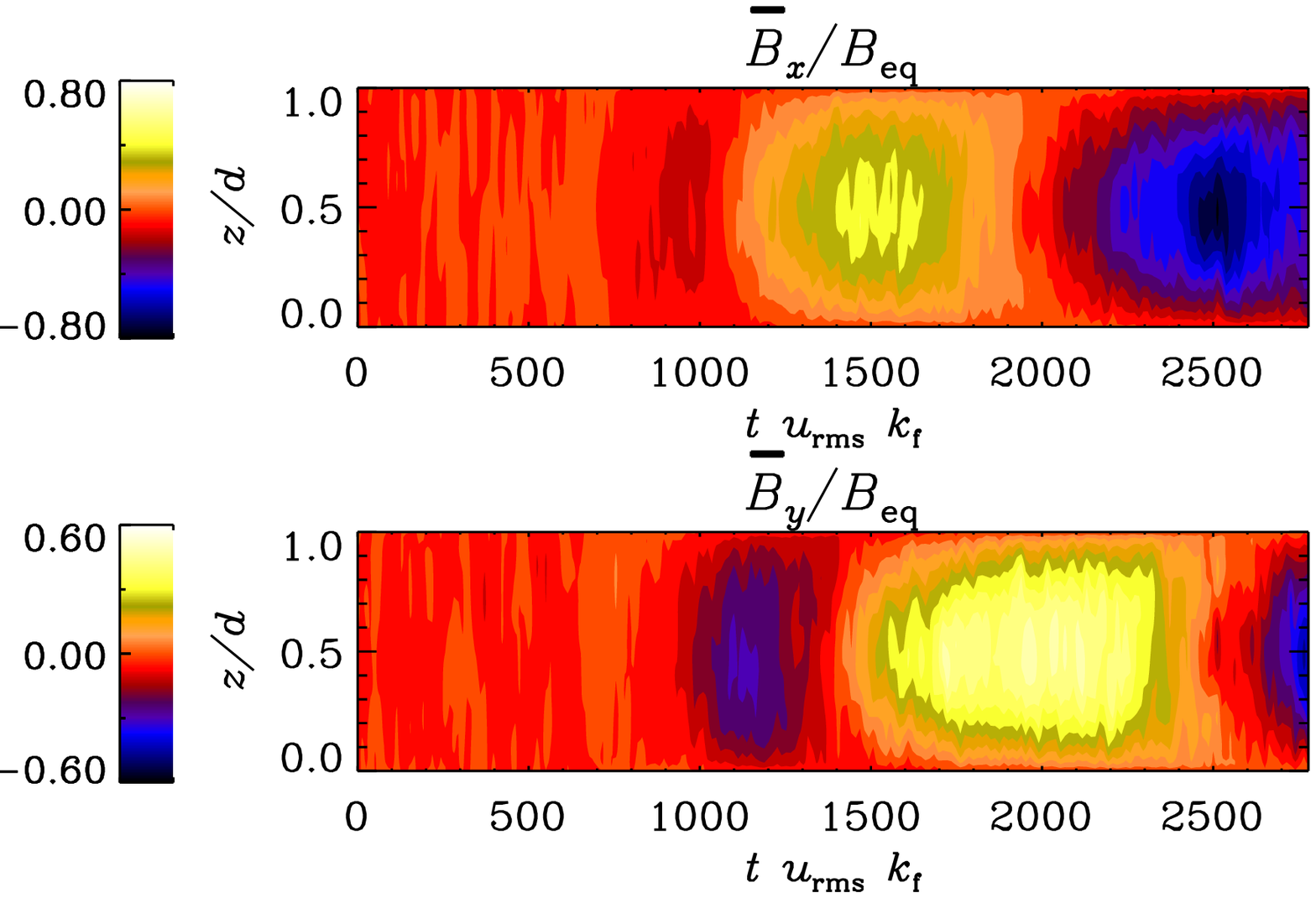}}
\resizebox{\hsize}{!}{\includegraphics{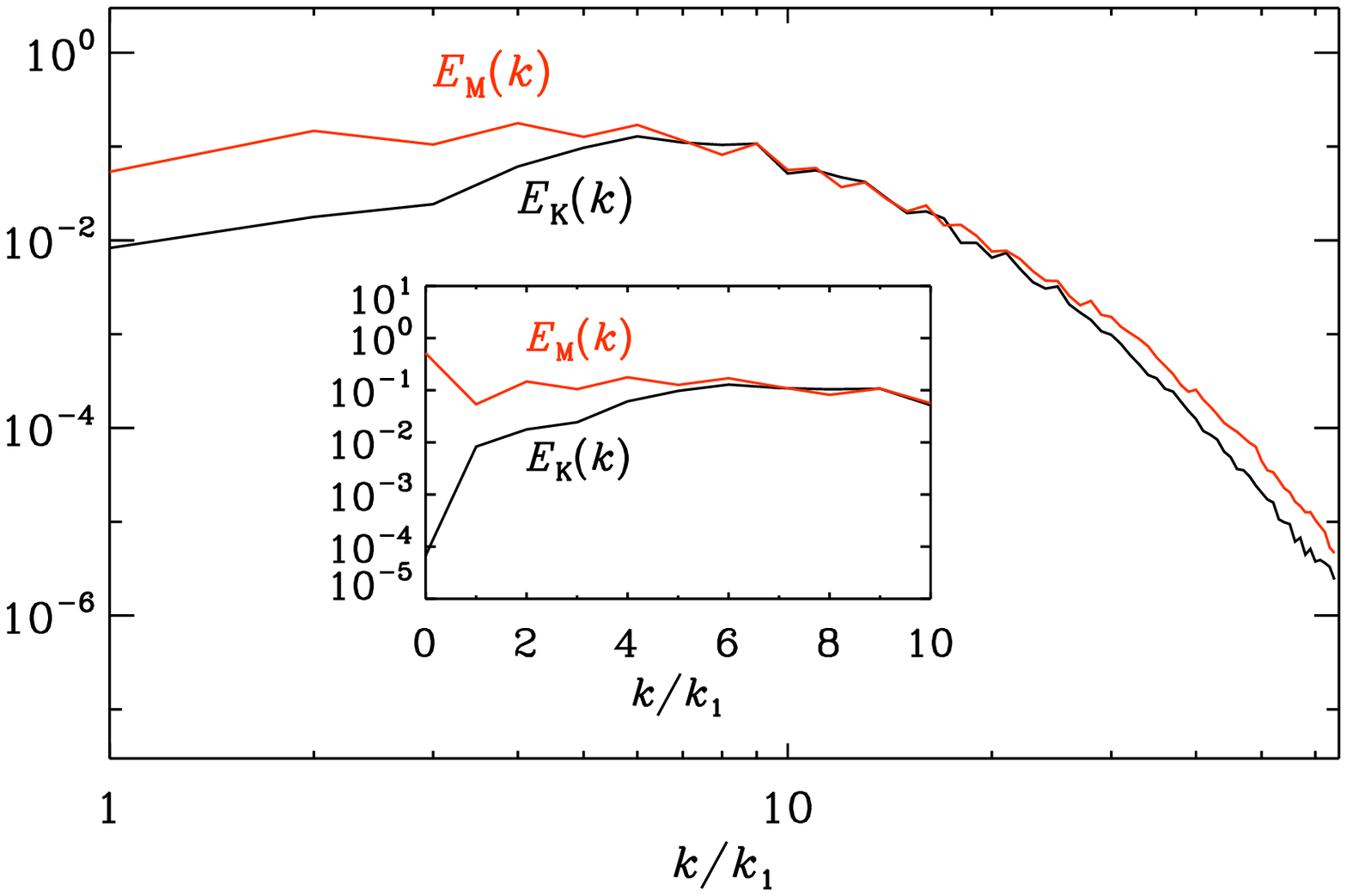}}
\caption{As Fig.~\ref{fig:fig3}, but here for the weakly 
stratified (quasi-Boussinesq) case, E1.
Note that the time-series here is relatively short, 
because the numerical time-step constraints make longer 
runs very difficult to carry out.
}
\label{fig:figE1}
\end{figure}

Whilst these large-scale dynamos do exhibit weak 
departures away from symmetry about the mid-plane, there
are no obvious suggestions that the stratification of the fluid
is playing a major role in the operation of the dynamo.
Indeed, as illustrated in Fig.~\ref{fig:figE1},
our quasi-Boussinesq case (E1) produces much the
same dynamo solution as the reference case.
Only a relatively short time-series is shown in this figure 
(in this quasi-Boussinesq regime, the time-step constraints 
that are associated with acoustic modes make long 
calculations prohibitively expensive), but it is clear that
the behaviour of this solution is qualitatively similar to that of 
the corresponding stages of the reference solution,
so it is very unlikely that this large-scale dynamo
will not persist.
In such cases, it is clearly more efficient to consider
a proper Boussinesq model -- whilst not shown here, we 
have confirmed that a Boussinesq calculation
\citep[using the code described by][]{Cattaneoetal2003} can 
produce a similar large-scale dynamo with the 
reference solution parameter values.
Due to the Boussinesq symmetries, 
$\overline{\mathcal{E}}_x$ and $\overline{\mathcal{E}}_y$ 
are anti-symmetric about the mid-plane, but the mean 
horizontal magnetic fields are again symmetric. 
The only significant change due to the reduction in the level
of stratification is in the cycle period.
The Boussinesq run suggests a cycle period (normalised by the 
convective turnover time) that is approximately twice 
that of the reference solution.
Whilst it is still evolving, the quasi-Boussinesq case E1 appears
to be consistent with this, with a longer cycle period compared
to that seen in the corresponding plots in Fig.~\ref{fig:fig3}.
The ohmic decay time in case E1 is approximately $2350$ 
turnover times, which is similar to that of the reference 
solution, so this longer cycle period is not simply a function
of normalisation.
Having said that, as indicated by Fig.~\ref{fig:period}, small 
changes in the convective driving can lead to large changes
in the cycle period in the vicinity of the reference solution, 
so this discrepancy between the cycle periods is probably
not that significant. 
If anything, it is more remarkable that this seems to be 
the only significant effect on the dynamo that can be 
attributed to a reduction in the stratification of the layer.

\subsection{Temporal modulation}

As noted in Table~\ref{table:1}, there are a number of cases in which 
the large-scale dynamo activity is described as ``intermittent''. 
Recalling that the critical Rayleigh number for the onset of 
convection is $\Ray_{\rm crit}=6.006\cdot10^6$ for this set 
of parameters, these cases are considerably more supercritical
(in terms of their convective driving) than the reference solution.
For cases B3a and B3b, $\Ray\approx 6.7\,\Ray_{\rm crit}$,
whilst $\Ray\approx 8.3\,\Ray_{\rm crit}$ for case B4 and
$\Ray\approx 10.0\,\Ray_{\rm crit}$ for case B5.
In each case, the large-scale vortex instability leads to 
vigorous convective flows.

\begin{figure}[t]
\resizebox{\hsize}{!}{\includegraphics{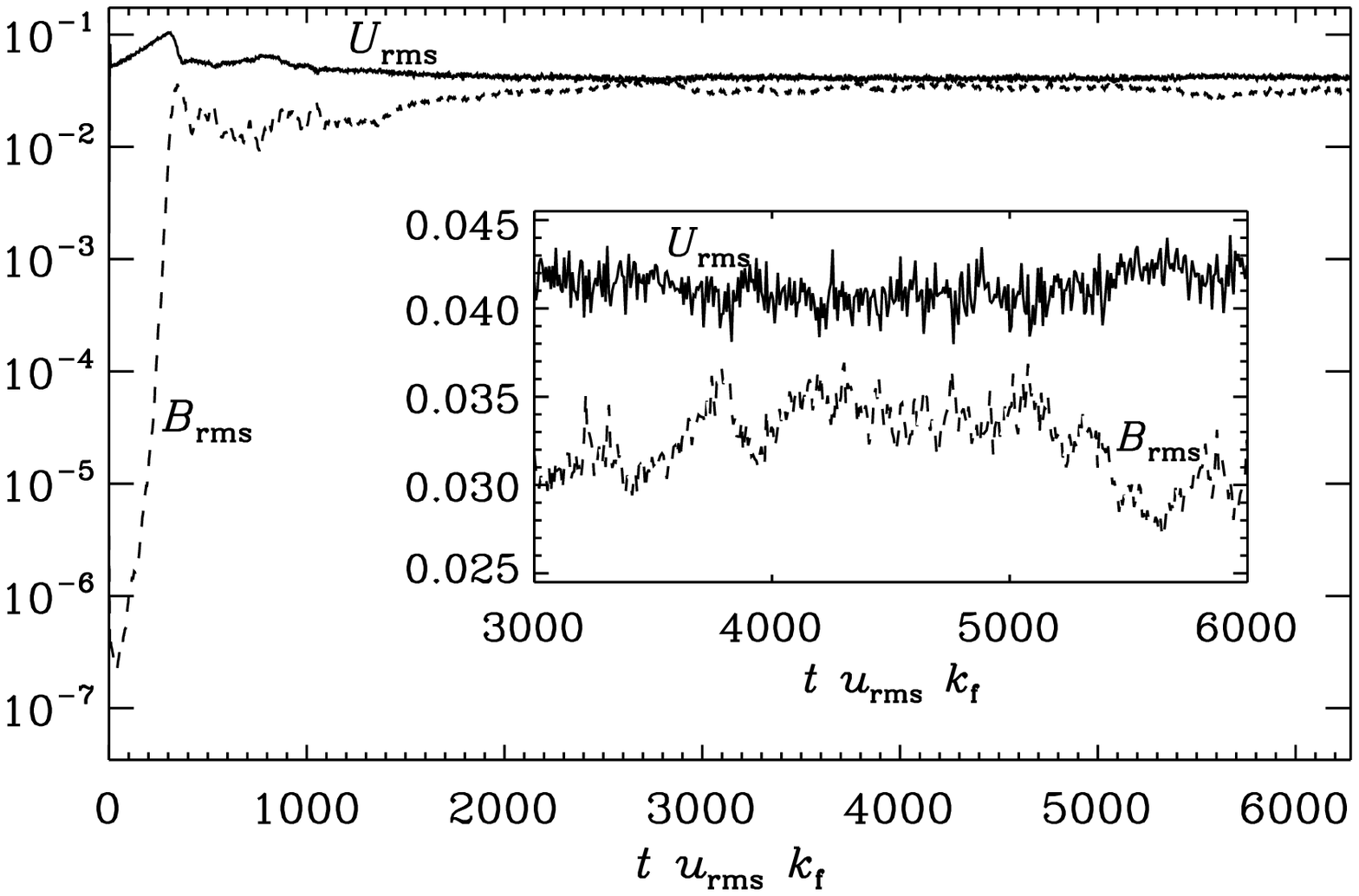}}
\resizebox{\hsize}{!}{\includegraphics{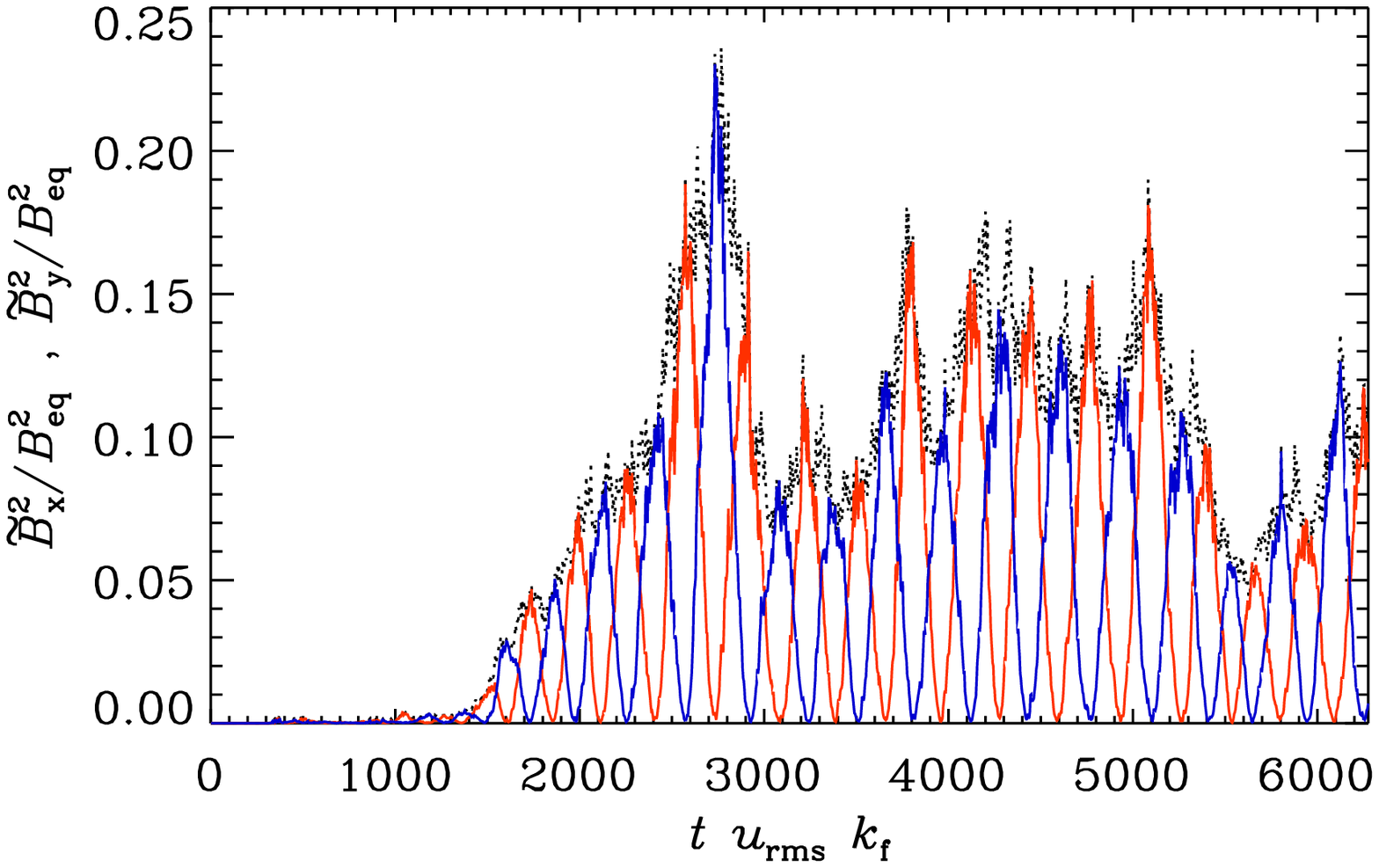}}
\resizebox{\hsize}{!}{\includegraphics{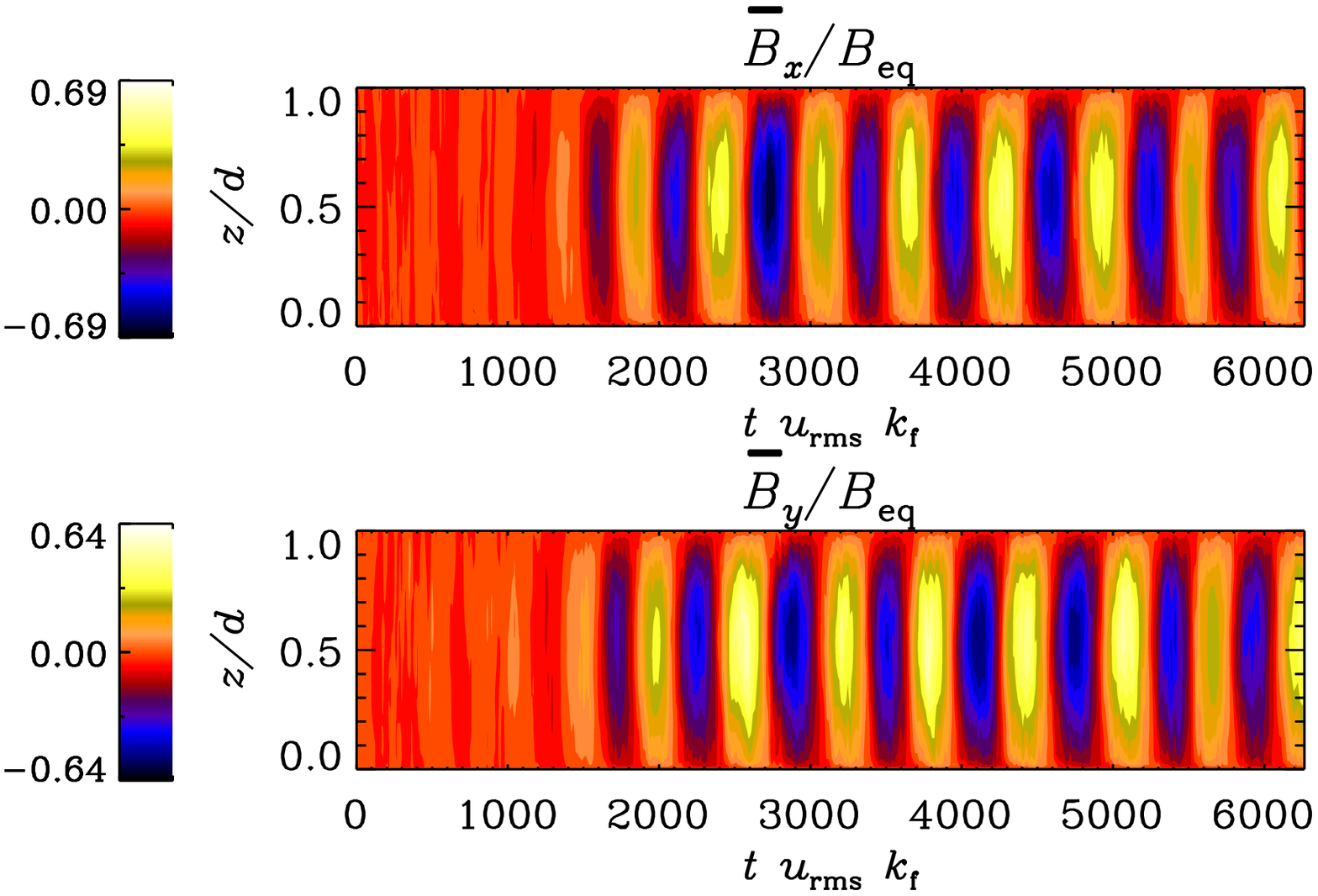}}
\caption{Dynamo evolution for Case B2. 
{\it Top:} time evolution of the rms velocity and magnetic 
field (the inset highlights the modulation during the nonlinear
phase).
{\it Middle:} volume-averaged squared horizontal fields,
$\tilde{B}_x^2/\Beq^2$ (blue) and
$\tilde{B}_y^2/\Beq^2$ (red), as functions of time;
the black dotted line shows 
$(\tilde{B}_x^2+\tilde{B}_y^2)/\Beq^2$.
{\it Bottom:} time and $z$-dependence of the mean
horizontal magnetic fields.}
\label{fig:b2}
\end{figure}

Even in the case of the large-scale dynamo in the reference solution,
there is some evidence of modulation in the mean magnetic field,
with the peak horizontally averaged magnetic field varying noticeably
from one cycle to the next. 
Small increases in the Rayleigh number amplify this effect.
This is illustrated in Fig.~\ref{fig:b2}, which shows the large-scale
dynamo for case B2 ($\Ray = 3\times10^7\approx 5.0\,\Ray_{\rm crit}$). 
In this simulation, the depth-averaged mean squared horizontal magnetic 
field (at $\sim2750$ convective turnover times) peaks at 
approximately $25$\% of the squared equipartition field strength.
The weakest cycles peak at approximately $5$\% of this 
equipartition value. 
Some fluctuations are visible in both the $\Urms(t)$ and $\brms(t)$ 
time-series, although these are fairly modest, particularly in the
case of $\Urms(t)$.

\begin{figure}[t]
\resizebox{\hsize}{!}{\includegraphics{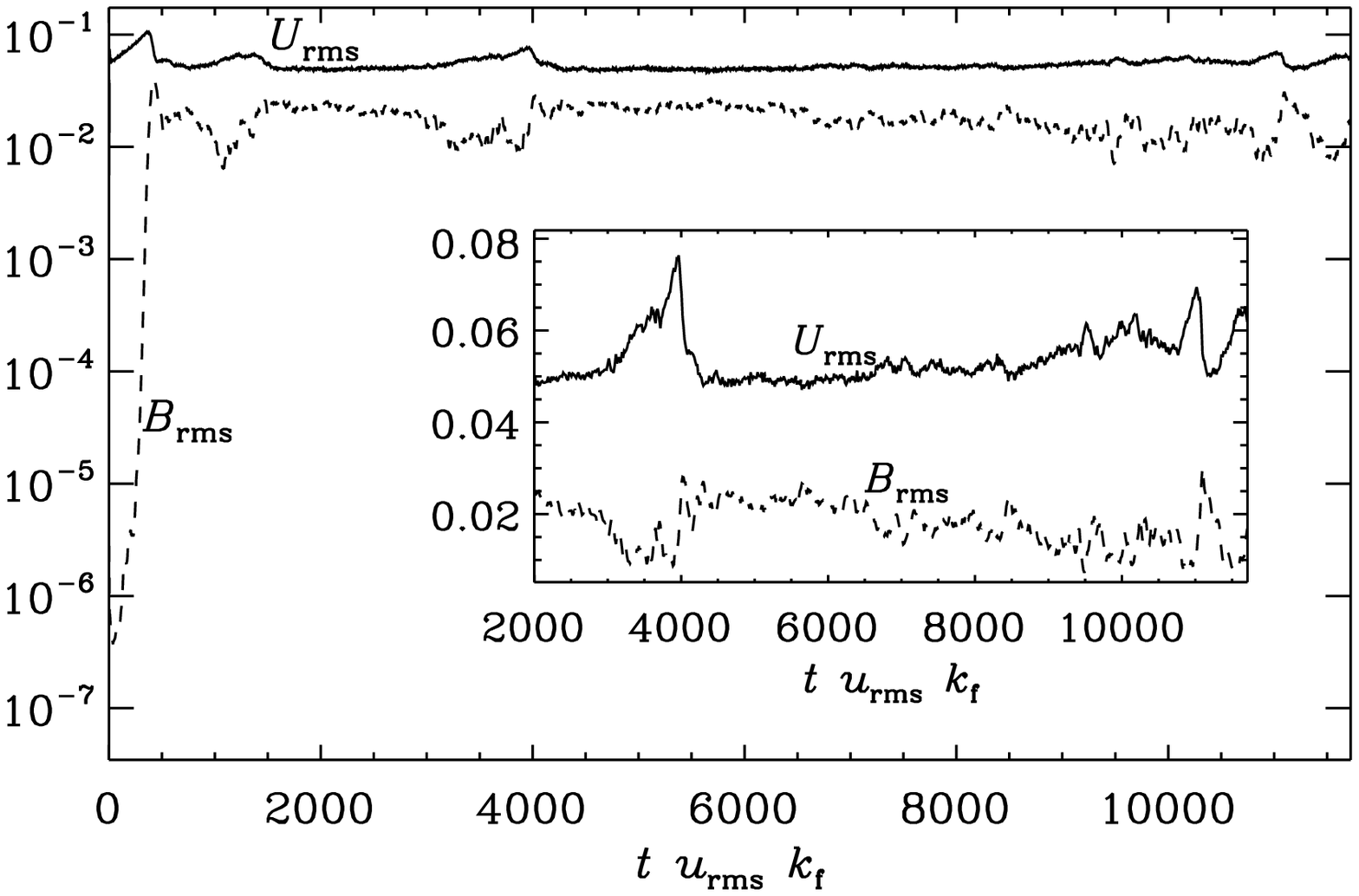}}
\resizebox{\hsize}{!}{\includegraphics{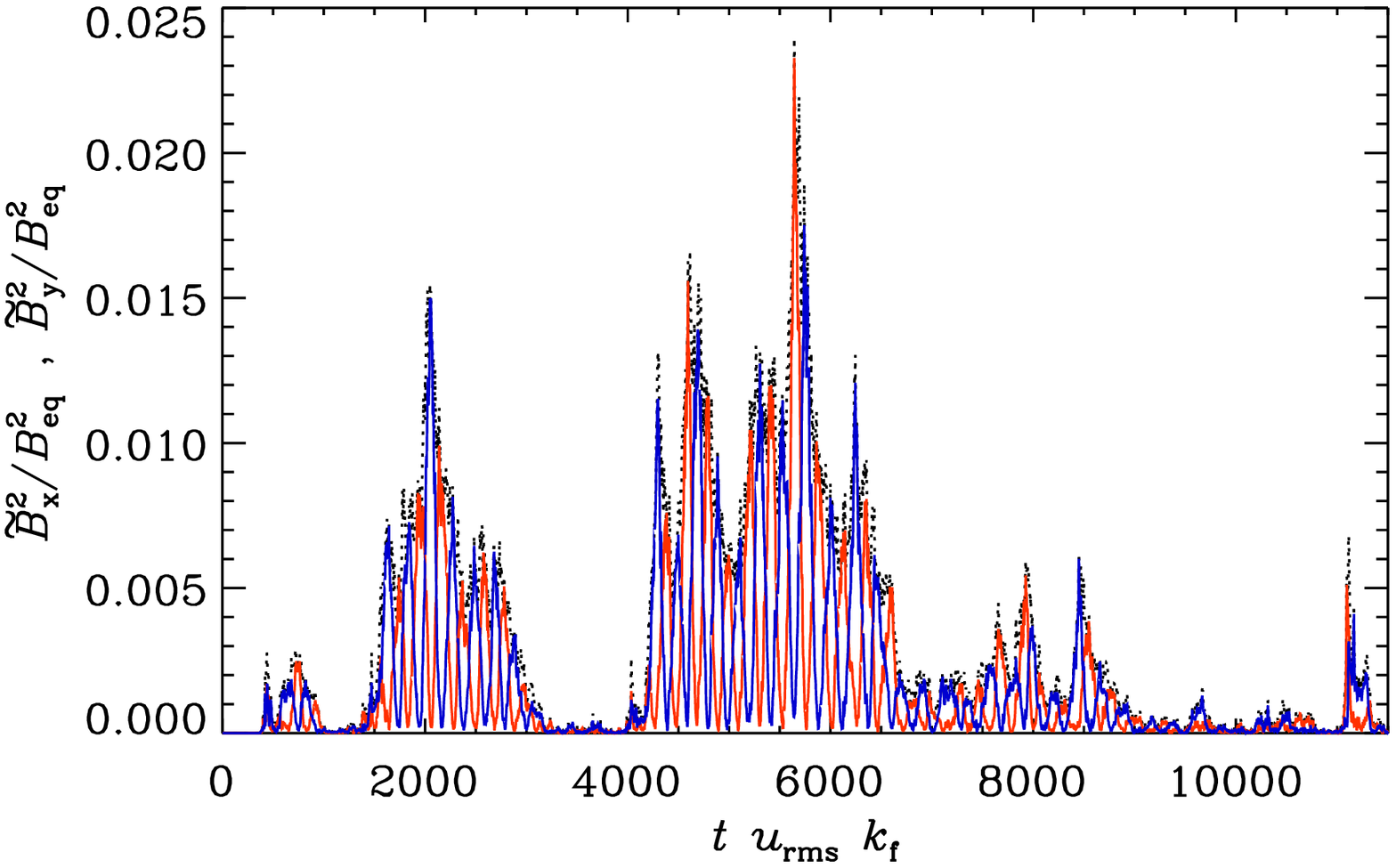}}
\resizebox{\hsize}{!}{\includegraphics{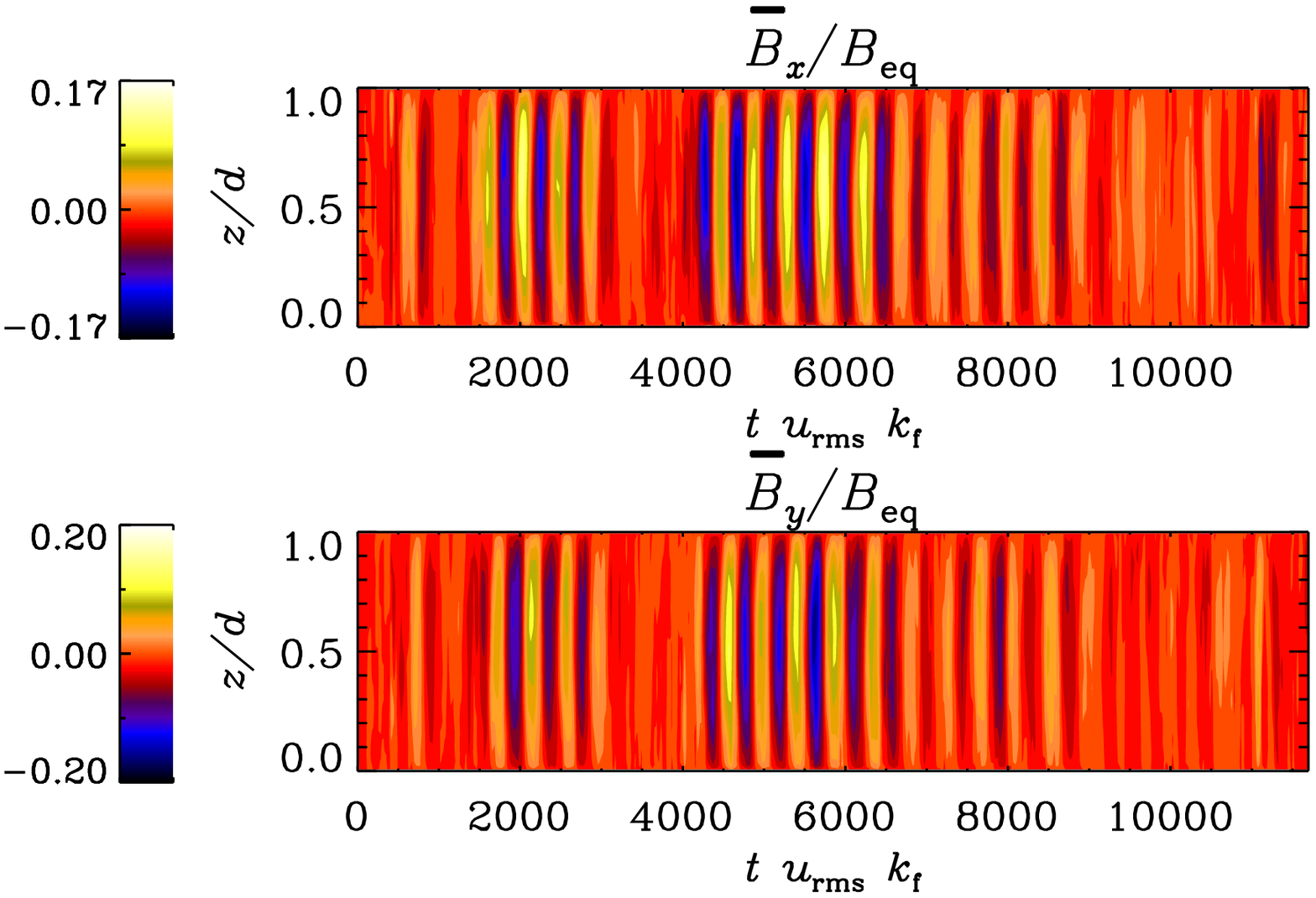}}
\caption{As Fig.~\ref{fig:b2}, but here for Case B3a.}
\label{fig:b3}
\end{figure}

\begin{figure}[t]
\resizebox{\hsize}{!}{\includegraphics{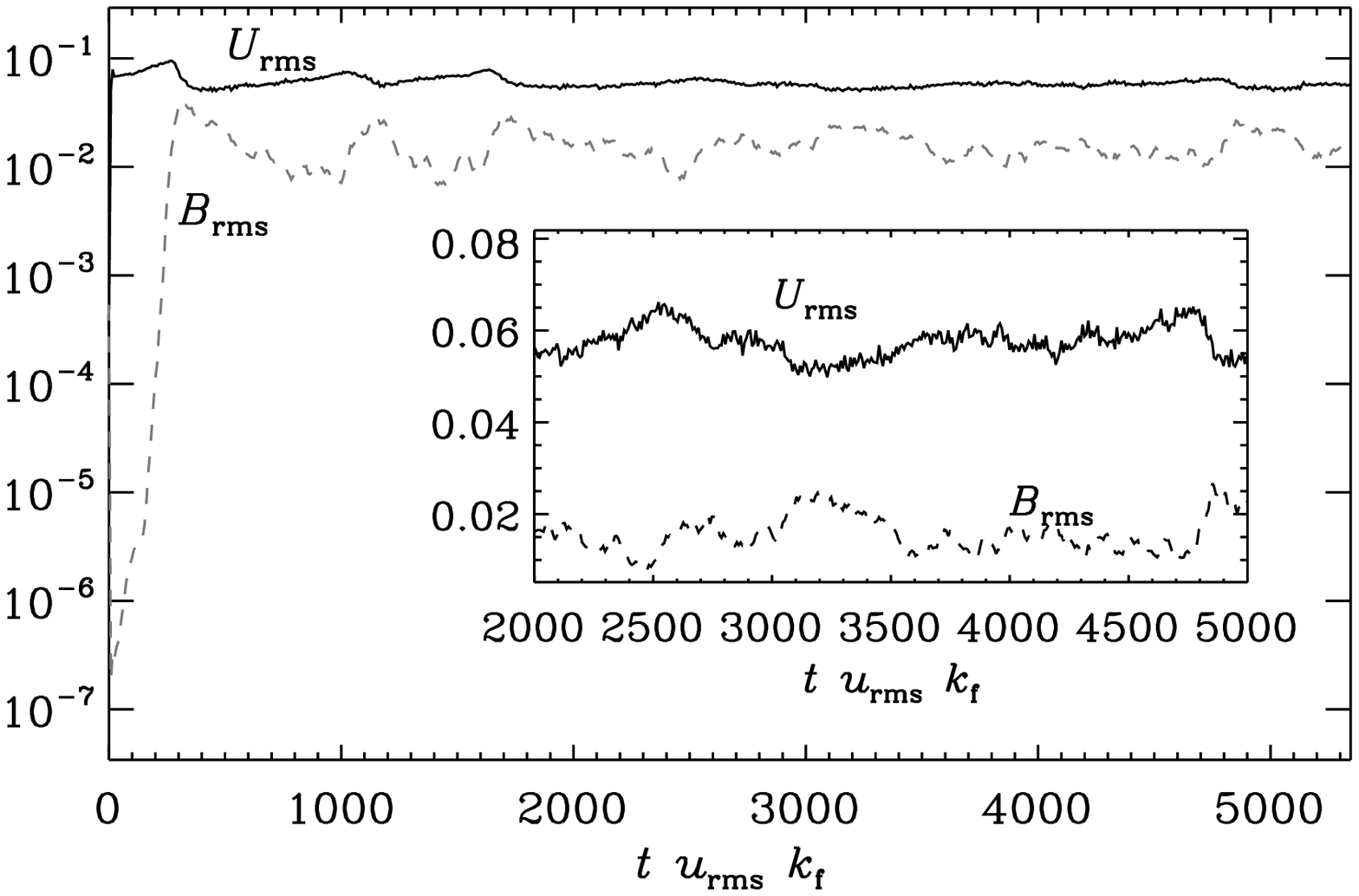}}
\resizebox{\hsize}{!}{\includegraphics{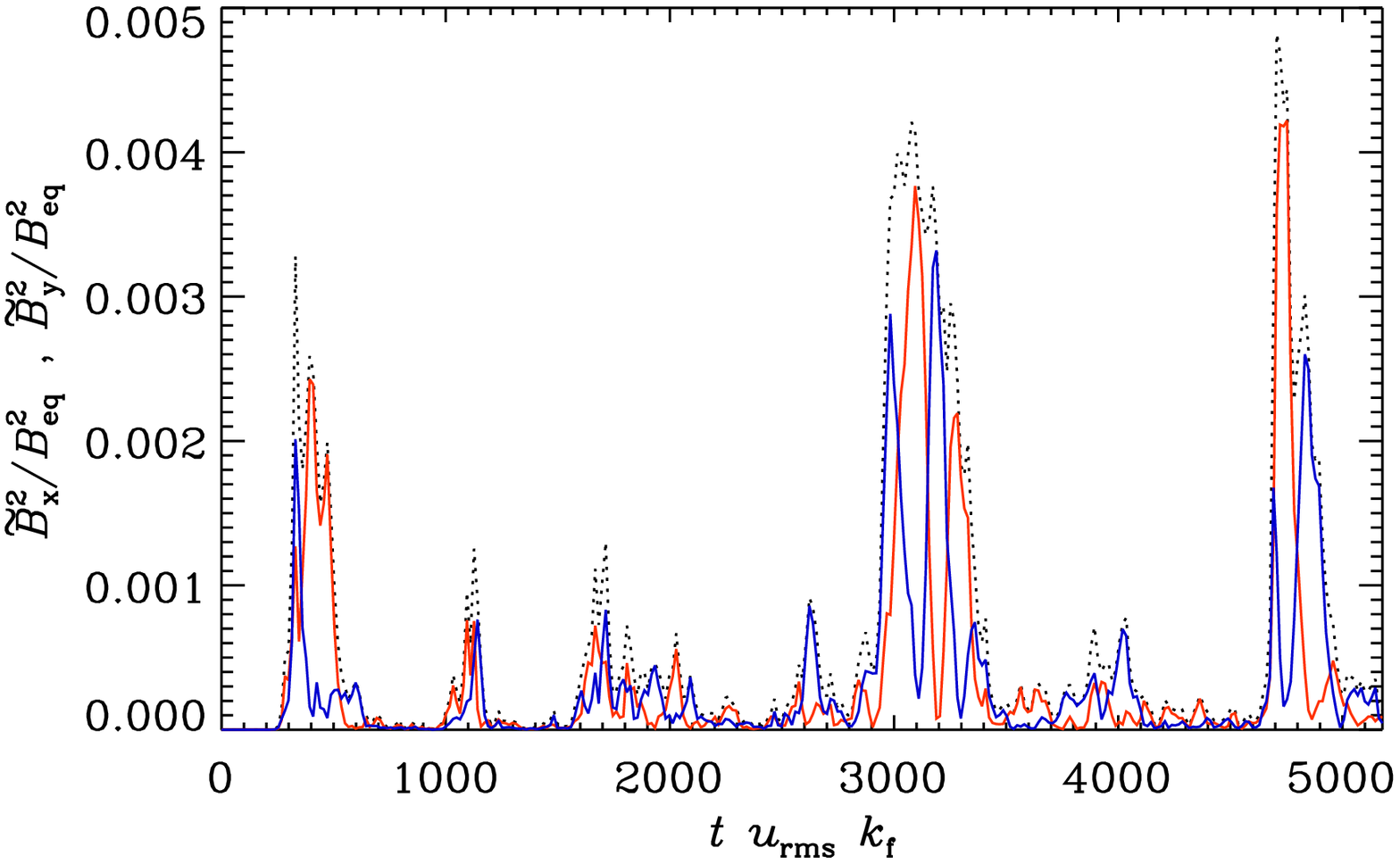}}
\resizebox{\hsize}{!}{\includegraphics{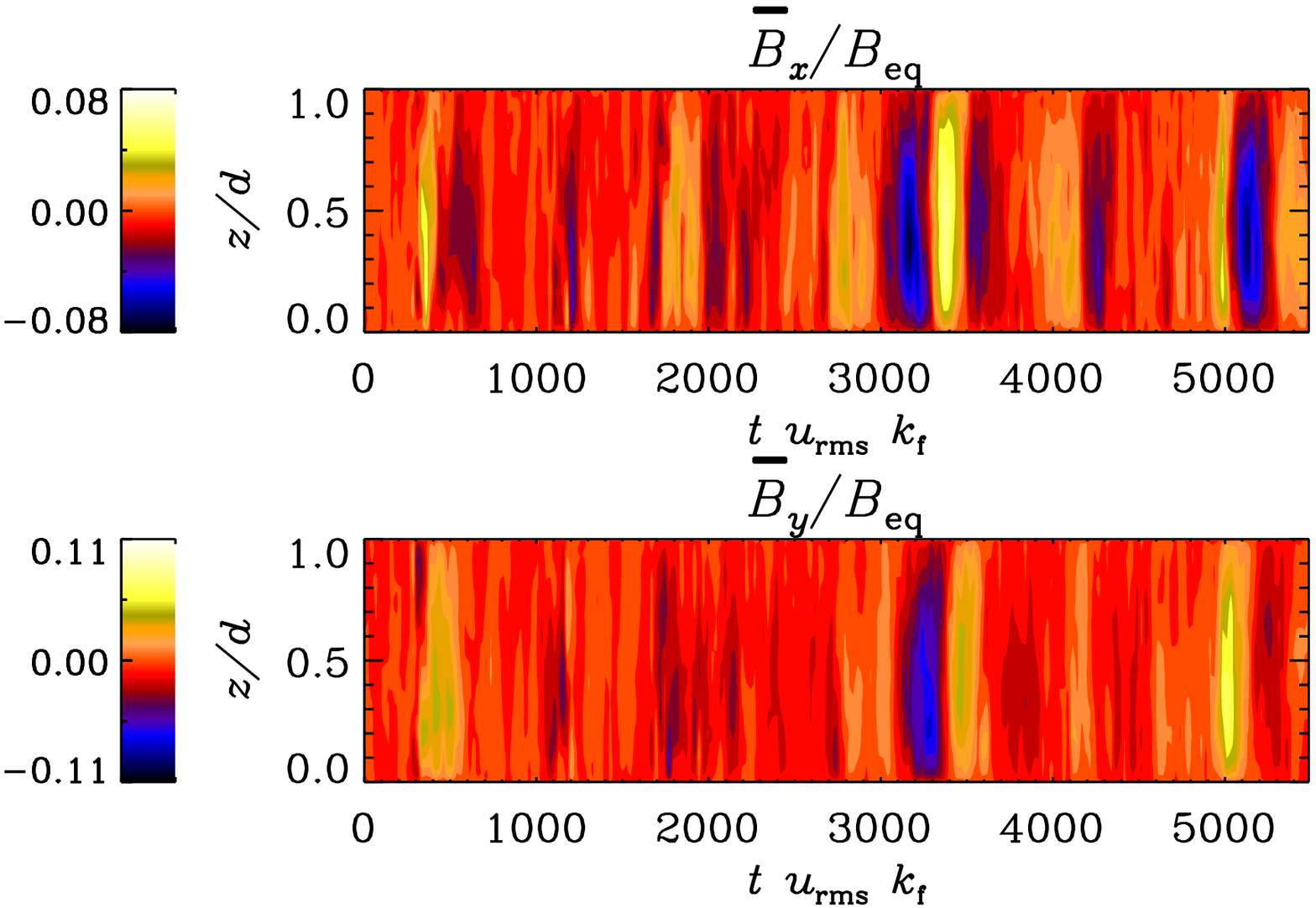}}
\caption{As Fig.~\ref{fig:b2}, but here for Case B4.}
\label{fig:but_B4PC}
\end{figure}

Moving to higher Rayleigh numbers, we see much more
dramatic modulation. 
Figure~\ref{fig:b3} shows the evolution of the large-scale
dynamo for case B3a.
Whilst the large-scale dynamo is generally weaker (in terms
of its equipartition value) than similar dynamos at lower Rayleigh 
number, it is still possible to identify well-defined oscillations
in the mean horizontal magnetic field.
However, the modulation is now extremely pronounced, with 
``active phases'' of large-scale cyclic behaviour punctuated 
by periods of negligible large-scale activity (during which the 
generated magnetic field is predominantly small scale).
At even higher Rayleigh numbers, 
this modulational pattern reverses, with bursts of large-scale activity
surrounded by long periods of relative inactivity (see 
Fig.~\ref{fig:but_B4PC}). 
For case B3a, large-scale activity cycles can
be seen over approximately two thirds of the time-series (mostly 
active phases, but with recurrent periods of inactivity). 
This behaviour becomes increasingly intermittent
at larger values of the Rayleigh number: B4 is active for about 
one third of the time
series, whereas B5 is active for approximately a quarter 
of the time.

Certainly in the case of B3a, this modulational behaviour is 
rather reminiscent of that observed in the solar cycle, where 
extended phases of reduced magnetic activity \citep[such as the 
Maunder Minimum, see, e.g.][]{Eddy1976}, often referred to as
``grand minima'', have been a recurrent
feature of the activity pattern over at least the last 10,000 years 
\citep[see, e.g.][]{McCrackenetal2013}.
We stress again that we do not claim to be modelling the solar dynamo
in this paper.
Nevertheless there may be some similarities between the
modulational mechanism in these simulations with that of the Sun.
As can clearly be seen in Figs.~\ref{fig:b3}
and~\ref{fig:but_B4PC}, the rms velocity in these simulations 
tends to increase during grand minima. 
This corresponds to the re-emergence of the large-scale vortex
instability. 
During these inactive phases, the magnetic field is no longer 
strong enough to suppress this large-scale flow (this can be seen in 
the time-series of the rms magnetic field), so it can again grow.
It then becomes temporarily suppressed again once the magnetic
field reaches dynamically significant levels. 
A number of authors have proposed mean-field models of the 
solar dynamo in which long-term modulation arises as the result
of the magnetic field inhibiting the large-scale differential rotation 
\citep[see, e.g.][]{Tobias1996,Brookeetal2002,Bushby2006}.
In these models, the occurrence of grand minima depended upon
there being a separation in timescales between viscous and 
magnetic diffusion, so that the perturbations to the flow velocity
relaxed over a much longer time-scale than the period of oscillation 
of the dynamo. 
There is a similar separation in time-scales in these modulated 
convective dynamo simulations, with the large-scale vortex growing 
on a much longer time-scale than the cycle period of the
large-scale dynamos. 
The exchange of energy between the magnetic field and the flow 
can therefore give rise to this modulational behaviour.

\section{Conclusions and discussion}

One of the great challenges for dynamo theorists is to explain the 
origin of large-scale astrophysical magnetic fields.
As anticipated from theory, helically-forced turbulence in an 
electrically conducting fluid can drive a large-scale dynamo in a 
Cartesian domain \citep{Br01}. 
Such idealised flows can never be realised in nature, although 
the effects of rotation are believed to give rise to helical convective 
motions in many astrophysical bodies, so similar large-scale 
dynamos might be expected in such cases.
However, the dynamo properties of rapidly rotating convection 
appear to be rather subtle. 
Near-onset rapidly rotating convection can drive a large-scale dynamo 
in a Cartesian domain \citep[][]{ChildressSoward1972} but,
unless there is an imposed shear \citep[which tends to promote 
large-scale dynamo action, see e.g.][]{KKB08,HP09}, dynamos tend to be
small-scale in the more turbulent regime that is relevant for astrophysics. 

Previous work \citep[][]{Kapylaetal2013,MasadaSano2014a,MasadaSano2014b}
has demonstrated that it is possible to find a large-scale dynamo in 
moderately turbulent, rapidly rotating convection in a Cartesian domain
(without shear). 
These simulations are in a 
parameter regime in which the large-scale vortex instability can operate.
Building on this previous work,
we have carried out a detailed analysis of the underlying dynamo mechanism
and have demonstrated that the large-scale dynamo is driven by
the components of the flow with a horizontal wavenumber in the range 
$3 \le k/k_{\rm 1} \le 5$. 
In particular, this confirms that the large-scale vortex itself (i.e.\ the 
$k=k_{\rm 1}$ mode, which becomes strongly damped) does not play
a direct role in sustaining the dynamo. 
Having said that, the structure of the flow at the driving scales is
still influenced (to some extent, at least) by the tendency for the 
large-scale vortex instability to transfer energy to larger scales. 
So, even though the large-scale vortex itself is damped, the effects of 
the underlying instability should not be discounted.
In particular the initialisation of the large-scale 
dynamo, in this parameter regime, may well depend upon there being an 
efficient large-scale vortex instability in the first place.
As a prelude to a possible benchmarking exercise, we have verified that 
this dynamo can be reproduced by three 
different codes, all of whom produce quantitatively comparable solutions 
in the nonlinear regime.

Provided that the magnetic boundary conditions (which appear to be 
very important in this context) allow magnetic flux to escape from the domain, we
have shown that this large-scale dynamo is robust to moderate changes in
the parameters.
The dynamo appears to be largely insensitive to the level of stratification
within the domain, although the calculations that we have carried out 
do suggest that large-scale dynamos in more weakly-stratified domains
may tend to have longer cycle periods.
Moving towards convective onset, the form of the dynamo remains largely
the same, although the cycle period of the large-scale 
dynamo increases (very dramatically at the lowest Rayleigh numbers).
As the level of convective driving is increased, the cycle period decreases, 
and the cycle becomes increasingly modulated.
At moderate Rayleigh numbers, this modulation is almost solar-like, with
active phases punctuated by inactive grand minima. 
At higher driving, the large-scale dynamo becomes increasingly intermittent.
The modulation is driven by an exchange of energy between the magnetic
field and the flow.

Future investigations in this field will focus on even more turbulent
regimes at higher Rayleigh numbers, exploring broad ranges of magnetic
and thermal Prandtl numbers and rotation rates.
The dynamo mechanism could also be analysed from the perspective
of mean-field dynamo theory.
We have seen that there is a positive correlation between the
components of the mean horizontal magnetic field and the mean horizontal
electromotive force, just as expected for an $\alpha^2$ dynamo located
in the northern hemisphere of a rotating star, but more could be done
to clarify this.
It would also be worthwhile to investigate possible connections between these
moderately-turbulent large-scale dynamos and the near-onset Boussinesq 
dynamos. 
We have shown that vertical field boundary conditions seem to promote 
this large-scale dynamo, but this does not rule out the possibility that
a similar dynamo could operate in a turbulent regime with the horizontal field 
boundary conditions adopted in most Boussinesq studies. 
Finally, there is the intriguing question of the extent to which these 
Cartesian dynamos are of relevance to particular astrophysical bodies. 
Is it possible to find analogous dynamos in a sphere or spherical
shell? 
Obviously the cycle periods of these large-scale dynamos are too 
long to be of direct relevance to solar-type dynamos, but we can 
speculate that there may be some possible application to planetary 
dynamos (where, at least in the case of the Earth, we know that there
are polarity reversals over long time-scales).

\begin{acknowledgements}

We would welcome any interest from the dynamo community in the 
benchmark exercise that is proposed in the appendix.
This work has been supported in part by
the National Science Foundation (grant AST1615100),
the Research Council of Norway under the FRINATEK (grant 231444),
the Swedish Research Council (grant 621-2011-5076),
the University of Colorado through its support of the George Ellery Hale
visiting faculty appointment (AB), the Academy of
Finland (grant 272157) to the ReSoLVE Centre of Excellence (PJK, MJK)
and the UK Natural Environment Research Council (grant NE/M017893/1, CG).
Some of the simulations were performed using the supercomputers hosted 
by CSC -- IT Center for Science Ltd.\ in Espoo, Finland, who are
administered by the Finnish Ministry of Education.
Much of this work made use of the facilities of N8 HPC Centre of Excellence, 
provided and funded by the N8 consortium and EPSRC 
(Grant No.EP/K000225/1). 
The Centre is co-ordinated by the Universities of Leeds and Manchester.
PB would also like to acknowledge a useful discussion with Michael 
Proctor that helped to clarify the nature of the underlying dynamo 
mechanism.
We would also like to thank the referee for their helpful
comments and suggestions.
\end{acknowledgements}

\bibliographystyle{aa}
\bibliography{bench} 

\begin{thebibliography}{62}
\expandafter\ifx\csname natexlab\endcsname\relax\def\natexlab#1{#1}\fi

\bibitem[{{Brandenburg}(2001)}]{Br01}
{Brandenburg}, A. 2001, \apj, 550, 824

\bibitem[{{Brandenburg} {et~al.}(1998){Brandenburg}, {Saar}, \&
  {Turpin}}]{Brandenburgetal1998}
{Brandenburg}, A., {Saar}, S.~H., \& {Turpin}, C.~R. 1998, \apjl, 498, L51

\bibitem[{{Brooke} {et~al.}(2002){Brooke}, {Moss}, \&
  {Phillips}}]{Brookeetal2002}
{Brooke}, J., {Moss}, D., \& {Phillips}, A. 2002, \aap, 395, 1013

\bibitem[{{Bushby}(2006)}]{Bushby2006}
{Bushby}, P.~J. 2006, \mnras, 371, 772

\bibitem[{{Calkins} {et~al.}(2015){Calkins}, {Julien}, {Tobias}, \&
  {Aurnou}}]{Calkinsetal2015}
{Calkins}, M.~A., {Julien}, K., {Tobias}, S.~M., \& {Aurnou}, J.~M. 2015, J.
  Fluid Mech., 780, 143

\bibitem[{{Calkins} {et~al.}(2016){Calkins}, {Julien}, {Tobias}, {Aurnou}, \&
  {Marti}}]{Calkinsetal2016}
{Calkins}, M.~A., {Julien}, K., {Tobias}, S.~M., {Aurnou}, J.~M., \& {Marti},
  P. 2016, Phys. Rev. E, 93, 023115

\bibitem[{{Cattaneo} {et~al.}(2003){Cattaneo}, {Emonet}, \&
  {Weiss}}]{Cattaneoetal2003}
{Cattaneo}, F., {Emonet}, T., \& {Weiss}, N. 2003, \apj, 588, 1183

\bibitem[{{Cattaneo} \& {Hughes}(2006)}]{CattaneoHughes2006}
{Cattaneo}, F. \& {Hughes}, D.~W. 2006, J. Fluid Mech., 553, 401

\bibitem[{{Chan}(2007)}]{Chan2007}
{Chan}, K.~L. 2007, Astron. Nachr., 328, 1059

\bibitem[{{Chan} \& {Mayr}(2013)}]{ChanMayr2013}
{Chan}, K.~L. \& {Mayr}, H.~G. 2013, Earth Plan. Sci. Lett., 371, 212

\bibitem[{{Chandrasekhar}(1961)}]{Chandrasekhar1961}
{Chandrasekhar}, S. 1961, {Hydrodynamic and hydromagnetic stability} (Oxford
  University Press)

\bibitem[{{Childress} \& {Soward}(1972)}]{ChildressSoward1972}
{Childress}, S. \& {Soward}, A.~M. 1972, Phys. Rev. Lett., 29, 837

\bibitem[{{Christensen} {et~al.}(2001){Christensen}, {Aubert}, {Cardin},
  {Dormy}, {Gibbons}, {Glatzmaier}, {Grote}, {Honkura}, {Jones}, {Kono},
  {Matsushima}, {Sakuraba}, {Takahashi}, {Tilgner}, {Wicht}, \&
  {Zhang}}]{Christensenetal2001}
{Christensen}, U.~R., {Aubert}, J., {Cardin}, P., {et~al.} 2001, Phys. Earth
  Plan. Int., 128, 25

\bibitem[{{Clarke}(1996)}]{Clarke1996}
{Clarke}, D.~A. 1996, \apj, 457, 291

\bibitem[{{Colella} \& {Woodward}(1984)}]{ColellaWoodward1984}
{Colella}, P. \& {Woodward}, P.~R. 1984, J. Comp. Phys., 54, 174

\bibitem[{{Courvoisier} {et~al.}(2009){Courvoisier}, {Hughes}, \&
  {Proctor}}]{Courvoisieretal2009}
{Courvoisier}, A., {Hughes}, D.~W., \& {Proctor}, M.~R.~E. 2009, Proceedings of
  the Royal Society of London Series A, 466, 583

\bibitem[{{Dobler} {et~al.}(2006){Dobler}, {Stix}, \&
  {Brandenburg}}]{Dobler_etal06}
{Dobler}, W., {Stix}, M., \& {Brandenburg}, A. 2006, \apj, 638, 336

\bibitem[{{Eddy}(1976)}]{Eddy1976}
{Eddy}, J.~A. 1976, Science, 192, 1189

\bibitem[{{Evans} \& {Hawley}(1988)}]{EvansHawley1988}
{Evans}, C.~R. \& {Hawley}, J.~F. 1988, \apj, 332, 659

\bibitem[{{Fautrelle} \& {Childress}(1982)}]{FautrelleChildress1982}
{Fautrelle}, Y. \& {Childress}, S. 1982, Geophys. Astrophys. Fluid Dynam., 22,
  235

\bibitem[{{Favier} \& {Bushby}(2012)}]{FavierBushby2012}
{Favier}, B. \& {Bushby}, P.~J. 2012, Journal of Fluid Mechanics, 690, 262

\bibitem[{{Favier} \& {Bushby}(2013)}]{FavierBushby2013}
{Favier}, B. \& {Bushby}, P.~J. 2013, J. Fluid Mech., 723, 529

\bibitem[{{Favier} \& {Proctor}(2013)}]{FavierProctor2013}
{Favier}, B. \& {Proctor}, M.~R.~E. 2013, \pre, 88, 053011

\bibitem[{{Favier} {et~al.}(2014){Favier}, {Silvers}, \&
  {Proctor}}]{Favieretal2014}
{Favier}, B., {Silvers}, L.~J., \& {Proctor}, M.~R.~E. 2014, Phys. Fluids, 26,
  096605

\bibitem[{{Guervilly} {et~al.}(2014){Guervilly}, {Hughes}, \&
  {Jones}}]{Guervillyetal2014}
{Guervilly}, C., {Hughes}, D.~W., \& {Jones}, C.~A. 2014, J. Fluid Mech., 758,
  407

\bibitem[{{Guervilly} {et~al.}(2015){Guervilly}, {Hughes}, \&
  {Jones}}]{Guervillyetal2015}
{Guervilly}, C., {Hughes}, D.~W., \& {Jones}, C.~A. 2015, \pre, 91, 041001

\bibitem[{{Hughes} \& {Proctor}(2009)}]{HP09}
{Hughes}, D.~W. \& {Proctor}, M.~R.~E. 2009, Phys. Rev. Lett., 102, 044501

\bibitem[{{Jones}(2011)}]{Jones2011}
{Jones}, C.~A. 2011, Ann. Rev. Fluid Mech., 43, 583

\bibitem[{{Jones} {et~al.}(2011){Jones}, {Boronski}, {Brun}, {Glatzmaier},
  {Gastine}, {Miesch}, \& {Wicht}}]{Jonesetal2011}
{Jones}, C.~A., {Boronski}, P., {Brun}, A.~S., {et~al.} 2011, \icarus, 216, 120

\bibitem[{{Jones} \& {Roberts}(2000)}]{JonesRoberts2000}
{Jones}, C.~A. \& {Roberts}, P.~H. 2000, J. Fluid Mech., 404, 311

\bibitem[{{Julien} {et~al.}(2012){Julien}, {Rubio}, {Grooms}, \&
  {Knobloch}}]{Julienetal2012}
{Julien}, K., {Rubio}, A.~M., {Grooms}, I., \& {Knobloch}, E. 2012, Geophysical
  and Astrophysical Fluid Dynamics, 106, 392

\bibitem[{{K{\"a}pyl{\"a}} {et~al.}(2008){K{\"a}pyl{\"a}}, {Korpi}, \&
  {Brandenburg}}]{KKB08}
{K{\"a}pyl{\"a}}, P.~J., {Korpi}, M.~J., \& {Brandenburg}, A. 2008, \aap, 491,
  353

\bibitem[{{K{\"a}pyl{\"a}} {et~al.}(2009){K{\"a}pyl{\"a}}, {Korpi}, \&
  {Brandenburg}}]{Kapylaetal2009}
{K{\"a}pyl{\"a}}, P.~J., {Korpi}, M.~J., \& {Brandenburg}, A. 2009, \apj, 697,
  1153

\bibitem[{{K{\"a}pyl{\"a}} {et~al.}(2013){K{\"a}pyl{\"a}}, {Mantere}, \&
  {Brandenburg}}]{Kapylaetal2013}
{K{\"a}pyl{\"a}}, P.~J., {Mantere}, M.~J., \& {Brandenburg}, A. 2013, Geophys.
  Astrophys. Fluid Dynam., 107, 244

\bibitem[{{K{\"a}pyl{\"a}} {et~al.}(2011){K{\"a}pyl{\"a}}, {Mantere}, \&
  {Hackman}}]{Kapylaetal2011}
{K{\"a}pyl{\"a}}, P.~J., {Mantere}, M.~J., \& {Hackman}, T. 2011, \apj, 742, 34

\bibitem[{{Kunnen} {et~al.}(2016){Kunnen}, {Ostilla-M{\'o}nico}, {van der
  Poel}, {Verzicco}, \& {Lohse}}]{Kunnenetal2016}
{Kunnen}, R.~P.~J., {Ostilla-M{\'o}nico}, R., {van der Poel}, E.~P.,
  {Verzicco}, R., \& {Lohse}, D. 2016, Journal of Fluid Mechanics, 799, 413

\bibitem[{{MacGregor} \& {Charbonneau}(1997)}]{MacGregorCharbonneau1997}
{MacGregor}, K.~B. \& {Charbonneau}, P. 1997, \apj, 486, 484

\bibitem[{{Mantere} {et~al.}(2011){Mantere}, {K{\"a}pyl{\"a}}, \&
  {Hackman}}]{Mantereetal2011}
{Mantere}, M.~J., {K{\"a}pyl{\"a}}, P.~J., \& {Hackman}, T. 2011, Astron.
  Nachr., 332, 876

\bibitem[{{Marti} {et~al.}(2014){Marti}, {Schaeffer}, {Hollerbach},
  {C{\'e}bron}, {Nore}, {Luddens}, {Guermond}, {Aubert}, {Takehiro}, {Sasaki},
  {Hayashi}, {Simitev}, {Busse}, {Vantieghem}, \& {Jackson}}]{Martietal2014}
{Marti}, P., {Schaeffer}, N., {Hollerbach}, R., {et~al.} 2014, Geophysical
  Journal International, 197, 119

\bibitem[{{Masada} \& {Sano}(2014{\natexlab{a}})}]{MasadaSano2014a}
{Masada}, Y. \& {Sano}, T. 2014{\natexlab{a}}, \pasj, 66, 2

\bibitem[{{Masada} \& {Sano}(2014{\natexlab{b}})}]{MasadaSano2014b}
{Masada}, Y. \& {Sano}, T. 2014{\natexlab{b}}, \apjl, 794, L6

\bibitem[{{Masada} \& {Sano}(2016)}]{MasadaSano2016}
{Masada}, Y. \& {Sano}, T. 2016, \apjl, 822, L22

\bibitem[{{Matthews} {et~al.}(1995){Matthews}, {Proctor}, \&
  {Weiss}}]{Matthewsetal1995}
{Matthews}, P.~C., {Proctor}, M.~R.~E., \& {Weiss}, N.~O. 1995, J. Fluid Mech.,
  305, 281

\bibitem[{{McCracken} {et~al.}(2013){McCracken}, {Beer}, {Steinhilber}, \&
  {Abreu}}]{McCrackenetal2013}
{McCracken}, K., {Beer}, J., {Steinhilber}, F., \& {Abreu}, J. 2013, \ssr, 176,
  59

\bibitem[{{Meneguzzi} \& {Pouquet}(1989)}]{MeneguzziPouquet1989}
{Meneguzzi}, M. \& {Pouquet}, A. 1989, J. Fluid Mech., 205, 297

\bibitem[{{Mizerski} \& {Tobias}(2013)}]{MizerskiTobias2013}
{Mizerski}, K.~A. \& {Tobias}, S.~M. 2013, Geophys. Astrophys. Fluid Dynam.,
  107, 218

\bibitem[{{Moffatt}(1978)}]{Moffatt1978}
{Moffatt}, H.~K. 1978, {Magnetic field generation in electrically conducting
  fluids} (Cambridge University Press)

\bibitem[{{Parker}(1993)}]{Parker1993}
{Parker}, E.~N. 1993, \apj, 408, 707

\bibitem[{{Rotvig} \& {Jones}(2002)}]{RotvigJones2002}
{Rotvig}, J. \& {Jones}, C.~A. 2002, \pre, 66, 056308

\bibitem[{{Sano} {et~al.}(1999){Sano}, {Inutsuka}, \& {Miyama}}]{Sanoetal1999}
{Sano}, T., {Inutsuka}, S., \& {Miyama}, S.~M. 1999, in Astrophysics and Space
  Science Library, Vol. 240, Numerical Astrophysics, ed. S.~M. {Miyama},
  K.~{Tomisaka}, \& T.~{Hanawa}, 383

\bibitem[{{Soward}(1974)}]{Soward1974}
{Soward}, A.~M. 1974, Royal Society of London Philosophical Transactions Series
  A, 275, 611

\bibitem[{{St.~Pierre}(1993)}]{StPierre1993}
{St.~Pierre}, M.~G. 1993, in Solar and Planetary Dynamos, ed. M.~R.~E.
  {Proctor}, P.~C. {Matthews}, \& A.~M. {Rucklidge}, 295--302

\bibitem[{{Stellmach} \& {Hansen}(2004)}]{StellmachHansen2004}
{Stellmach}, S. \& {Hansen}, U. 2004, \pre, 70, 056312

\bibitem[{{Stellmach} {et~al.}(2014){Stellmach}, {Lischper}, {Julien}, {Vasil},
  {Cheng}, {Ribeiro}, {King}, \& {Aurnou}}]{Stellmachetal2014}
{Stellmach}, S., {Lischper}, M., {Julien}, K., {et~al.} 2014, Phys. Rev. Lett.,
  113, 254501

\bibitem[{{Stix}(2002)}]{Stix2002}
{Stix}, M. 2002, {The sun: an introduction} (Springer)

\bibitem[{{Stone} \& {Norman}(1992)}]{StoneNorman1992}
{Stone}, J.~M. \& {Norman}, M.~L. 1992, \apjs, 80, 791

\bibitem[{{Tilgner}(2014)}]{Tilgner2014}
{Tilgner}, A. 2014, \pre, 90, 013004

\bibitem[{{Tobias}(1996{\natexlab{a}})}]{Tobias1996b}
{Tobias}, S.~M. 1996{\natexlab{a}}, \apj, 467, 870

\bibitem[{{Tobias}(1996{\natexlab{b}})}]{Tobias1996}
{Tobias}, S.~M. 1996{\natexlab{b}}, \aap, 307, L21

\bibitem[{{Tobias} {et~al.}(2011){Tobias}, {Cattaneo}, \&
  {Brummell}}]{Tobiasetal2011}
{Tobias}, S.~M., {Cattaneo}, F., \& {Brummell}, N.~H. 2011, \apj, 728, 153

\bibitem[{{van Leer}(1979)}]{vanLeer1979}
{van Leer}, B. 1979, J. Comp. Phys., 32, 101

\bibitem[{{Williamson}(1980)}]{Williamson}
{Williamson}, J.~H. 1980, J. Comp. Phys., 35, 48

\end{thebibliography}

\begin{appendix}
\section{A possible dynamo benchmark}
\label{BenchmarkComparison}

This appendix contains further details on the quantitative comparison that 
has been carried out between the three codes that are described in 
Sect.~\ref{subsec:methods}. The aim here is to assess whether or
not this large-scale dynamo solution could form the basis for a 
community-wide nonlinear dynamo benchmark.
We choose to present the details here to encourage other researchers 
with similar codes to try this calculation. 
If there is sufficient enthusiasm from the community to carry out
a detailed benchmark study, we would agree on a common set of 
initial conditions and carry out a detailed analysis of the early 
evolution of the system 
(e.g.\ comparing the mean growth rate of $\Urms(t)$ during the large-scale
vortex phase) as well as the final nonlinear state.
Whilst this solution is relatively complicated, we would (at the very least) 
expect different codes to produce quantitatively comparable statistics.
Here, we focus upon the simpler questions of whether or not the final 
nonlinear solution is robust to small changes in the initial conditions, 
confirming that three independent codes can produce quantitatively 
comparable nonlinear dynamos.

\begin{table}[b]
\caption{\label{table:2}Details on the benchmark simulations.} 
\centering
\begin{tabular}{cccccc}
\hline
\hline
Case & Code & Grid & $\urms$ & $b_{\rm rms}$ & $\tau_{\rm cyc}$ \\
\hline
A1 & 1 & $256^3$ & 0.0355 & 0.0364 & $\sim$1050  \\
A2 & 2 & $192^3$ & 0.0359 & 0.0346 & $\sim$1060 \\
A3 & 3 & $256^3$ & 0.0364 & 0.0327 & $\sim$970 \\
\hline
\end{tabular}
\end{table}

Fixing $\lambda=2$,  $\Pra=\Pm=1$ and $\Tay=5\cdot10^8$, it was first 
confirmed that all three codes 
agree on the critical Rayleigh number, $\Ray_{\rm crit}=6.006\cdot10^6$, 
with exponentially growing (decaying) solutions being obtained for 
values of $\Ray$ that are fractionally above (below) this value. 
Once this agreement was confirmed, nonlinear dynamo runs were carried out 
for the reference solution value of $\Ray=2.4\cdot10^7$ (see the upper 
three rows of Table~\ref{table:1}). 
As has already been described, this reference solution exhibits highly 
non-trivial behaviour, in which the initial convective instability is 
subject to a secondary hydrodynamical instability (corresponding to the 
large-scale vortex). 
The resulting flow then drives a nonlinear large-scale dynamo 
in which the total magnetic and kinetic energies are in a state of 
near-equipartition.  
This solution is therefore an excellent test of all aspects of any 
compressible Cartesian MHD code.  
Although they were all evolved from the same initial polytropic state,
random initial perturbations were applied.
Some quantitative differences between the codes are therefore to be 
expected during the early stages of evolution.
If, however, it is possible to confirm that all of the codes converge 
upon the same nonlinear dynamo, then this comparison 
can be deemed to be successful.

\begin{figure}
\resizebox{\hsize}{!}{\includegraphics{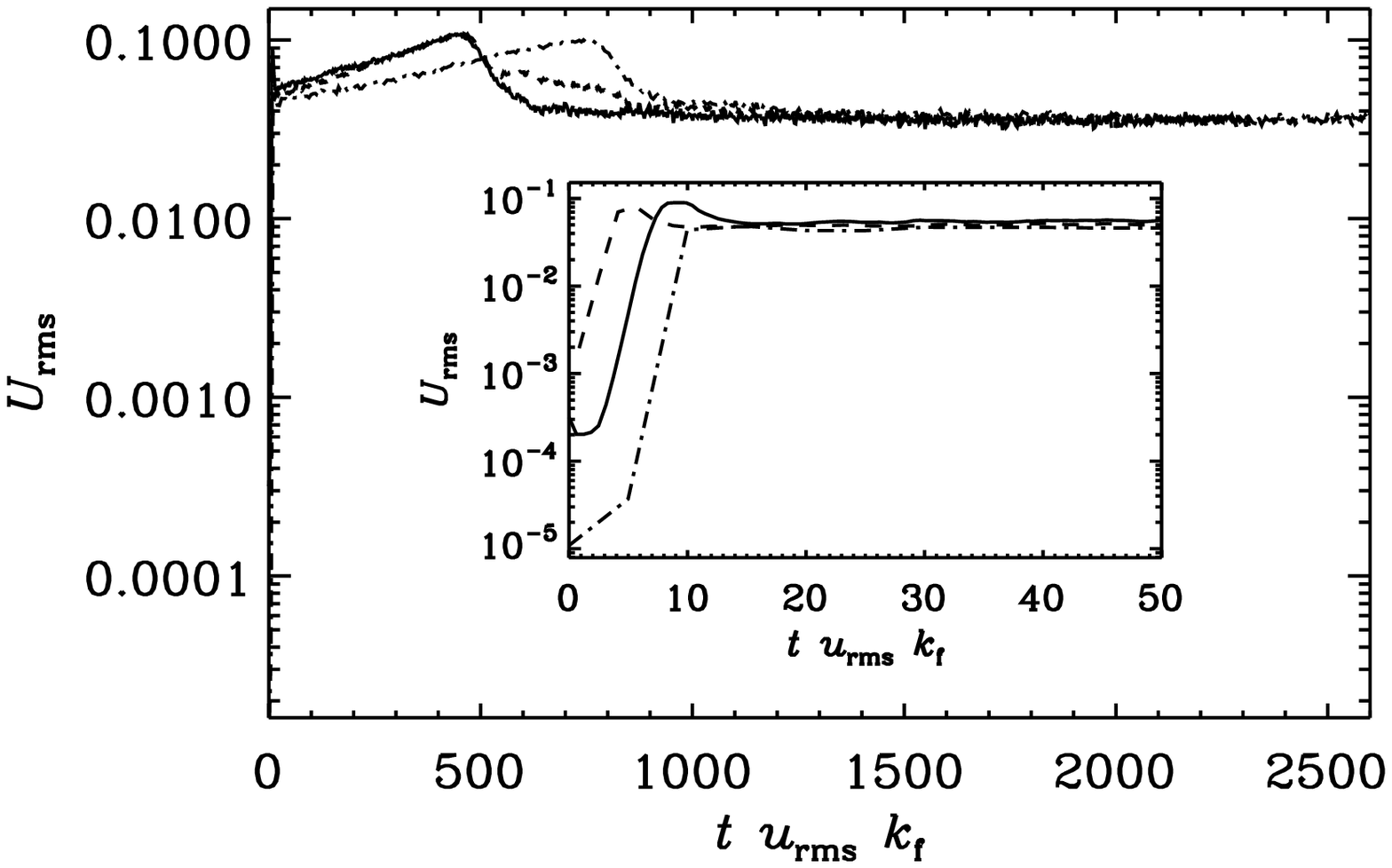}}
\resizebox{\hsize}{!}{\includegraphics{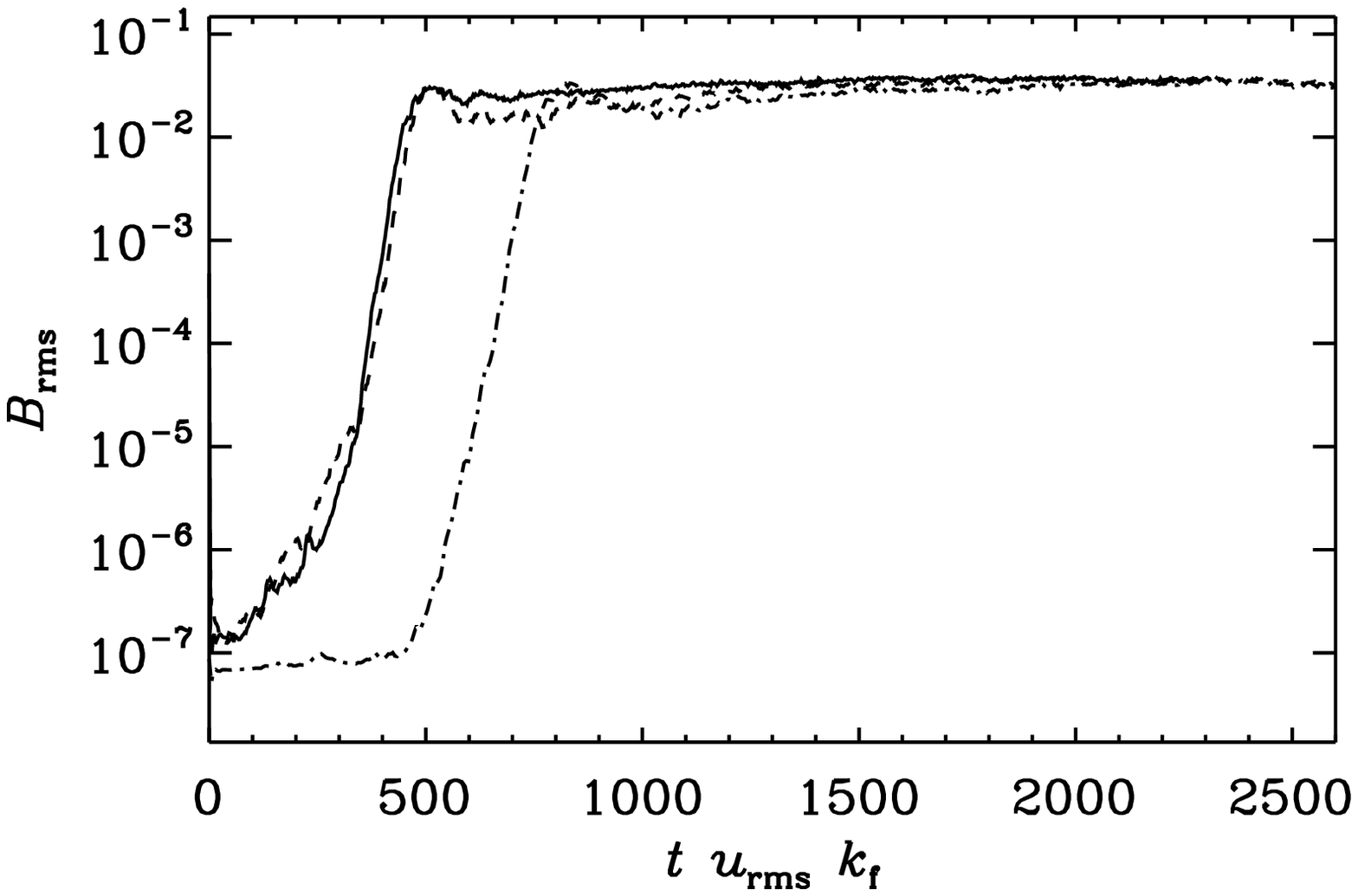}}
\caption{Root mean square velocity ({\it top}) and magnetic field
({\it bottom}) for Code~1 (solid line), Code~2 (dashed line) and 
Code~3 (dash-dotted line). The inset in the upper plot
shows the early time behaviour, highlighting the different
initial conditions that have been used.} 
\label{fig:figA1}
\end{figure}

\begin{figure*}[t]
\includegraphics[width=.32\textwidth]{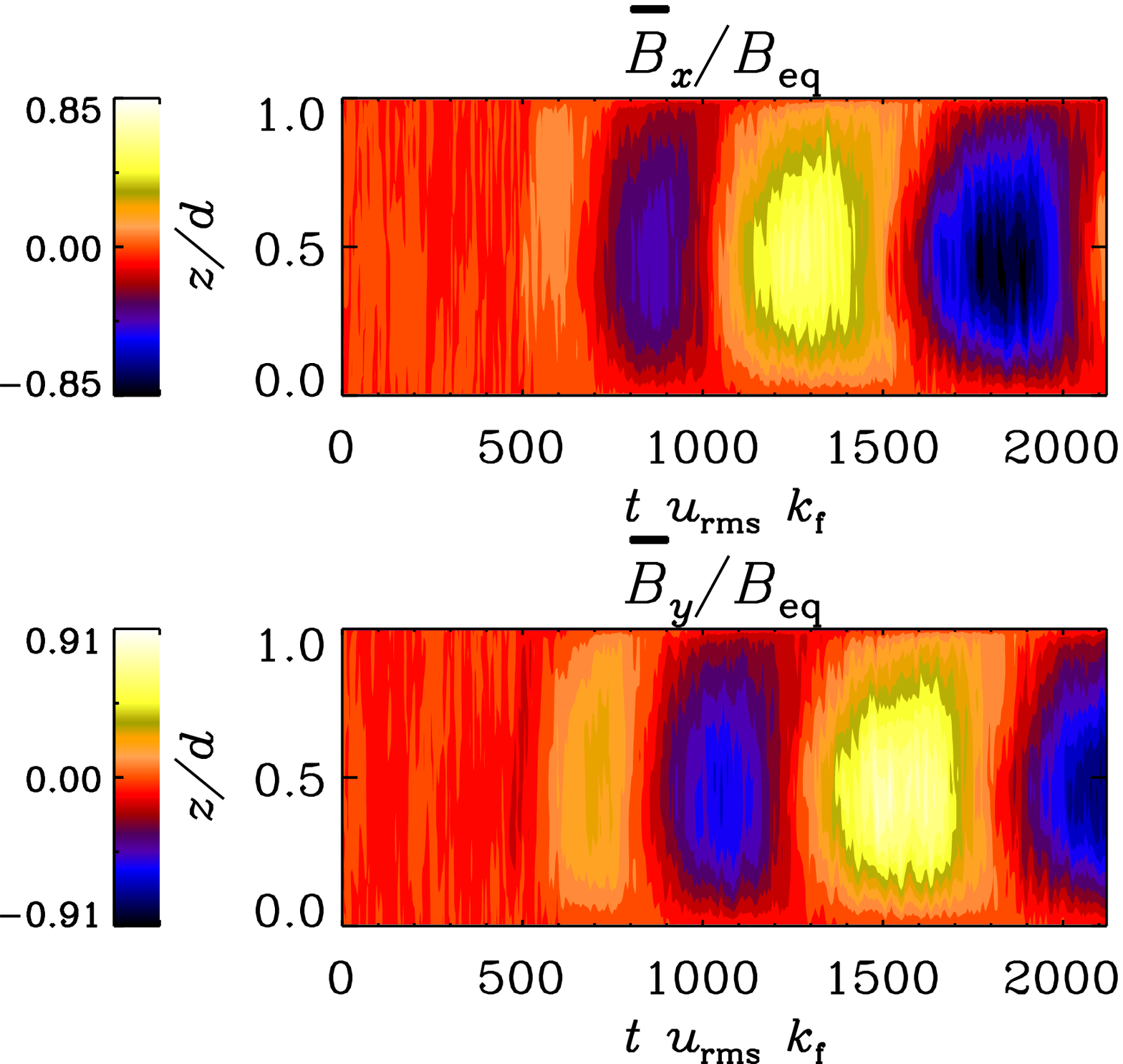}
\includegraphics[width=.32\textwidth]{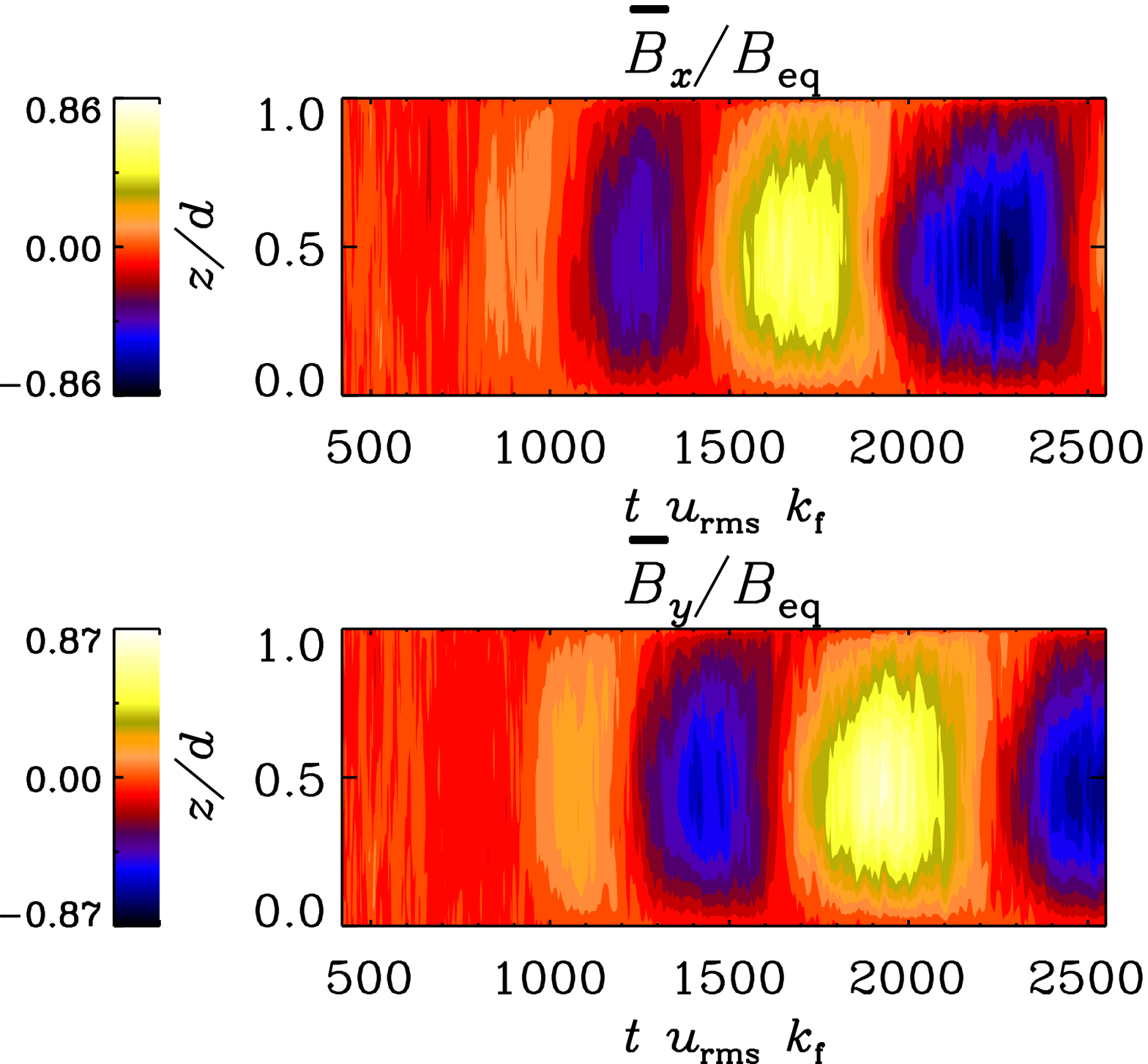}
\includegraphics[width=.32\textwidth]{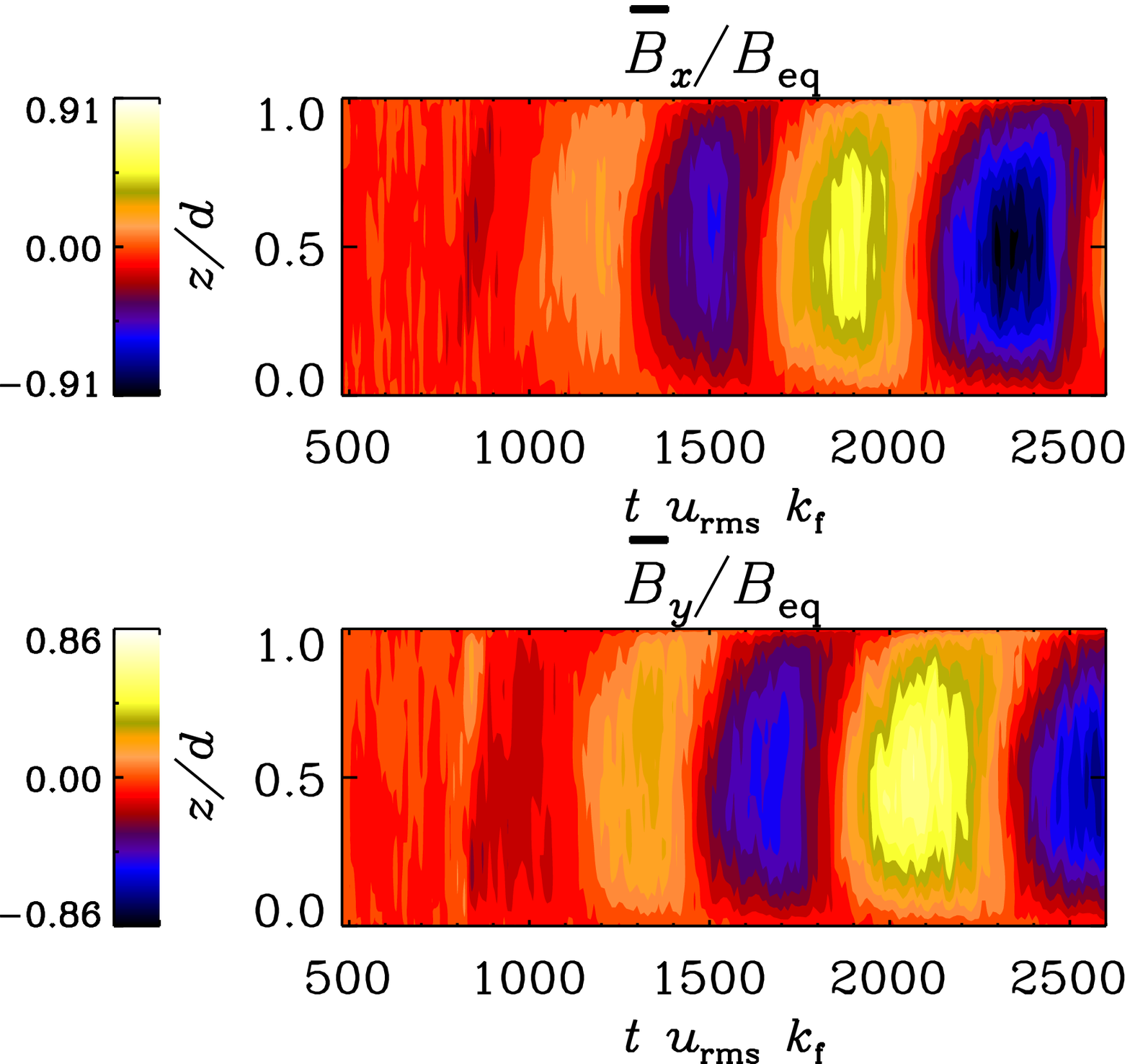}
\caption{Horizontally averaged horizontal magnetic fields (normalised by 
the equipartition field strength) as functions
of $z$ and time for Code~1 ({\it left}), Code~2 ({\it middle}) and Code~3 ({\it right}). 
Note that the time-axes for Codes~2 and 3 have been shifted by $420$ turnover times
and $480$ turnover times (respectively) for ease of comparability with Code~1.}
\label{fig:figA2}
\end{figure*}

\begin{figure*}[t]
\includegraphics[width=.32\textwidth]{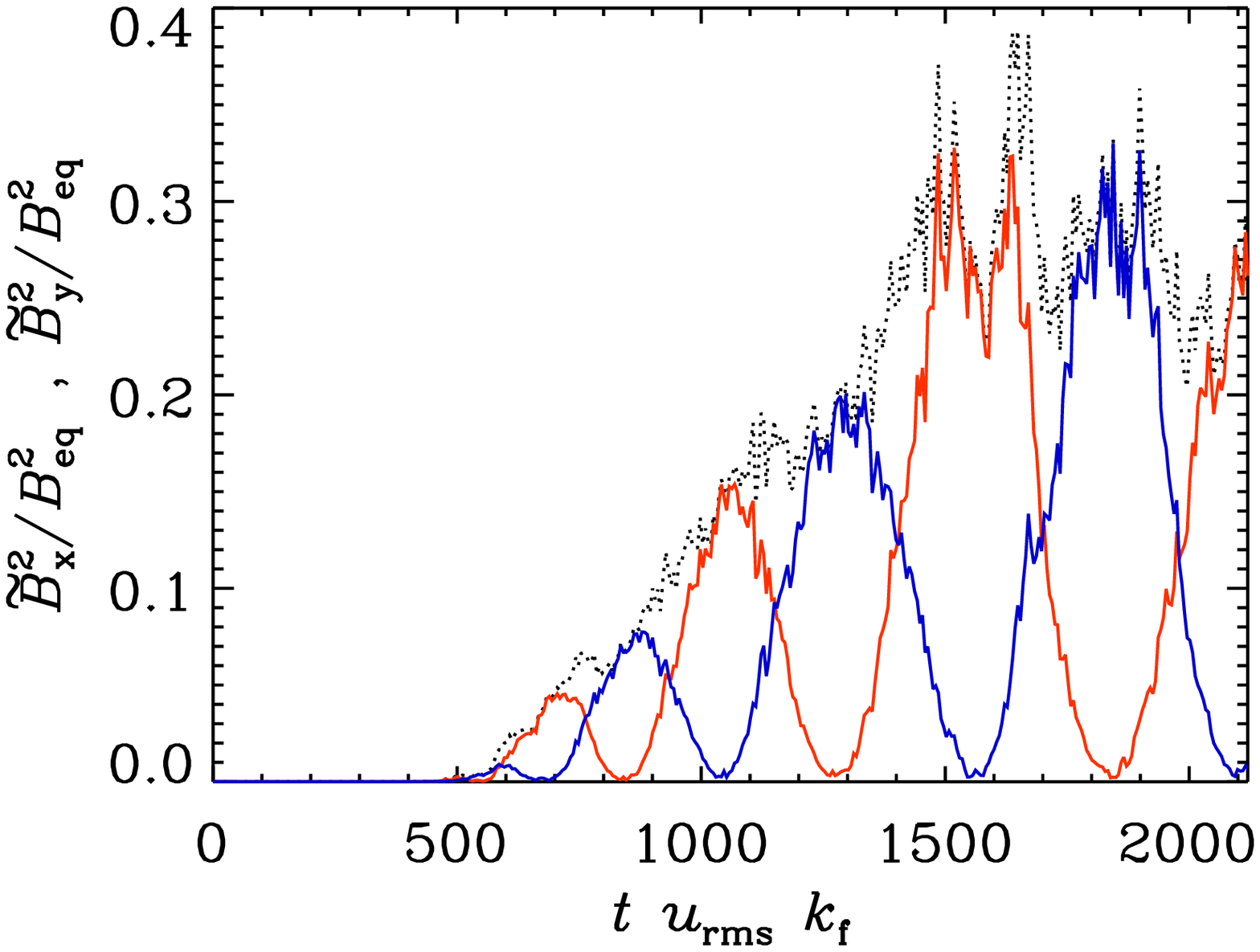}
\includegraphics[width=.32\textwidth]{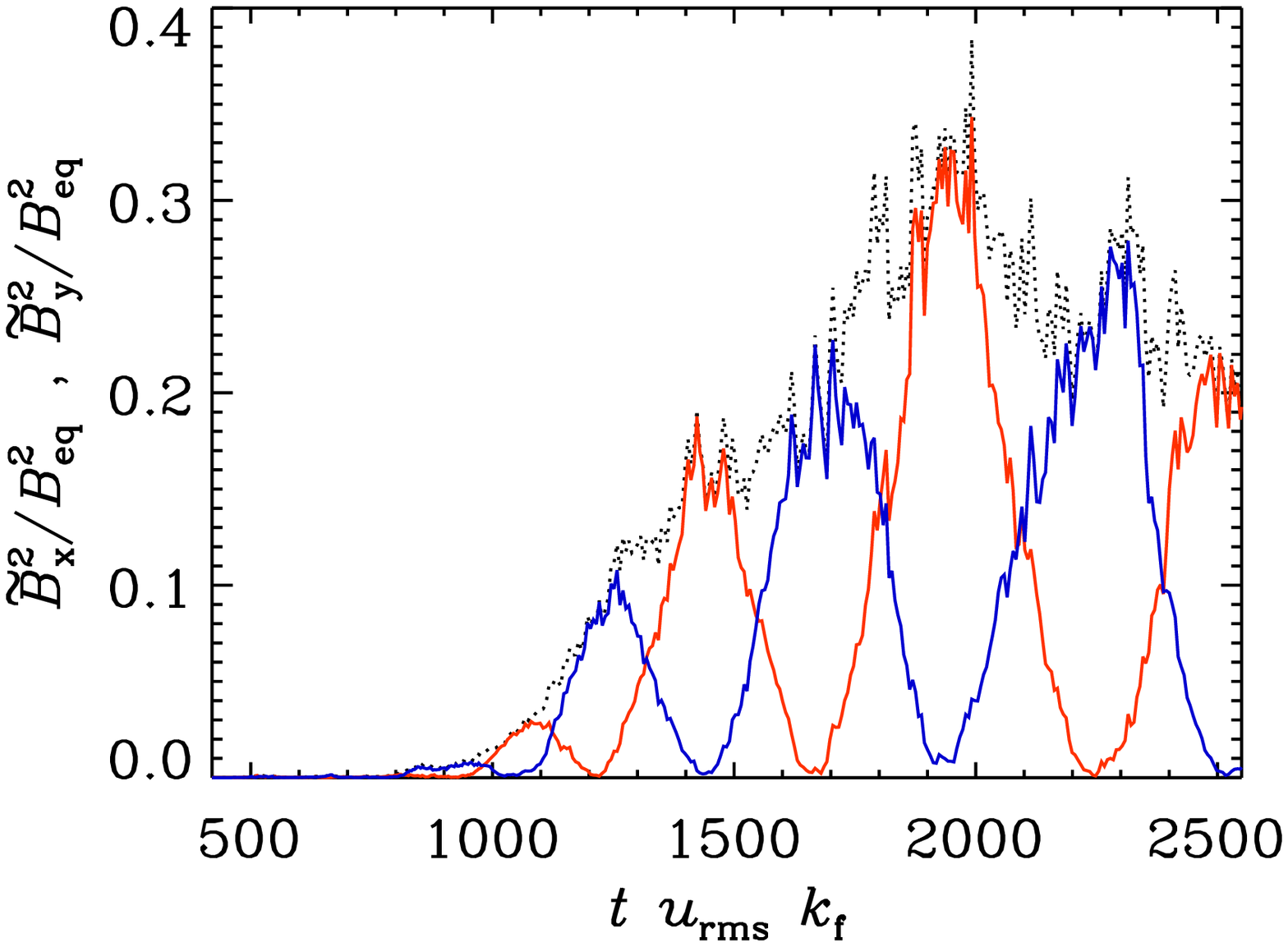}
\includegraphics[width=.32\textwidth]{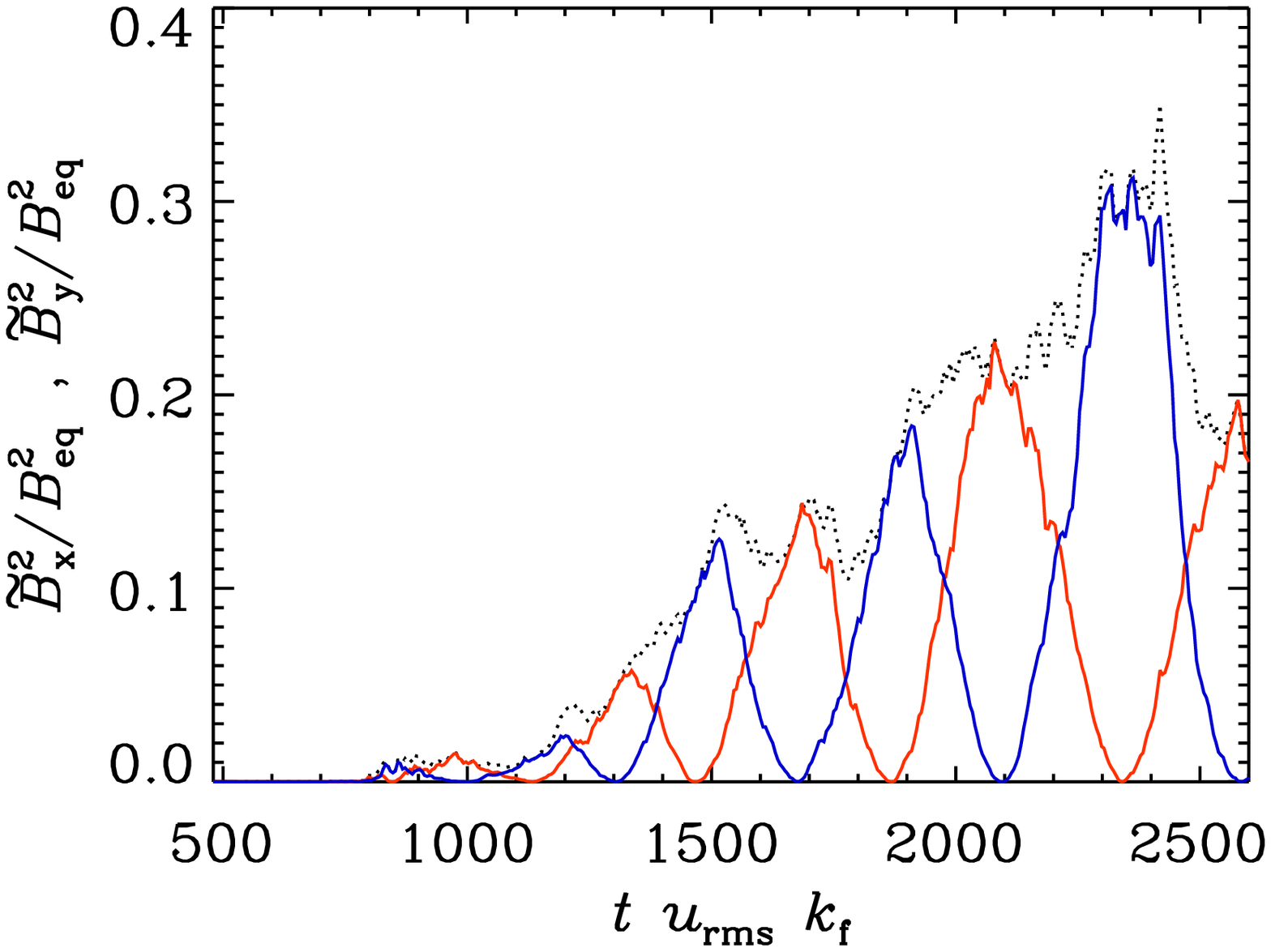}
\caption{Volume-averaged horizontal (squared) magnetic fields,
$\tilde{B}_x^2/\Beq^2$ (blue) and 
$\tilde{B}_y^2/\Beq^2$ (red), as functions of time for Code~1 ({\it left}), 
Code~2 ({\it middle}) and Code~3 ({\it right}). 
For each code, the black dotted line shows $(\tilde{B}_x^2+\tilde{B}_y^2)
/\Beq^2$ as a function of time.
Note that the sampling time for the mean fields is less frequent in the right-hand 
plot than in the other two.}
\label{fig:figA3}
\end{figure*}

\begin{figure*}
\includegraphics[width=.32\textwidth]{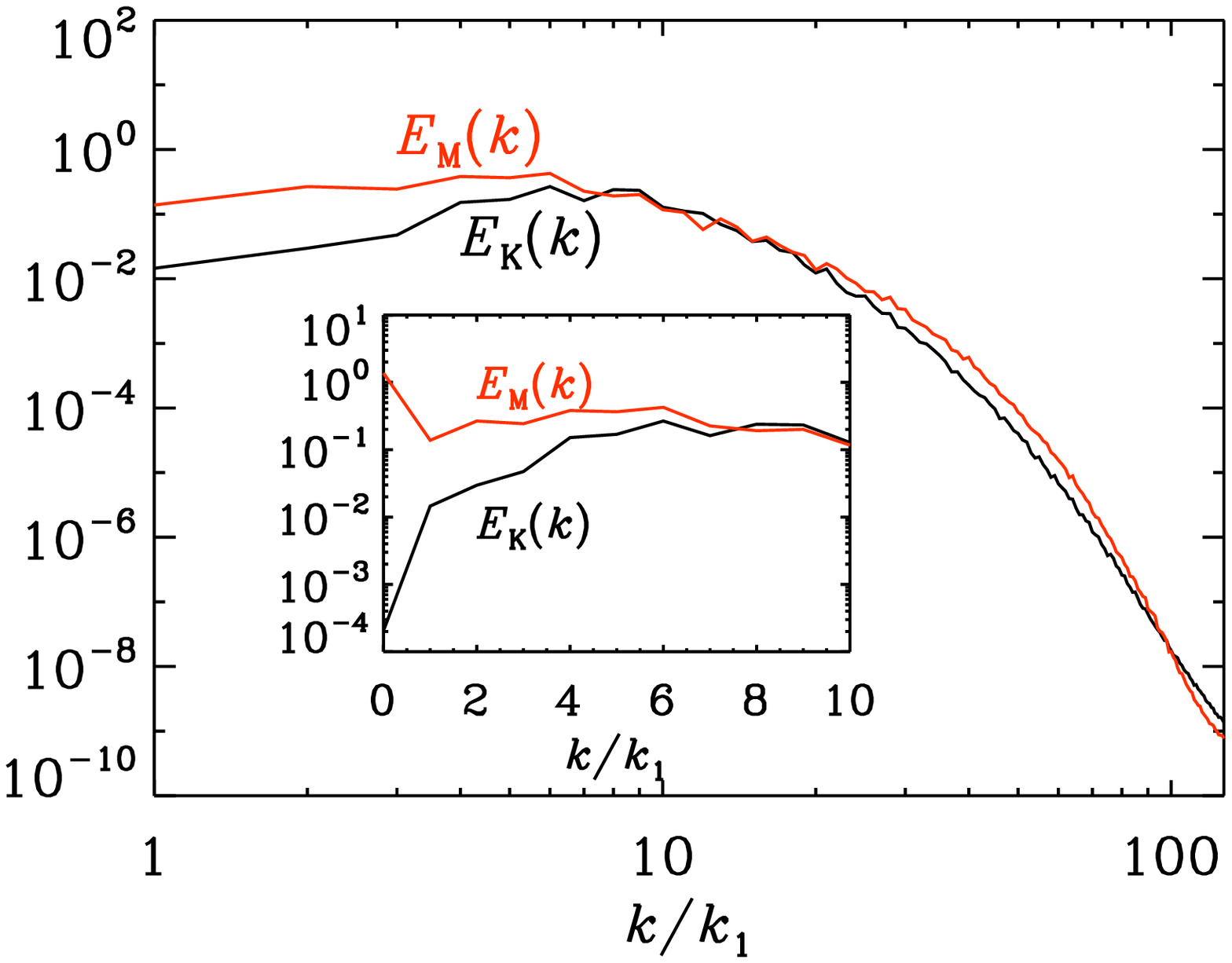}
\includegraphics[width=.32\textwidth]{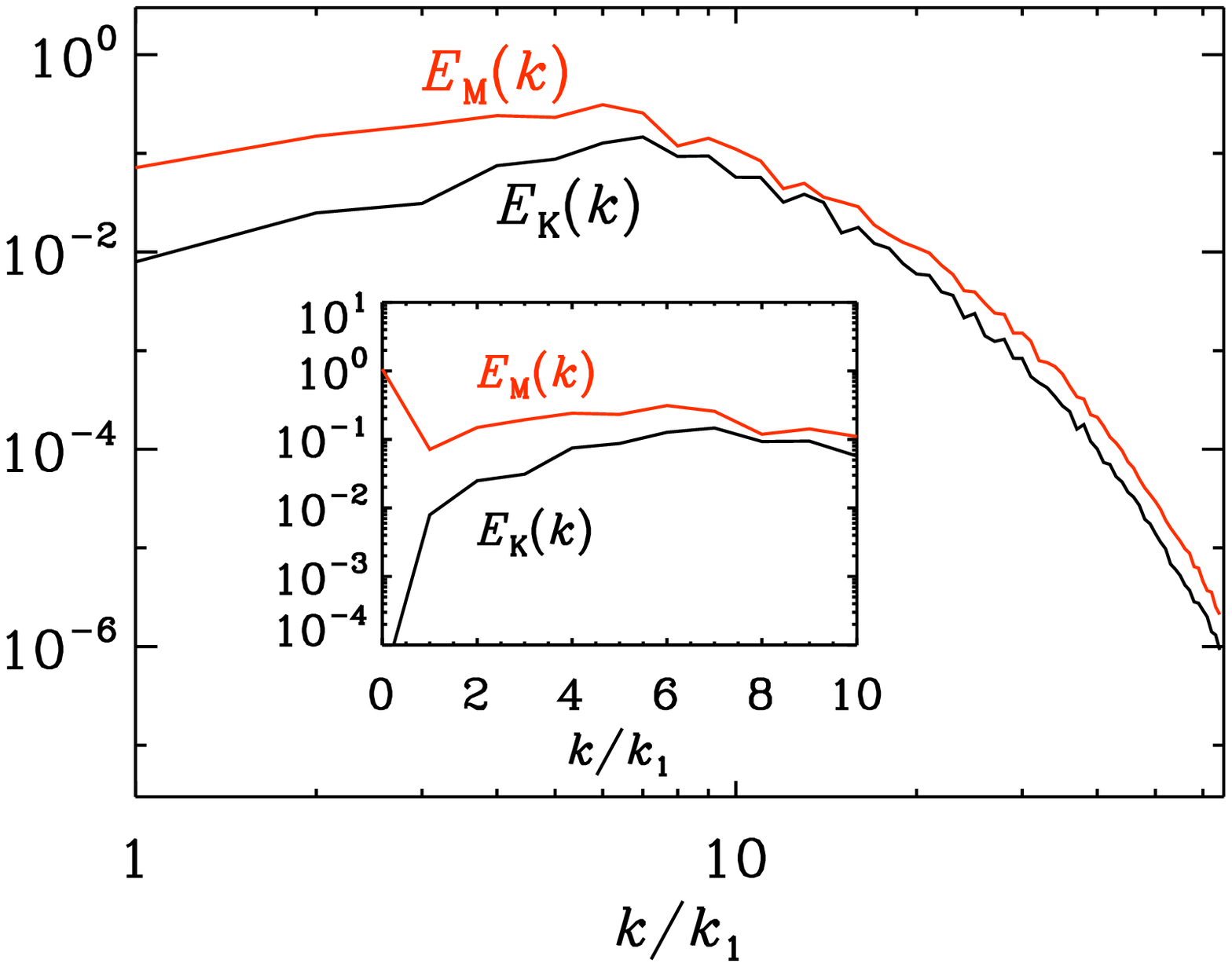}
\includegraphics[width=.32\textwidth]{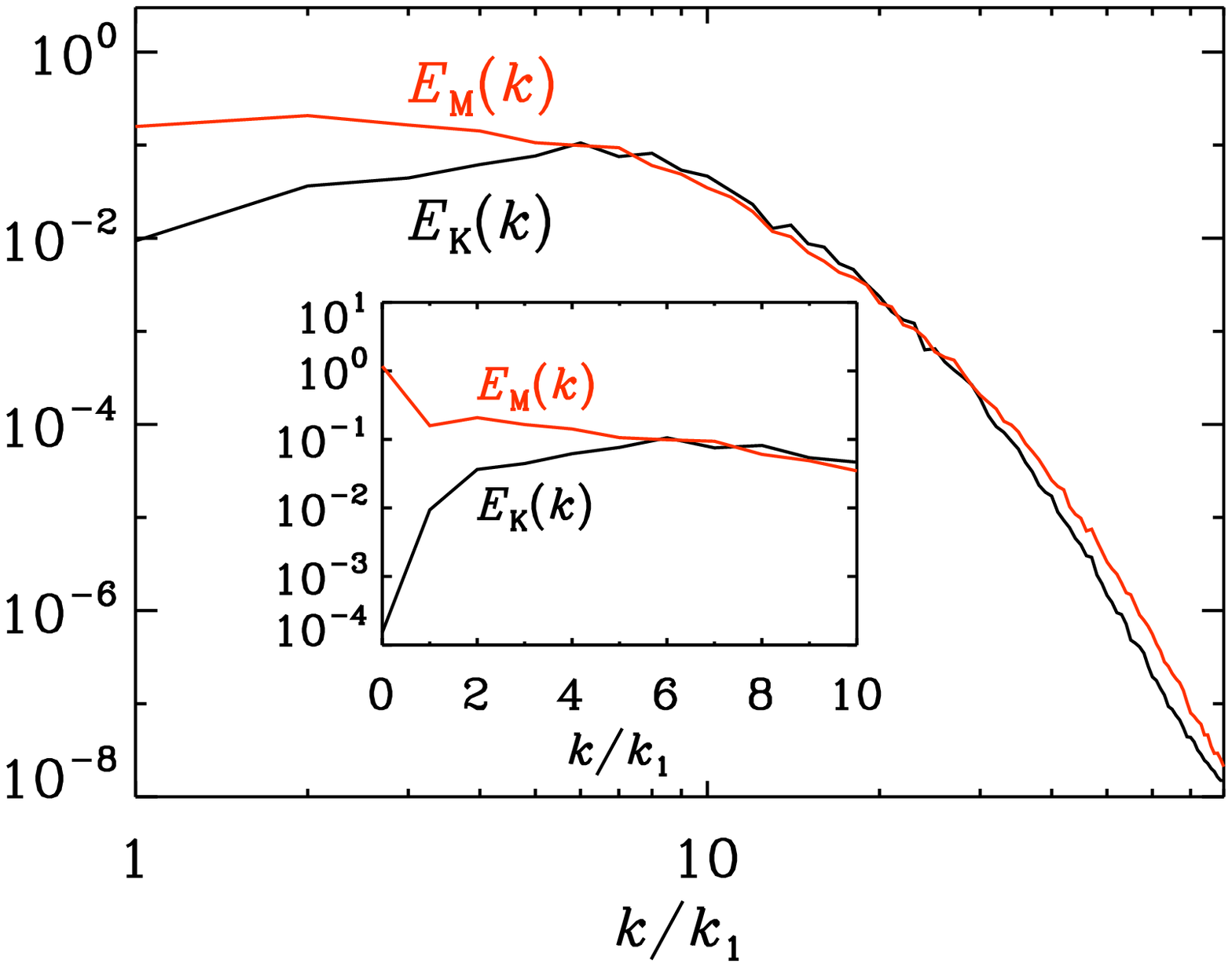}
\caption{Typical power spectra of velocity (black lines) and magnetic field
  (red lines) from the middle of the convective layer for Code~1 ({\it left}), Code~2 
  ({\it middle}) and Code~3 ({\it right}). In each case, the inset shows the smallest
  wavenumbers to illustrate that the magnetic field peaks at
  $k=0$. All spectra quantities are normalised by the average kinetic energy
  at $z_{\rm m}$ during the nonlinear phase.}
   \label{fig:figA4}
\end{figure*}
 
Figures~\ref{fig:figA1}--\ref{fig:figA4} show the evolution of this
system for each of the three codes. 
Figure~\ref{fig:figA1} shows the time evolution of the rms velocity
and magnetic field.
The time- and $z$-dependence of the mean horizontal magnetic
field components is illustrated in Fig.~\ref{fig:figA2}, whilst 
Fig.~\ref{fig:figA3} shows the corresponding vertically-averaged 
values. 
Finally, Fig.~\ref{fig:figA4} shows the mid-layer kinetic and
magnetic energy spectra at the end of the three runs.  
It is immediately apparent that all three codes are producing 
quantitatively comparable nonlinear dynamos. 
There are similar pronounced peaks at $k=0$ in the 
magnetic energy spectra, whilst the kinetic energy spectra all
have broad peaks at around $k/k_{\rm 1}=7$--$9$. 
The peak amplitudes of the large-scale horizontal
magnetic field components show similar levels of agreement 
(as indicated by Fig.~\ref{fig:figA3}).
Table~\ref{table:2} shows the (time-averaged) values of 
$\urms$ and $b_{\rm rms}$, 
the rms velocity and magnetic field (respectively),
from the nonlinear phase of the dynamo.
It is clear that there is quantitative comparability for both
of these quantities across the three codes.
Whilst there is some variability in terms of the measured
cycle periods ($970$, $1050$ and $1060$ convective 
turnover times), it should be stressed that there is some 
intrinsic variation in the cycle duration as the dynamo 
progresses. 
This is one of the main contributors to the error bars in 
Fig.~\ref{fig:period}, and the variation across the three
codes could simply reflect this variability.
This variability in the cycle period may complicate a 
future benchmarking exercise, but this can probably
be addressed satisfactorily by running longer 
calculations to average the cycle period over more
cycles.  

We should also note some other differences between the
three cases. 
As shown by Fig.~\ref{fig:figA1}, the initial large-scale 
vortex growth phase is longer in Code~3 (with a lower 
growth rate) than it is in the other two codes.  
This can be probably attributed to differences in the initial 
conditions: the initial rms velocity starts from a much lower level
in the case of Code~3, which apparently delays the onset
of the large-scale vortex instability. 
It is also worth noting that the seed magnetic field is weaker 
in Code~3 than it is in the others, so it takes longer to grow to 
a level where it is able to influence the flow.
This dependence upon the strength of the seed field clearly
indicates that it is the Lorentz force that is eventually
suppressing the large-scale vortex instability rather than 
a geometrical constraint due to the finite box size. 
Further differences can be seen in Figs.~\ref{fig:figA2}
and~\ref{fig:figA3}, where the large-scale dynamo emerges 
at different times for the different codes.
Again, this can be explained by differences in the initial 
conditions. 
The important point to note is that this  
large-scale dynamo is robust to small variations in the initial
configuration. 
This nonlinear dynamo solution is therefore an excellent
benchmark candidate.

\end{appendix}
\end{document}